\documentclass[10pt,preprint]{aastex}
\def\etal{{et al.~}}
\begin{document}
%\noindent
%\vskip .2in
%\baselineskip 20pt
\title{The Burst and Transient Source Experiment (BATSE) \\
Earth Occultation Catalog of Low-Energy Gamma-Ray Sources \\
Short title: BATSE Earth Occultation Catalog}
\author{B. A. Harmon\footnote{Present address: NE50, U.S. Department 
of Energy, 19901 Germantown Road, Germantown, MD 20874}, 
C. A. Wilson, \& G. J. Fishman}
\affil{NASA Marshall Space Flight Center, 
SD50, Huntsville, AL 35812} 
\email{colleen.a.wilson-hodge$@$nasa.gov}
%\vskip .2in
\author{V. Connaughton, W. Henze, \& W. S. Paciesas}
\affil{University of Alabama in Huntsville, Huntsville, AL 35899}
%\vskip .2in
\author{M. H. Finger, M. L. McCollough\footnote{Present address:
Harvard-Smithsonian Center for Astrophysics, 60 Garden Street, 
Cambridge, MA 02138}, M. Sahi, \& B. Peterson}
\affil{Universities Space Research Association, SD50,
NASA Marshall Space Flight Center, \\ 
Huntsville, AL 35812}
\author{C. R. Shrader}
\affil{Universities Space Research Association, 
Code 661, NASA Goddard Space Flight Center, \\ 
Greenbelt, MD 20771}
\author{J. E. Grindlay}
\affil{Harvard-Smithsonian Center for Astrophysics, 60 Garden Street,  
Cambridge, MA 02138}
\author{D. Barret}
\affil{Centre d'Etude Spatiale des Rayonnements, 9 Avenue du Colonel Roche, \\
31028 Toulouse Cedex 04, France}
\begin{abstract}

The Burst and Transient Source Experiment (BATSE), aboard the {\it Compton 
Gamma Ray 
Observatory} (CGRO), provided a record of the low-energy gamma-ray sky
($\sim$20-1000 keV) between 1991 April and 2000 May (9.1y).
%BATSE exhibited unique capabilities for studying a wide range of astrophysical
%phenomena, including gamma-ray bursts, black holes, pulsars, and solar flares.
BATSE monitored the high energy sky using the Earth occultation technique (EOT)
for point
sources whose emission extended for times on the order of the CGRO orbital
period ($\sim$92 min) or greater.  Using the EOT to extract flux information, a
catalog of sources using data from the
BATSE large area detectors has been prepared. The first part of the catalog 
consists of results
from the all-sky monitoring of 58 sources, mostly Galactic, with intrinsic 
variability 
on timescales of hours to years.  For these sources, we have 
included tables of
flux and spectral data, and outburst times for transients.
Light curves (or flux histories) have been placed on the world wide web. 

We then performed a deep-sampling of these 58 objects, plus a selection of
121 more objects,  
combining data from the entire 9.1y BATSE dataset.  Source types considered 
were primarily accreting binaries, but a small number of representative
active galaxies, X-ray-emitting stars, and supernova remnants were also included.
The sample represents a compilation of sources 
monitored and/or discovered with 
BATSE and other high energy instruments between 1991 and 2000, known sources  
taken from the HEAO 1 A-4 \citep{Levine84} and
\citet{MacombGehrels99} catalogs. The deep sample results include definite detections of
83 objects and possible detections of 36 additional objects.  The  
definite detections spanned three classes of sources: accreting black hole
and neutron star binaries, active galaxies and 
supernova remnants. The average fluxes measured for the fourth class, the X-ray emitting
stars, were below the confidence limit for definite detection.
%We excluded EGRET sources from 
%the Macomb \& Gehrels (1999) catalog due to the relatively large uncertainty 
%of reported locations. 
%We included transient sources discovered with 
%other X-ray monitors operating in the 
%CGRO era such as the {\it Rossi X-Ray Timing Explorer} (RXTE) All-Sky Monitor
%and {\it BeppoSax}, and searched for previous outbursts in the BATSE database.

Flux data for the deep sample are presented in four energy bands: 20-40, 40-70,
70-160, and 160-430 keV.  The limiting average flux level 
(9.1 y) for the sample varies from 3.5 to 20 mCrab (5$\sigma$) between 20
and 430 keV, depending on systematic error, which in turn is primarily 
dependent on the sky location.
%The best sensitivity is achieved for sources outside the Galactic center 
%region.  
To strengthen the credibility of detection of weaker sources 
($\sim$5-25 mCrab), we generated Earth occultation
images, searched for periodic behavior
using FFT and epoch folding methods, and critically evaluated the 
energy-dependent emission in the four flux bands. 
%for a subset of sources. 
The deep sample results are
intended for guidance in performing future all-sky surveys or pointed 
observations in the hard X-ray and
low-energy gamma-ray band, as well as more detailed studies with
the BATSE EOT. 
%Finally, an estimate of the number of transient binary
%systems in
%the Milky Way is made using the results of a search for previously unknown 
%transient activity.
%The search limits are for transients to an intensity (duration) of 
%$\sim$35 mCrab ($\ge$30d) and $\sim$80 mCrab ($\ge$5d) in the 20-100 keV band.
%These results are compared to a Galactic population model. 
\end{abstract}


\section*{}
\twocolumn
\section{Introduction}
The {\it Compton Gamma Ray Observatory} 
(CGRO)\footnote{Acronyms and abbreviations used throughout the text and tables
are listed in Appendix A.}, one of the four missions in
NASA's ``Great Observatories" series, was launched
in 1991 April and operated in
low Earth orbit until its controlled re-entry
in 2000 June.  CGRO was 
responsible for many discoveries in the study of gamma-ray bursts,
accreting binaries, active galaxies, and pulsars \citep[see general
reviews by][]{GehrelsShrader97, KniffenGehrels97, LeonardWanjek00}.  The 
quest for the origin of 
gamma-ray bursts \citep{Meegan92, FishmanMeegan95}
led to the development and flight of the Burst and Transient Source
Experiment (BATSE) on
the CGRO. 
%BATSE pointed the way to the extragalactic origin of gamma-ray bursts
%through mapping of burst location and number-brightness distributions
%(Meegan \etal 1992).
In addition to BATSE's primary science goals, its nine years of nearly
continuous operation
and all-sky capability allowed monitoring of the low-energy gamma-ray
sky using the
Earth occultation technique (EOT). Several new objects 
were discovered in the BATSE dataset with the EOT, and
numerous outbursts of transient and periodic sources were
detected during the mission.  Here we present a compilation of BATSE 
EOT observations that can be used as the basis for more detailed
studies on individual sources, for which it is important
to have a knowledge of the long-lived emission in the BATSE energy range.
It is our hope that these results will stimulate and guide
future observations of the low-energy gamma-ray sky.
In this work, we performed a careful assessment of 
systematic error in deriving results for the entire mission.
The intent was to push the instrumental and technical limits of 
the present application of the BATSE Earth occultation technique
(EOT). These data are made available on-line 
to other investigators through the High Energy Astrophysics Science Archive 
Research Center (HEASARC) at NASA Goddard Space Flight Center, 
the web site of the BATSE
Earth Occultation Team at the NASA Marshall Space Flight Center, and 
the Compton Observatory Science 
Support Center (COSSC).

To make the presentation of this work
tractable, we divided the catalog into two parts.  First, we emphasized
the all-sky monitoring capability of BATSE. For the bright sources that vary 
on timescales greater than a few hours, we give an overview of activity and 
representative
flux and spectral behavior.  We also briefly make note of the science
investigations performed with the EOT during the mission.
Second, we performed a ``sky survey" of 
a comprehensive set of low-energy gamma-ray sources.  
We note that this is not a true survey, in that it is 
biased by preselection of sources.  
The emphasis is on obtaining good quality results along the 
Galactic plane where accreting sources dominate the  
low-energy gamma-ray sky.

We note that other missions with all-sky or wide-field coverage 
were operational between 1991 and 2000. A non-exhaustive list includes the   
Japanese {\it Ginga} satellite, with its All-Sky Monitor, functioning between
1987 February and 1991 October \citep{Tsunemi89}, and covering the bandpass
of 1$-$20 keV, the scanning and
imaging instruments aboard the French-Russian spacecraft {\it Granat} 
\citep{Brandt90, Roques90, Sunyaev90, Paul91},     
which operated between 1989 December and 1997 October, 
and the U.S. {\it Rossi X-Ray Timing Explorer} (RXTE) mission, with its
All-Sky Monitor (2-12 keV) \citep{Bradt93, Levine96},
which began operation in 1995 December and remains operational. {\it
Gra\-nat}/Sig\-ma
\citep{Roques90}, a
coded mask, sodium iodide imaging detector system, provided considerable
coverage and detailed images of the central region of the Galaxy in the 
35$-$1300 keV bandpass. There were also scanning instruments of common 
design on board 
{\it Granat} and Danish {\it EURECA} spacecraft known as 
WATCH \citep{Lund86, Brandt94, Castro94} that monitored a portion of the X-ray 
sky in the 5-120 keV band.  The powerful Wide Field Cameras \citep{Jager97} 
on the Italian-Dutch {\it BeppoSax} satellite 
(1.8-28 keV) provided coded aperture imaging for both surveying
and monitoring of large areas of the sky.
BATSE complemented these instruments by 
full-sky monitoring of the energy
range from 20 keV up to about 1 MeV.  Much potential remains for 
investigations using combinations of these datasets. 

%This covers the lower energy portion of the
%nonthermal regime, where the emission is produced by processes such as 
%Compton up-scattering of soft photons by energetic particles, bremsstrahlung, synchrotron
%radiation and nuclear decay.  Although we have a reasonably good
%understanding of the radiation processes that produce high energy emission, it remains a
%challenge to reconcile the physical changes that occur in 
%accretion disks, coronal plasmas, or the intense magnetic fields near a neutron star,
%with the highly variable and predominantly nonthermal emission that is readily seen. 

BATSE served as an all-sky monitor using the Earth occultation technique
\citep{Harmon02} between 1991 April 23 and 2000 May 28 ($\sim$9.1y).  This capability was 
used to trigger numerous pointed observations with the CGRO and other
observatories. 
We extracted Earth occultation measurements from the 
BATSE dataset at a 
maximum rate of twice per 92m orbit of CGRO in sixteen energy 
channels between 20 keV and 1 MeV.  Intensity data for any part of the
sky, subject to the geometric constraints of occultation, were available 
for the duration of the mission with the exception of a few brief intervals  
(see Appendix B).  Although the BATSE EOT was limited in sensitivity 
by high backgrounds
and modest spatial resolution ($\sim$1$^{\circ}$), we
could investigate intensity and spectral variations over a dynamical range from 
days to years. Although
$\sim$daily timescales have been accessible with pointed instruments, BATSE was
the first all-sky monitor since those of {\it Vela 5B} \citep{Conner69} and 
{\it Ariel 5} \citep{Holt76}, to cover the upper end of this time range. 

%We also performed a search for transient 
%activity, and with some 
%assumptions on efficiency of time and spatial coverage, 
%make source population estimates.

On the other hand, no low-energy gamma-ray full-sky surveys have been performed since the pioneering 
HEAO 1 A-4 experiment in 1977-79 
\citep{Levine84}.  That experiment covered the energy range between 13 and
180 keV, and resulted in the detection of about 70 sources (7 of them extragalactic).  
Such a survey with BATSE has 
been attempted by our group in the past
\citep{Grindlay96, Harmon97} using the Earth occultation 
imaging technique \citep{Zhang93, Harmon02}. This did not prove 
to be an efficient way to
survey the sky due to the limitations on computer time combined with
residual features and the small size (a few degrees) of the images. 
A new effort, incorporating more sophisticated background modeling techniques
developed for the {\it International Gamma-Ray Astrophysics
Laboratory}(INTEGRAL)\citep{Lei99} and detector point spread functions for limb 
geometry, holds much 
promise for achieving this goal \citep{Shaw01, Shaw03}. In Fig. 1 we compare the 
sensitivity of the BATSE deep sample (this work) with that of the HEAO 1 survey and 
a proposed future low-energy gamma-ray sky survey mission  
%{\it Swift} (Barthelmy \etal 2001) and 
the {\it Energetic X-ray Imaging
Survey Telescope} (EXIST) \citep{Grindlay01, Grindlay03}. 
The binning of the BATSE sensitivity curve  
reflects the maximum energy resolution obtainable with the EOT.

In addition to the list of monitored sources accumulated during the operation 
of BATSE,
we reviewed the HEAO 1 A-4 hard X-ray \citep{Levine84} catalog and the recent 
gamma-ray source compilation of \citet{MacombGehrels99} for completeness of the
known high energy Galactic sources, and performed
a literature search of other high-energy 
objects above the
BATSE sensitivity threshold.  We did not include the EGRET ``unidentified" sources
from \citet{MacombGehrels99} due to the probable extragalactic nature of many of them, and the 
relatively large uncertainty in location ($\sim$1$^{\circ}$) that would strongly
affect the measured flux using the EOT.  
%We also compared our list of gamma-ray selected 
%objects to the exhaustive catalogues of low mass (LMXBs) (Liu, van Paradijs \&
%van de Heuvel 2001) and high mass X-ray binaries (HMXBs) (Liu, van Paradijs \&
%van de Heuvel 2000).  We find that 94 of the 150 LMXBs and \_\_ of the 
%130 HMXBs in these catalogs are included in this work. Most of the omitted binary sources are not 
%known to be gamma-ray emitters; however, a sky-location dependent 
%systematic error is derived to account for unresolved point source emission
%(see Sections 4.1 and 4.2).  

The sources addressed
in this catalog are shown on a sky map in Galactic coordinates in Fig. 2.
We have included 179 sources in our deep sample of the sky.
The majority of sources are accreting binaries; however, we include 
the brighter sources from the other object classes for comparison.  
%We do not
%include new candidate sources discovered with the transient search on this map.
Of the 179 sources, most (76\%) are accretors, i.e., 
black hole or non-pulsed neutron star binaries 
(97), or binaries containing pulsars (41). Again we stress
that our deep sampling of the known gamma-ray sources should not be considered
as a true background-limited survey to a 
given brightness threshold. 
However, by virtue
of our comparisons to HEAO 1 A-4 survey data, we believe that 
we have a reasonably complete sampling of high
energy accreting systems in the Milky Way Galaxy down to an average
flux level of $\sim$30 mCrab (20-100 keV). 

We also note that to date, samples of 
objects have been performed with BATSE 
for active galaxies \citep{Connaughton99, Malizia99, 
Bassani00}, supernova remnants \citep{McCollough97}, globular clusters 
\citep{Ford96}, X-ray stars \citep{White94},
and prompt emission from nearby novae \citep{Hernanz00}.  The most
promising classes of sources, other than accreting binaries, 
for which BATSE has shown positive detections, are active galaxies, such as 
Seyferts and blazars, 
and supernova remnants. As part of our 179-source sample, we
included selected supernova remnants (17) (some of which contain pulsars 
or soft gamma-ray repeaters (SGRs)),
active galaxies (12), and X-ray emitting stars (12), that overlap these other
studies. 

In Sec. 2, we give a brief overview of BATSE and the use of the Earth 
occultation technique.  We then discuss
production details for the catalog. In Sec. 3 we present 
highlights of the all-sky monitoring effort,
and give an overview of science obtained for
the different classes of detected sources. Sec. 4
presents the deep sample results and how they are organized.  
We describe a sky-dependent systematic error model and apply it 
to the measured source fluxes.  We then examine the consistency
of source detections by focusing on some of the weak sources in
the few mCrab intensity range.  This is done by testing with methods 
that are independent
of the occultation step fitting routines.  Finally, in Sec. 5, 
we briefly discuss applications of the catalog data.

%The unabsorbed high energy emission
%from these objects can potentially settle questions about the
%nature and populations of these objects, and a deep survey is therefore 
%very important.  As we show in our assessment of systematic error, the relatively quiet 
%sky in the uncrowded regions outside 
%the galactic bulge facilitates surveys of high latitude sources. 

\section{Analysis Methods and Treatment of Point Source Data}

\subsection{The BATSE Instrument and the Earth Occultation Technique}

%Pre-launch discussions of the method can also be
%found in Paciesas \etal (1985) and Fishman \etal (1989). Discussions 
%concerning the
%post-launch performance and expanded capabilities can be found in Harmon
%\etal
%(1992), Wilson \etal (1992), Zhang \etal (1993a), and Zhang \etal (1994a). 
%Preliminary
%compilations of detected sources can be found in Harmon \etal (1993a) and Robinson
%\etal (1997).  
%An imaging capability was added after launch, based on the
%Radon transform, which greatly improved our ability to locate
%and identify gamma-ray sources (Zhang \etal 1993a, 1994a).  

The Earth occultation analysis technique 
utilizes the large area detectors (LADs) on BATSE, which are
sensitive to photons above 20 keV.  The {\it Compton Gamma Ray Observatory}
including placement
of the BATSE detector modules on the spacecraft is illustrated 
pictorially in Fig. 3.
The LADs are composed of sodium iodide (NaI(Tl)) crystals, 1.27 cm (0.5 in) thick by
50.8 cm (20 in) across (2025 cm$^{2}$ total area of one detector).  Eight 
modules 
are mounted on
the corners of the CGRO with normal vectors perpendicular to the faces
of a regular octahedron.  Any point on the
sky can viewed by four detectors at angles less than 90$^{\circ}$ to
the source direction.  
%This gives BATSE a
%capability of obtaining crude locations to within a few  degrees using the 
%count rates from a combination of detectors and the known response
%of each detector from photons directly from the source, and source photons
%scattered off
%the atmosphere (Pendleton \etal 1995a).  This is routinely done to 
%locate gamma-ray bursts to within an accuracy of a few degrees 
%(Pendleton \etal 1999).   
For the EOT, the measurement sensitivity
of the LADs is maintained by combining statistics from two, three and 
ultimately four
detectors at successively larger angles from any one of the detector normals.  
A full description of 
the BATSE detectors and the experiment can be found, for example, 
in \citet{Fishman84, Fishman89a, Fishman89b}. In a companion paper 
to this catalog \citep[see also references therein]{Harmon02}, we 
discussed the use of the EOT
in various forms, and described in detail 
a number of aspects of the
technique such as source identification, sensitivity, and systematic error.
The Earth occultation technique was also developed and applied to BATSE data in
a separate, parallel effort at the Jet Propulsion Laboratory (JPL) 
\citep{Skelton94, Ling96, Ling00, LingWheaton03}.  
We discuss our results in comparison to this work in Appendix C.  
%Another powerful 
%application of BATSE as an all-sky monitor was the ability to
%to detect and monitor pulsars using Fourier
%analysis or epoch folding techniques. A detailed description of these
%methods and results can be found in Bildsten \etal (1997). 

In one orbit around the Earth,
two occultation step features, a rise and set pair, will be superimposed on
the background counting rate as each point source is occulted.  
A measurement can be made of the intensity of a source in each energy
channel at rise or set.  In practice,
two measurements per orbit were not always achieved.  The most common
reasons
were passages through the lower Van Allen radiation belt at the
South Atlantic Anomaly when the detector high voltage was off,
or that CGRO was out of line-of-sight contact with the NASA Tracking \& Data
Relay Satellites,
or data were flagged and not available for analysis.  In addition, high
declination
sources ($\mid\delta\mid\geq$41$^{\circ}$) experience
an interruption of occultations near the orbital poles
\citep[see][]{Harmon02}, and source confusion,
where occultations of one
source are indistinguishable from another, limits the number
of usable occultation steps.  The combined impact of these effects caused 
Earth occultation coverage averaged over one precession cycle ($\sim$52d)
to range between 80\%-90\%, at best, and at worst, about 50\%. 
%The attenuation of gamma-rays by the Earth's atmosphere and the
%variation in thickness of the air mass along the line of sight to the
%X-ray source produce the step-like features in the detector count rate as
%a function of time.  The attenuation is 50\% for 100 keV photons that 
%pass through the atmosphere along a line of sight with minimum altitude of
%70 km.
%For a typical orbital speed of 8 km/s, the duration
%of the occultation step for a source rising or 
%setting in the plane of the spacecraft orbit is about 10 seconds.
The count rate for a source in the LAD was
extracted by
simultaneously fitting occultation step features with terms
for each source in the fit and a quadratic polynomial to represent
the detector background.  The fit was performed independently for each
LAD and in each energy channel.   

Occultation steps (lasting $\sim$10s)
are relatively sharp features superimposed
on the slower background variations caused
by orbital motion around the Earth. 
We assumed that the background was smooth and adequately
fit by a second order polynomial.  Occasional failures of this assumption were usually
caused by the presence in the data of bright pulsars, weak bursts,
solar flares, and other disturbances on the timescale of the fitting window. 
Generally, in the absence of unusual background fluctuations, we have found that 
the LADs were consistently reproducible in the measured fluxes based on 
tests using the Crab Nebula 
fluxes in a given detector, angle and energy bin.  The only exception to this was 
the late mission (approx. last two years) loss of gain stabilization in LAD 7.  
We took this considerably further, however, to reduce measurement 
inconsistencies between LADs that might be intermittent (unusual background
fluctuations), and LAD-response related (time-in\-de\-pen\-dent effects).  We did this by 
(1) performing a flagging procedure for outliers 
at the single step level \citep[see][Sec 2.2, page 152]{Harmon02}, and also outlined in Sec 2.3 
of this manuscript), and 
(2) developing an empirical correction to the detector response based on the Crab data 
in channels 1-10 in all eight LADs at angles between
0 and 90 degrees \citep[see][Sec 3.5, pages 167-170]{Harmon02}. 
All light curves and spectra in the catalog were subjected to these 
two procedures except for a few sources as noted in Sec. 2.3.  Any remaining non-Poissonian
components in the data must then be accounted for in estimating systematic errors,
which we have done on an average basis using a grid-interpolation procedure
as discussed in Sec 4.2.  More details of the properties of
Earth occultation features and how the EOT was used can be
found in \citet{Harmon02}.

\subsection{Treatment of Catalog Data}

A raw count rate history for each point source was accumulated from fitting 
occultation steps on successive days (usually about 15-30 steps per
day). The instrument response was deconvolved in a following
step to produce intensities
in units of photon number, energy flux, or units
relative to a reference source, such as the Crab Nebula. 
%The octahedral 
%geometry of BATSE insures that at least four detectors
%Furthermore,
%the natural timescale for which spectral measurements are most easily obtained
%without changing detector combinations is the
%observation, or pointing period, of CGRO.  Observations, where the orientation
%of the LADs were kept fixed with respect to celestial coordinates,
%lasted two to three weeks.   The specific orientation of the X and
%Z axes, which were set by the CGRO observing schedule, determined the
%LAD combination for viewing a point on the sky.  
The process of generation and analysis of the occultation count 
rate histories is shown schematically in Fig. 4.  The raw light curves, 
uncorrected for detector response, were
generated with a software package known as
OF\_W.  The code carries or collects all information on source
location, spacecraft orientation and trajectory, detector gains, and
a limited amount of data on relative source intensity, required to 
extract source count rates
from occultation step features
in the BATSE CONT (2.048s, 20-1000 keV in 16 energy channels) data
\citep[see][for a more complete description of datatypes]{Fishman89a,Fishman89b}.

In Table 1, we show the standard input
parameters for OF\_W.  The definitions of these parameters are: (1) 2$\tau$,
length of the background data sample in seconds.  The data that define
the {\it fitting window} were selected
to be symmetric about the occultation step of the source of interest (SOI).    
From an analysis of 
relative size of systematic error to the statistical error using the 
second-degree polynomial for the background \citep{Shaw01}, a value for
$\tau$ of $\sim$110s was found to be optimal.  
We also allowed an additional non-adjustable time of 20s for
the width of the occultation step itself. (2) $\theta$$_{cut}$
is the maximum detector aspect angle of a source (i.e., the angle between the detector
normal vector and the vector toward the source) allowed for source measurement.
Since there was no improvement in the signal-to-noise ratio by 
combining data from detectors beyond about 60$^{\circ}$, this was a convenient 
cutoff for choosing which detectors were used.  This value produced a 
combination of one to four LADs depending on CGRO orientation.  The
only exceptions to this were cases where a small range of 
azimuthal viewing angles were occasionally blocked by the CGRO high 
gain antenna.  When this occurred we excluded these LADs from the 
measurement. (3) $\theta$$_{int}$ is the maximum detector aspect
angle for a source, with an occultation step time in the fitting window of the
SOI, that must be simultaneously fit with the SOI 
\citep[see for fitting model details]{Harmon02}.  A value of 70$^{\circ}$ was 
selected because bright sources
such as the Crab or Cygnus X-1 still have substantial flux even at very large
aspect angles.  (4) $\delta$ is the minimum time separation between occultations
of two sources in the fitting window that were allowed separate terms in the
fit.  If the times of the SOI and an interfering source fell within the fitting
window and satisfied the aspect angle criteria, but were within $\delta$=10s,
then the fit was rejected by OF\_W.  (5) $\lambda$ is the minimum fraction of data
packets accepted for an occultation step fit.  The acceptance criterion helped to
minimize potentially singular matrices and other instabilities in the fitting routines. 
For this work, $\lambda$ was set
at 95\%.   Finally, (6) $\phi$$_{cr}$ is the threshold
photon flux, which if exceeded, caused OF\_W to add a fitting term
for an occultation step of a source within the fitting window of the SOI,
provided criterion (4) was not invoked.  A database
was kept of sources that were found from previous passes
through the BATSE dataset to exceed
$\phi$$_{cr}$ for intervals of more than a few minutes, and typically
a day or longer.  The value of  
 $\phi$$_{cr}$,
0.01 photons cm$^{-2}$s$^{-1}$, refers to the
energy band of 20-100 keV, which we took as our standard band for this
work.
			
In addition to a list of sources for measurement, OF\_W requires 
the following databases as also noted in Fig. 4. 

\subsubsection{\it Flare and Eclipse Databases}
  
To optimize the sensitivity of the EOT, we tried to
minimize the number of source terms (SOI + potentially interfering sources) in the fitting window. 
A database of source outburst times, and intensity bands provides information 
to determine whether terms for other sources should be included
in the fit.  This database was built up as new
sources were found either through occultation analysis (light curves and
images) or observations with other instruments. 

Information about source outburst
intensity levels as a function of time was read from the database by OF\_W before 
the flux measurement was performed. For
a first complete pass through the BATSE dataset (which we refer to as the
``first iteration"), $\phi$$_{cr}$ was set at 0.02 photons cm$^{-2}$s$^{-1}$ 
(20-100
keV, $\sim$75 mCrab).  Sources dimmer than this threshold, however,  
were found to introduce significant interference in our measurements.
For this work (the second iteration), we lowered $\phi$$_{cr}$ to 
0.01 photons cm$^{-2}$s$^{-1}$
(20-100 keV, $\sim$35 mCrab).  In Table 2, we show the sources that
consistently exceeded $\phi$$_{cr}$ (identified
as ``persistent sources"), along with their average flux, and for
Table 3, we show those that exceeded $\phi$$_{cr}$
for an identifiable, but limited period of time.
Although a wide variety
of source behavior was observed with BATSE (See also the assessment of
the RXTE ASM data by \citet{Bradt00}), 
Table 3 sources were broadly classified as ``transient sources", although 
duty fractions could range from $<$1\% up to 90\%.  
If the beginning and ending times of the fitting window fell within the outburst
interval of a potentially interfering transient source, and it met 
the usual geometric criteria, a fitting term was included.  
For this particular iteration of the
catalog, we have added a number of low-level outbursts for the transient 
sources in Table 3 not previously reported in other publications.
Some special notes regarding
the detection and monitoring of transients are given in Appendix A.

A few bright X-ray binaries containing neutron stars, Her X-1, 
4U 1700-377, Vela X-1, 
Cen X-3, and OAO 1657-415, undergo complete eclipses by their
companion stars and have well-documented e\-phem\-er\-i\-des (see Fig. 5).   We took 
advantage of this in the fitting process by eliminating
interfering source terms for these sources during eclipse.  
%If one or more of
%the above five sources had occultation times within the fitting window of the SOI,
%and met the usual geometric criteria, we checked to see if
%the beginning and ending times of the fitting window are 
%completely within the eclipse interval.  If they were we dropped
%this source from the fit, otherwise, it is fit as a persistent source.
Comparison of the eclipse times to the fitting window times were made
in the solar system barycentric frame \citep{Standish92}.

\subsubsection{\it CGRO Orientation History}

CGRO was 3-axis stabilized, and maintained the same orientation for
extended periods, typically 2-week intervals.
The CGRO orientation was defined by the right ascension and 
declination of X and Z-axes of the spacecraft
in J2000 coordinates, and was used to determine the
optimum detectors for the occultation measurement of the SOI and 
potentially interfering sources.

\subsubsection{\it BATSE CONT Channel Energy Binning Lookup Table}      

The sixteen channel CONT data had programmable energy bins.  These were usually kept
to a standard set of values (see Appendix B), except for occasional changes that were made 
to allow higher resolution measurements of 
sources with strong spectral cutoffs such as pulsars and SGRs.
Energy calibration of the bin boundaries or ``edges" are required to obtain
photon fluxes.  They are also used to compute small electronic deadtime corrections 
in OF\_W, which depend on the total observed counting rate (source + background). 

\subsection{Data Flagging and Computation of Photon Fluxes}
 
%The raw source data, in counts
% s$^{-1}$ per energy channel per LAD, 
%without correction for the detector response, are stored in a 
%time-ordered file with a beginning and ending time specified
%by the user (see Fig. 4).   Energy spectra (flux per energy channel) and light curves 
%showing the history of the source's intensity
%as a function of time can be generated from the raw history file.  

Before deconvolution of the detector response, we
flagged (i.e., excluded) any remaining large outliers from the single step raw light
curve files.  The two reasons for doing this were to place a tighter
limit on obviously unphysical fitting results, and to exclude  
data contaminated by background fluctuations that 
flagging by daily visual inspection failed to remove \citep{Harmon02}. 
Both of these situations were relatively rare; however, they could produce
spurious daily flux averages.    
We settled on a scheme that could be implemented without special treatment
of individual sources, 
%One is based on the quality of fit, 
and was based on a filter for variations from an average flux level.  
%There is a quality of
%fit parameter generated by OF\_W for each of the 16 CONT channels.  
%During the data processing, a flag is placed
%in the raw count rate file when the reduced chi-square per channel for the fit
%exceeds
%1.5, with the intention to use the quality of fit to determine if the criterion
%for smoothness of the background is met. This scheme turned
%out to be sensitive to energy calibration changes and intrinsic source
%behavior and relatively insensitive to
%the longer timescale background variations (on the order of the fitting window
%sampling time).  The resulting chi-square
%is therefore not particularly well-correlated with the appearance of outliers
%in the data.
In this scheme, we averaged the count rate from a 
source for each CGRO observing period,
and subtracted the average from the rates (channels 1-10) over the same period.
We then made a cut on the rates at $\pm$3.5$\sigma$ to remove
outliers.  This approach worked well for all sources except those that
showed strong variations of 0.5-1 Crab on timescales of minutes to a few hours.  
The persistent sources that fell into this category were Cyg X-1,
Sco X-1, 4U 1700-377, and Vela X-1.  A few of the brighter, highly variable transient
sources also exceeded this variability criterion during outburst.  Typically, 
the percentage of rates rejected in this way was less than 5\%
for the majority of sources, and did not
significantly affect the measured fluxes and uncertainties.  However
Cyg X-1 could, at times, have rejection rates
as high as 13\%.   We therefore
converted count rate data to photon fluxes for these four persistent 
sources without this flagging procedure.  
For the bright transient sources, we visually inspected the raw light 
curves during
the times listed in Table 3, and removed flags on apparently good data that
were set by the automated flagging procedure.       

Two methods were developed to obtain flux and spectral information from the BATSE 
occultation count rate
data.  The first method is a fitting procedure used commonly in high energy 
astronomy called ``for\-ward-fol\-ding" \citep{Briggs95}, where 
we assumed a spectral model that was a reasonable representation of a 
source flux in the LAD energy range.  Then the selected model
was folded through the instrument response \citep{Pendleton95} for the illuminated detectors to determine model count 
spectra.  The count data were fit, by varying the model parameters,
in a chi-square minimization procedure.  Finally the best fitting model 
was integrated over the desired energy 
range to determine the photon flux.   This method 
avoids a potentially unstable inversion of the BATSE response matrices, 
which have large off-diagonal elements due to the LAD's relatively shallow detection
depth and lack of Compton suppression.  It lends itself well to generation
of multi-channel spectra, obtaining photon fluxes in a desired energy range, and
performing broad-band spectral analyses.  For the spectral analysis results  
in Tables 2 and 3, we used a widely available code known as XSPEC
(Arnaud 1996), and for the flux history generation, a code developed
locally for specific use with Earth occultation data known as 
HISGEN (see Fig. 4).  Both codes use forward-folding to estimate photon fluxes. 
The second method uses the Crab \citep[see for an example spectrum]{Harmon02} as a 
standard candle by
obtaining a ratio of the count rate for a given source relative to that of the Crab
in the same detector orientation and energy channel.  This is convenient for 
computing hardness ratios, epoch folding and investigating the deep sample 
results discussed later.  The resulting fluxes were less model dependent
than those using the forward-folding method.

Four energy bands in Crab-relative units were selected to report the results of the 
deep sample in Sec. 4: 20-40, 40-70, 70-160, and 160-430 keV.  The energy binning is 
somewhat constrained by the CONT data channel binning scheme (see Appendix B, Table B1). 
And, although the LADs have some sensitivity to fluxes in excess of 430 keV, we chose not
to include a higher energy band due to statistics (a relative lack of positive detections 
at higher energy), and the self-consistent treatment of systematic error which was based 
on a study only in the 20-100 keV range. Conversion factors from the reported four-band 
Crab relative intensities to photon and energy fluxes can be found in Appendix B, Table B2.

\section{BATSE All-Sky Monitor Results}

High voltage was first applied to the Large Area Detectors on Truncated 
Julian Date 
(TJD) 8362, or 16 April 1991. Final shutdown occurred on
TJD 11691, or 26 May 2000.  Data were available between TJD 8362 and 11690, except
for a few brief periods due to orbital reboosts and occasional mission anomalies as
listed in Appendix B. Light curves and representative spectra for the sources 
considered persistent
or transient by OF\_W (a total of 58 objects) can be found on the web at \\
\url{http://gammaray.msfc.nasa.gov/batse/occultation} \\  in 
visual as well as tabular form. 
Table 2 is a list of the ``persistent (P)" sources known by the EOT flux history generator
OF\_W (see Fig. 4). Likewise, transient (T) sources in the BATSE flare 
database appear in Table 3.
The tables carry information about source type, 
characteristic spectral data, outburst times, and energy flux for these sources 
in the 20$-$100 keV band.  

Of the four general classes of sources
represented in this catalog (x-ray binaries, supernova rem\-nants/pul\-sars/SGRs,
AGNs and x-ray stars), three are represented in Tables 2 and 3.
The brightest sources in the BATSE database are clearly dominated by
X-ray binaries (both HMXBs and LMXBs), but we also
found one supernova 
remnant/pulsar (the Crab), and three galaxies (Centaurus A, 
Markarian 501, and NGC 4151).  No X-ray stars were bright enough to
be included. In the following sections we will briefly describe science
investigations of these source classes using BATSE EOT results.

\subsection{Black Hole Systems}

Among the persistent black hole candidates, Cyg\-nus X-1 
\citep[e.g.,][]{Ling97, Paciesas97, Zhang97a, Brocksopp99b, Zdziarski02},  
Cygnus X-3 \citep{McCollough99}, GX 339-4 \citep[e.g.,][]{Harmon94, Rubin98}
1E 1740-29 and GRS 1758-258 \citep[e.g.,][]{Zhang97c, Smith02} have 
been studied most extensively
using the BATSE EOT. Emission in the low-energy gamma-ray range is understood
to be primarily upscattered low-energy photons by the inverse Compton process. 
BATSE studies have illustrated both spectral and intensity variations 
associated with state transitions (e.g, high (soft), low (hard) and off states) of black
hole binaries. Such state transitions are thought to be driven by 
changes in the mass accretion
rate. A prime example of this phenomenon is the 1996 hard-soft-hard state transition 
in Cyg X-1, which was monitored
in both X-rays (2-12 keV) with RXTE/ASM and low-energy gamma-rays with BATSE
\citep{Zhang97a, Esin98, Zdziarski02}.  These sources have also been
found to exhibit rather closely correlated radio flux variations: 
1E 1740-29 \citep{Sunyaev91, Mirabel92}, GX 339-4 \citep{Fender99}, 
and Cyg X-3 \citep{McCollough99}.  
In Fig. 6, we show the 9.1y flux histories for four of the more
persistent black hole candidates.  It can be seen, that, in addition to
the periods of low gamma-ray flux (soft state) in Cyg X-1 around TJD 9400 and
10270, all of these sources show extended 
periods of low level activity.
The low intensity or ``quenched" state of Cyg X-3, in particular, 
precedes violent radio flaring and jet activity \citep{McCollough99}. 

Bright transients
($\sim$200 mCrab or more) such as GRO J0422+32 \citep{Shrader94,
Callanan95, vanderHooft99, LingWheaton03}, 
GRO J1655-40 \citep{Harmon95, Zhang97b}, GRS 1915+105 
\citep{Foster96, Paciesas96, Harmon97}, and XTE J1550-564 
\citep{Corbel01, Wu02}
have been detected at an average rate of about 1 per year.  In Fig. 7, we show the 
flux histories of several bright black
hole transients.  Some have confirmed optical mass functions with lower
limits exceeding the theoretical mass of a neutron star. Most
have very bright high energy spectral tails extending to several hundred
keV, and thus, even without firm mass limits, 
are considered likely to be black holes.  These transients
reach maximum brightness within a day to a few days with highly variable
and generally unpredictable multiple outbursts over an
interval of months to years following the initial outburst.  The overall brightness 
of the secondary outbursts tends toward lesser intensity over time (GRS 1915+105
being an exception).  Sometimes there is an 
apparent return to X-ray quiescence between outbursts, e.g., GRO J1655-4
\citep{Orosz97}.  Transitions occur between
high, low, very high and/or intermediate states \citep{Corbel01}, with 
spectral shapes and luminosity levels that appear to be the same as the
persistent black hole systems.  Radio spectral
density measurements and long baseline interferometry of black
hole transients at times reveal compact radio sources,
and, with lesser regularity, 
extended high velocity ($\sim 0.3-0.9$c) radio-emitting jets
\citep{MirabelRodriguez94, HjellmingRupen95, Tingay95}.  The jet formation is 
sometimes associated with 
high energy outbursts \citep{Harmon95, Harmon97}. 
 
\subsection{Low Mass X-Ray Binaries with Neutron Stars}

Low mass X-ray binaries containing neutron stars detected by BATSE fall
into two general categories, usually known as ``atoll" and ``Z" sources.  The
names are based on the X-ray spectral behavior (generally below the sensitive energy range of
the LADs) where four energy bands are plotted as paired hardness ratios
on a two-dimensional diagram (a color-color
diagram).  As the source intensity varies with time, the combined ratios will 
tend to trace out a Z or
atoll-like track on the diagram.  The physical differences between the
two NS LMXB categories are not completely known. In particular, atoll 
sources tend to emit X-ray bursts, and have spectral states that resemble those
of black hole binaries.

The realization that many LMXB X-ray burst sources  
produce significant emission above a few keV came as
a result of observations with {\it Granat}/Sigma \citep{BarretVedrenne94} and 
BATSE \citep{TavaniBarret97, Barret01}.  In pulsars, high energy radiation is produced 
by acceleration of particles in the strong magnetic fields 
($\ge$10$^{12}$G) 
near the polar regions of the neutron star.  The bursters are
thought to have substantially weaker fields and should be inefficient particle
accelerators.  In addition, the presence of
the neutron star surface is thought to generate a large flux of soft photons,
effectively cooling surrounding plasma, and suppressing high energy emission.  
Nevertheless,
a number of burst sources, e.g., KS 1731-260, 1E 1724-308 
\citep{BarretVedrenne94}, Aql X-1 \citep{Harmon96}, 4U 1608-522 \citep{Zhang96b}, 4U 0614+091
\citep{Ford97}, and GX 354+0 \citep{Claret94}, show 
significant gamma X-ray flux, at times, in excess of 100 keV.  
They are considerably less 
luminous in the low-energy gamma-ray range relative to black hole 
LMXB systems,  $\sim$0.05-0.1{\sl L$_{Edd}$}
\citep{Barret00}, which is probably 
related to the presence of the neutron star.  

The spectra of LMXB bursters are similar to those of black hole systems, 
typically
consisting of a soft, disk component with temperatures of
a few keV and a low-energy gamma-ray tail.  The spectra of these sources sometimes show a cutoff of 50-60 keV,
but not always, and may extend to higher energy.  
Comparisons with soft X-ray measurements indicate that these sources exhibit 
anticorrelations between X-ray luminosity and spectral hardness 
\citep{vanParadijsvanderKlis94, Ford97}, and thus, as in BHCs,
spectral behavior of the LMXB NS systems should be governed by mass 
accretion rate.  Whether or not NS binaries have equivalent spectral
states to BH binaries is not clear, but 
these observations support the picture of a neutron star surrounded by an 
accretion disk as in BH systems, differing only at the innermost
regions of the disk.  Models involving boundary layers
between the disk and neutron star surface have been devised, although broad-band
spectral observations do not as yet strictly require conditions that are unique to neutron
star binaries.

In Fig. 8, we show examples of persistent and transient LMXBs containing
neutron stars detected with BATSE.  BATSE observations have shown that 
the gamma-ray flux from neutron star LMXBs varies over intervals of a few days or longer.  
Aquila X-1 was observed to have extended outbursts
of many months in the low-energy gamma-ray band \citep{Harmon97} similar to
LMXB BHCs.  These outbursts 
were correlated with an increase in brightness of its optical counterpart
{\it V1333 Aquilae}.  4U 1608-522 shows similar duration outbursts, 
one of which was bright enough to establish that the high energy spectrum had a high
energy cutoff of about 60 keV \citep{Zhang96b}.  GS 1826-238, a source once thought to 
be a black hole candidate, 
was observed to produce X-ray bursts with {\it BeppoSax} \citep{Ubertini97}, and 
therefore is clearly
a neutron star system.  The source reached a level of $\sim$0.01 
ph cm$^{-2}$s$^{-1}$ 
around TJD 10000, and remained roughly a factor of 2 brighter as of mission end. 

There is
also mounting evidence that the Z-type LMXBs, e.g.,
Sco X-1 \citep{StrickmanBarret00, DAmico01}, GX 17+2 \citep{DiSalvo00}, and 
GX 349+2 \citep{DiSalvo01},
exhibit transient high energy radiation as do the atoll sources 
discussed above.  These
spectral tails carry only a few percent of the total luminosity
of the source \citep{Barret01}, and it is unclear whether the high energy emission in the Z sources is the same spectral
component more commonly seen in the atoll sources.  So far, the presence of such
hard tails has not been significantly investigated in the BATSE database, although
the two bright Z-sources Sco X-1 and GX 17+2 were detected in the deep sample in 
excess of 160 keV (see Sec. 4). 

\subsection{Pulsed Sources}

Accreting X-ray pulsars were primarily studied with BATSE using
pulse frequency and pulsed flux measurements \citep{Bildsten97,  
Finger99, WilsonHodge99, Wilson02, Wilson03} that in 
general are more sensitive than EOT
data. However, EOT fluxes been used to obtain pulse fractions, study
correlations between the spin-up rate and 
total flux, and for the
characterization of apastron outbursts. 
EOT light curves for several accreting X-ray pulsars are
shown in Fig. 9. 
%Figure~\ref{fig:pulsed_sources}.

The most common use of EOT measurements has been to 
estimate the pulse fraction, i.e., the ratio of the pulsed flux to the 
mean flux (pulsed + unpulsed) in conjunction with simultaneous or 
approximately simultaneous pulsed flux measurements. These measurements,
along with orbital parameters, can be used to investigate changes
in the accretion environment, which can depend on a number of system
properties such as disk-induced torques on the neutron star
and the circumstellar environment of the companion.
%This has been generally done in one of two ways, either 
%formally, where ratios of count rates are taken for carefully selected 
%simultaneous data, or less formally, where the model dependent flux is computed separately for the
%EOT data and the pulsed flux (using the same model) for data that are 
%approximately simultaneous. 
Formal pulsed fractions, using simultaneous data, have been computed for
several sources including:
GRO J1744--28, 25\%, 20-40 keV, TJD 10092-10098 \citep{Finger96a}, 
GX 301--2, $\sim 50$\%, 20-55 keV \citep{Koh97}, GRS 0834--430, 10-15\%, 
20-70 keV, outbursts in 1991 and 1992 \citep{Wilson97a}, 
EXO 2030+375, 36(5)\%, 30-70 keV, TJD 9120-9131 \citep{Stollberg99}, 
GRO J1008--57, 67(4)\%, 20-70 keV, TJD 9186-9195 \citep{Bildsten97},
A 0535+262, varied from $>80$\% at lowest flux to 30\% at highest flux in 
giant outburst
\citep{Finger96b, Bildsten97}, and GRO J2058+42,
36(3)\%, 20-70 keV, TJD 9971-10018 \citep{Wilson98}. Less  
rigorously determined pulse 
fractions, with approximately simultaneous data and some model-dependency,
have been computed for 4U 1145--619, widely varying from 28(4)-70(20)\%, 
20-50 keV, TJD 8678-10742
\citep{Wilson94, WilsonHodge99}, GS 1843+00, 7\%, 20-50 keV, 
TJD 10515-10516 \citep{Wilson97b}, and 4U 0115+63, 40-60\%, 20-50 keV, 
TJD 11231 \citep{Wilson99}.

Simple accretion theory predicts a correlation between bolometric flux $F$ and
spin-up rate $\dot \nu$ in X-ray pulsars, described by a power law $\dot \nu 
\propto F^{6/7}$. Using EOT data, this correlation has been fitted for the giant
outbursts of GRO J2058+42 
%with a slope of $\sim 1.2$ 
\citep{Wilson98} and 
%the giant outburst of 
A0535+262 
%with a slope of 0.951(26) 
\citep{Finger96b}.  This correlation  
indicates the likely presence of an accretion disk, because disk
accretion is more efficient at transferring angular momentum than 
wind accretion.  Furthermore, a
correlation between the frequency of
quasi-periodic oscillations and the 20-100 keV flux measured with the EOT 
was
consistent with predictions of either the beat frequency or Keplerian frequency
models, confirming the presence of an accretion disk in A0535+262  
\citep{Finger96b}.

Using dynamic power
spectra of BATSE EOT and RXTE ASM data, \citet{Clarkson03} showed that
the $\sim 60$-day superorbital period previously known in SMC X-1 
changed smoothly from 60 days to 45 days and then returned
to its former value on a time\-scale of approximately 1600 days, varying in a
coherent, almost sinusoidal, manner.  
%Results from the
%RXTE ASM and BATSE EOT data agree during the 4.4 year stretch of simultaneous
%coverage. 
They interpreted this variation as consistent with a
radiation-driven warping model of the accretion disk. 
%Other periodicity
%searches, primarily aimed at confirming known periods in high-mass X-ray
%binaries, were reported in Laycock \etal (2003), who detected known orbital periods
%for EXO 2030+375, Cen X-3, 4U 1145--619, and GX 301--2 using 1-day average
%fluxes measured using the EOT. In addition, they detected the known
%30.4 day superorbital period for LMC X-4, that had previously been detected in
%BATSE data by Zhang \etal (1996a). Further, Laycock \etal (2003) reported a 
%marginal
%detection at approximately twice the previously reported orbital period for 
%4U 1907+097. 
%Additional periodicity searches using EOT data are described in
%Section 4.4.

\citet{Pravdo95} and \citet{PravdoGhosh01} folded
EOT intensity histories at the 41.498 day orbital period of GX 301--2 from 1993 September -
1994 July (TJD 9258-9561). They discovered that the near-periastron flare was asymmetric and
that a much weaker near-apastron flare was also present. In later data, after
the secular spin-up trend reversed, they found that the apastron flare became vanishingly small. 
\citet{Laycock03} folded GX 301--2 EOT data for the entire
BATSE mission and found increasing flux levels consistent with orbital phase
0.5, that were about 10\% of the peak near-periastron flux. For EXO 2030+375,
\citet{Laycock03} found evidence for an increase in flux near apastron,
averaging 30-40\% of the peak flux in EOT data folded for TJD 8363-9540, prior
to the change from spin-up to spin-down \citep{Wilson02}. Direct comparison of 
folded BATSE and RXTE ASM light curves from TJD 10114-10724 suggested that the 
apastron emission is softer than the periastron emission in EXO 2030+375. 

Two X-ray pulsars, discovered with BATSE, were initially located using EOT data
in conjunction with pulsed measurements: GRO J1008--57 \citep{Stollberg93} 
and GRO J2058+42 \citep{Wilson95, Wilson98}. GRO J1849--03
\citep{Zhang96a,Zhang96b}, 
which was later found to be identical with GS 1845-024 \citep{Finger99}, 
was located using Earth occultation imaging (see also Appendix A).

\subsection{Active Galaxies}

Several AGN are sufficiently bright in hard x-rays to permit spectral
and/or
variability studies using BATSE Earth occultation data alone. 
In Fig. 10 we show intensity histories for four of the brightest AGN
in the BATSE database.  An extensive
analysis of the Seyfert galaxy NGC 4151 by \citet{Parsons98} covered six
years of BATSE monitoring data. \citet{Paciesas94} analyzed BATSE
observations of the radio-loud quasar 3C273 during 1991-1993. Using BATSE 
data, \citet{Malizia00} discovered hard x-ray emission from the 
high-redshift quasar 4C71.07 and presented evidence for long-term source 
variability. \citet{Dean00} discussed BATSE observations of variability
in 
4C71.07 and two other blazars, Mrk 501 and PKS 2005-489. \citet{Wheaton96} 
reported BATSE spectra of the nearby radio galaxy Cen A from 1991-1992
derived 
using the methodology developed by JPL \citep[see also Appendix C]
{Skelton94, Ling96, Ling00}.

In several cases the BATSE data were useful in multi-wavelength studies
of
AGNs. BATSE observations were included in a radio to gamma-ray spectral
analysis of 3C273 by \citep{Lichti95}. \citet{Courvoisier99} and
\citet{Turler99}
investigated correlations of BATSE results on 3C273 with optical and
infrared
data during a six-year period. BATSE observations of the 
synchrotron component of the TeV blazar Mrk 501
during
its 1997 high state were used in a multi-wavelength study by \citet{Petry00}.

Finally, the BATSE light curves were useful as context information for
observations of AGNs by other instruments. \citet{Steinle98} used the
BATSE
long-term history of Cen A to characterize the source state during COMPTEL
observations in 1991-1995. Similarly, \citet{Haardt98} used the BATSE
light
curve in interpreting {\it BeppoSAX} observations of 3C273 in 1996-1997.

\section{BATSE Deep Sample Results}

A comprehensive deep sampling of the low-energy gamma-ray sky using the 
present formulation of the EOT is predicated on the sensitivity curve
in Fig. 1.  The improvement in sensitivity over the HEAO 1 A-4 survey
is only fully realized if systematic error can be minimized.
In \citet{Harmon02} we identified systematic errors in the application
of the EOT to BATSE.  Improvements to the detector response model, and 
multiple passes through the BATSE dataset with a better picture of
the bright sources in the sky (Tables 2 and 3), have allowed us to
reduce some contributers to systematic error.  These results
encouraged us to combine all the mission data together to 
achieve maximum sky exposure.  Later in
this section we incorporate a systematic error model
in assessing the detection confidence of
the sources in our sample.    
Limited applications of the approach we 
take here can
be found in \citet{White94} and \citet{Zhang96a}.  

%The iterative method used
%in identifying the bright sources that must be fit by OF\_W is
%especially relevant.  From successive passes through the BATSE dataset,
%the high duty fraction sources that contribute most of the total point 
%source flux contribution were established. 
%Those sources and their
%times of outburst (for transients) then became part of the 
%``bright source model" used by
%the software.  Prior to the current pass through the dataset, the
%model was represented by the information given in Tables 2 and 3.

In Table 4, we present the 179 objects in the BATSE deep sample in 
order of increasing right ascension ($\alpha$).  
Firm detections, which include the systematic error assessment 
described later, are indicated in boldface text.
The columns from left to right are (1) the BATSE 
catalog name, (2) an alternate name recognized by the SIMBAD catalog if the 
BATSE name was not recognized, or other well-known name for the source,  
(3) identification of the type of source, if known, given the classes defined in
Fig. 2. (4) An assigned category (A, B, C, N or I, see definitions below) 
based on the significance of detection after, 
and before (in parentheses) correction
for systematic error and, (5-6) its celestial coordinates ($\alpha$,$\delta$) (J2000) and 
(7-8) 
Galactic coordinates ({\it b,l}).  The next columns from left to right are
the (9) detection significance for the 20-100 keV average photon flux 
averaged over the mission (9.1y) before (in parentheses) and
after correction for systematic error
\footnote{To compute the
20-100 keV flux for sources for which we did not have the spectral fitting
information given in Tables 2 and 3, we assumed a power law spectral
shape with a photon index of -3, typical of accreting systems  
(an intermediate value between 
spectral slopes seen in black holes and those of softer sources such
as pulsars).  Because the computed 20-100 keV photon flux is sensitive
to the assumed power law spectral shape, we only report significance.}, 
and finally, the flux in mCrab (columns 10-13) of each source in the
four energy bands 20-40, 40-70, 70-160, and 160-430 keV.  
The Crab relative fluxes, we found, are the most accurate 
method
of displaying flux information for the deep sample in the absence of 
precise spectral fitting information for each source. 

It is helpful to classify the sources in our sample according to a 
scheme based on their significance of detection and variability, 
and taking into 
account the systematic effects present in our results.  We use the following 
categorization:

A.      (Definite detections) Persistent sources in category A
have an average flux (9.1y) that exceeded a 
threshold of 0.01 photons cm$^{-2}$s $^{-1}$ in the 20-100 keV band.
Transient sources in category A (denoted with a dagger ($\dagger$) 
in column (4))
exhibited at least one identifiable outburst
with an average daily flux 
exceeding a threshold of 0.01 photons cm$^{-2}$s $^{-1}$ in the
20-100 keV band.  Typical category A sources show variability on 
timescales of at least hours to a few days, have a readily 
identifiable period by 
FFT, or variability that correlates well with behavior in other 
wavelength bands. Because category A source identifications are  
independent of duty fraction, the detection significance can vary
from a statistical zero over 9.1y for brief transients, to very 
high values for bright, persistent sources.
54 sources fall
into category A: 21 accreting PSR, 29 BH or NS binaries, 1 SNR (the Crab 
Nebula), 3 AGN and no x-ray emitting stars. 

B.      (Definite detections) Sources with a 9.1y-av\-er\-age flux that 
exceeded 10$\sigma$, but with average flux 
$<$0.01 photons cm$^{-2}$s $^{-1}$ in the 20-100 keV band.   A periodicity 
search or epoch folding analysis may show a period 
associated with the source.  After correction for systematic error, 29 sources
fall into category B: 6 accreting PSR, 18 BH or NS binaries, 2 SNR, 3 AGN and 
no x-ray emitting  stars.  In later sections, we test the category B sources
in several ways for consistency of detection.

C.      (Possible detections) Sources with a 9.1y-av\-er\-age flux
(20-100 keV) with a significance $\ge$3$\sigma$ and $<$10$\sigma$.  
Variability was not investigated; thus, transient activity was
not considered and could be present in the BATSE dataset.  After correction 
for systematic error, 
36 sources fall into category C: 6 accreting PSR, 16 BH or NS binaries, 4 SNR,
1 SGR, 5 AGN and 4 x-ray emitting  stars. 

N.      (Non-detections)  Sources with a 9.1y-av\-er\-age flux
(20-100 keV) of significance between $\pm$3$\sigma$.  Transient 
activity from the sources was not considered, and could be present in the
BATSE dataset.  After correction of systematic 
error, 45 sources fall into category N: 7 accreting PSR, 21 BH or NS
binaries, 2 SNR,
2 SGRs, 4 isolated PSR, 1 AGN and 8 x-ray emitting  stars. 

I.      (Indeterminate) Sources with a 9.1y-average flux
(20-100 keV) of significance $<$-3$\sigma$, indicating
confusion with nearby sources, or a poorly characterized systematic error
in the surrounding sky
region.  The category I sources represented, for the most part, our
inability to separate crowded source regions where bright sources known to 
OF\_W were close by. Transient activity from category I sources was not considered
and could be present in the BATSE dataset.  After 
correction of systematic 
error, 15 sources fall into category I: 1 accreting PSR, 13 binary BH or NS, 
1 SGR, and no SNR, isolated PSR, AGN, or x-ray emitting stars.  Since these 
sources were indeterminate, we do not report 4-band fluxes for these in Table 4.

\subsection{Determination of Systematic Errors Due to Sky Location}

In addition to the set of 179 known sources, we produced light curves 
for a grid of 162 test or control ``source" locations over the entire mission.  
The grid covers the entire Galactic plane, with the highest density
of test locations in the central bulge region, as shown in Fig. 11. 
The grid has points every 3$^{\circ}$ along the
plane between -60$^{\circ}$ and +60$^{\circ}$ longitude 
with two sets of grid points of the same spacing at +6 and -6$^{\circ}$ 
latitude. Beyond $\pm$60$^{\circ}$ longitude,
there are grid points at 6$^{\circ}$ spacing and 0$^{\circ}$ latitude. 
The grid light curves were subjected to the same flagging
process for outliers as performed for the known source set. Deconvolution of 
the detector response for the grid points was performed using an assumed 
power law of the spectrum with a photon index of -3. Ideally, with no 
systematic error (a
perfect sky model) there should
be no residual flux.  However, a few points fall very close 
(less than a degree) to 
known bright sources, which is desirable, as it allows a more accurate 
assessment
of systematic error due to the effects of finite spatial resolution.  
On the other hand, some grid points may fall very close to unknown 
sources that are not in our bright source model.

In Fig. 12, we show the average fluxes for the 
9.1y-long light curves at each grid point as a function of Galactic longitude.
The samples at -6$^{\circ}$, 0$^{\circ}$, 6$^{\circ}$ latitude are 
shown together for longitudes between -60$^{\circ}$ and 60$^{\circ}$.
We also show the standard deviation (in sigmas) for the one-day flux averages
\footnote{An earlier version of this figure, including the
first $\sim$7y of data, appears in \citet{Harmon02}.}.
Average values (absolute) for the fluxes are approximately 0.0015 photons cm$^{-2}$s$^{-1}$ 
($\sim$5 mCrab), with the exception of outliers near the Galactic center and 
points such as ({\it l,b}) of (-24,0),
(42,0).  We found no global trend in longitude or 
latitude, but there clearly were regions
where known bright sources are concentrated, or  
unknown point source emission has not been well-characterized in our 
bright source model.  Artifacts due to subtraction of point source fluxes can be
manifest either as positive or negative measurements in the grid
\citep{Harmon02},
which are allowed by the fitting method. In specific cases we have found that 
simultaneous fits made to 
weak sources or empty fields with nearby bright sources can sometimes
generate significantly negative measurements.  Unknown sources 
could also produce negative residuals, but their actual effect is not separable from the 
subtraction of known sources.  

We found that some of the positive grid outliers mentioned above fall 
near sources not known in iteration 2 but revealed very significant flux over at 9.1y
average: 4U 1907+097, Cir X-1 and H 1624-490.  
Source confusion in the Galactic
center region (GCR), as expected, also showed a large residual flux from 
unresolved point source
emission. It was not possible to attribute the excess GCR flux
to specific sources. 
We treat the GCR sources, along with the full source set, using a model for 
systematic error obtained from the grid analysis. However
extra care should be used in interpreting 
fluxes from weak
sources (category B and below) in the few degrees (2-4$^{\circ}$) surrounding this region.
There are deep images of the GCR with {\it Granat}/Sigma \citep{Paul91}, for 
example, that are better suited
for determining the average emission characteristics of these sources.  
A similar situation is likely to exist for the Magellanic Cloud regions,
which have not been evaluated extensively using BATSE data.

The second moments, or standard deviations had a completely different behavior.
Unlike the average fluxes, standard deviations were a relatively smooth 
function of sky location.
These are plotted in Fig. 12 in units of Gaussian statistical error for the
average daily fluxes, i.e., a value of 1 in Fig. 12 means there was no contribution
from any source of error except Poisson statistics (which, of course, is never
the case!).  The standard deviations were clearly broader than predicted
for a Gaussian distribution about zero flux due to systematic errors
(1 $\sigma$ = $\sim$0.01 photons cm$^{-2}$s$^{-1}$).  
We found broadening factors of about 20\% (between 70$^{\circ}$ and
180$^{\circ}$ longitude
and peaking near a factor of 2 at about -105$^{\circ}$) in excess of normal
statistics.  The standard deviations were largest near bright, variable sources 
such as Vela X-1(${\it l}$=-96.9$^{\circ}$) and Cygnus X-1 
(${\it l}$=71.3$^{\circ}$).  These effects are caused by well-known
properties of the two sources:  the long period pulses from Vela X-1, and
shot noise from Cyg X-1. 
 
%In Fig. 11, we show the grid points that had average fluxes more than 3$\sigma$
%positive or negative.  

\subsection{Application of the Systematic Error Model}

Correction factors for the flux average (additive) 
and flux average error (multiplicative) for the 179-source set
were derived as follows:
Using the flux and standard deviation for the 162-point grid, 
we interpolated correction factors both for the flux and its associated error
corresponding to the locations of the known 179 source set from the 
results shown in Fig 12. If the known source was within Galactic 
longitudes of -60$^{\circ}$ (300$^{\circ}$) and +60$^{\circ}$ longitude and 
within $\pm$6$^{\circ}$ latitude, 
we applied an interpolation based on the four nearest 
neighbors on the grid.  Outside
of the bulge region, -60$^{\circ}$  
longitude and +60$^{\circ}$ longitude, we applied a two-point 
interpolation of the nearest neighbors on the grid. For any point sources
less than $\pm$60$^{\circ}$ 
longitude, and
latitudes between 6$^{\circ}$ and 10$^{\circ}$ and 
between -6$^{\circ}$ 
and -10$^{\circ}$, we also applied a two-point interpolation.   
Outside of these regions, well above and removed from the Galactic plane, 
there 
was no correction to the flux average, and a 1.4 multiplicative factor was
applied to the average error. This value is a conservative estimate based
on the observed standard deviations in Fig. 12.

The detection significance and relative fluxes in Table 4, corrected for 
systematic error, were then calculated as

\begin{equation}
\Pi_{cor} = \frac{F-F_{cor}}{\epsilon_{cor}\delta F}
\end{equation}
where $\Pi_{cor}$ is the corrected significance in column 9, and 
F$_{cor}$ and $\epsilon_{cor}$
are the interpolated correction factors 
derived from the grid analysis, for the 20-100 keV flux ($F$) and error ($\delta${\it F}), 
respectively,
and 
\begin{equation}
B_{i,cor} = B_{i} - \frac{F_{cor}}{F}B_{i}
\end{equation}

where the B$_{i,cor}$ are the corrected Crab relative rates in columns 10-13. 
The statistical errors for the relative rates were also multiplied by the 
$\epsilon_{cor}$.  

In Fig. 13, we show the distribution of the 9.1 yr, 20-100 keV flux averages in order of
brightness, beginning with the brightest persistent source, the Crab Nebula.
The curve has been computed with and without systematic corrections.  
We also
show the assigned category, as determined after the corrections were applied.    
We note that transients, and especially those with low duty fractions, 
may have an average flux that is much less 
than the peak flux, and can appear anywhere on the curve.  
%In Fig. 13, we show 
%the standard deviations also ordered by brightness.  
Of the 179 sources, about 
110-120 sources were above the 3$\sigma$ flux limit.  This agrees well
with the combined number of sources (119) in categories A through C. 
Note that category N sources with a slightly negative,
non-significant flux, and all of the 
category I sources, whose fluxes are $<$ -3$\sigma$ negative, are 
not visible in the log plot. The main effect of the systematic corrections is to smooth the
trend of the lower part of the intensity curve, which shows a rollover at the
low flux end.  The physical interpretation of the rollover is that we are no longer able to 
distinguish significantly measurable flux due to systematic and statistical
effects at low source intensity.  We are also prone to missing sources 
in our biased sample such as transients
or persistent sources with fluxes comparable to those in category C.  

There is a possibility that the rollover of the intensity curve
reflects the true Galactic distribution of sources.
The depth of our sample in flux
is significantly beyond the $\sim$70 sources detected in the HEAO 1 A-4
survey \citep{Levine84}, even with BATSE's higher energy threshold. 
For example, HMXBs and LMXBs are thought, respectively, to be formed in
binaries with population I and II progenitor stars, and thus have sharply
differing Galactic distributions \citep{Grimm02}. 
However,
to extrapolate information about source populations in our
data is difficult due to the missing sources and the relatively small
size of the sample.
A deep, unbiased survey with an instrument such as EXIST, with excellent
background rejection and spatial resolution characteristics,
could produce a much more precisely determined extension
and shape for the curve shown in Fig. 13. 
 
%Although not presented, we calculated standard deviations from the one-day average
%light curves for all 179 sources.  Approximately the 30 brightest
%sources showed significant variability on timescales of one day, which is about half 
%the number of category A sources, plus a few remaining sources
%with significant one-day variability 
%due to transient behavior. We also found that a few sources in our sample with very 
%low average brightness showed significant variability.  These sources turned
%out to be in the
%proximity of the GCR where we could not assign the observed 
%variability to any particular object. Such
%source-specific studies are beyond the scope of this work, and for short timescale
%variability in the GCR, beyond the capability of the BATSE EOT.

\subsection{Consistency of Deep Sample Detections} 

Before concluding this work, it is interesting to look at the robustness
and consistency of the reported detections
in Table 4 without going into great detail.
To do this, we examined sources at flux levels 
close to the sensitivity limit of the EOT, i.e., just above the level where 
systematic effects dominate.  
%Examination of the flux averages
%and the light curves also reveal information about the bright 
%source information in Tables 2 and 3.  

We saw from the grid analysis 
that there were positive
as well as negative flux offsets at the measured control locations,  
as well as a few significant outliers.
Ideally, interpolation of the offsets
should allow us to estimate systematic errors in the point source set.  
However, we have noted some shortcomings of the approach:
there are category I sources remaining after the
model was applied. Another imperfection 
is that there are cases where a reasonably bright source is 
omitted from our model (Tables 2 and 3), and we may erroneously subtract 
``real" flux from the statistically-derived average. 

From the analysis of the grid outliers and in the results of Table 4,
we found that the HMXB PSR 4U 1907+097 \citep{Chitnis93}, and the  
LMXBs Cir X-1 \citep{Iaria01} and H 1624-490 \citep{BalucinskaChurch00} have
a raw significance in the 20-100 keV band sufficient to
be considered as category B, but in failing to identify these sources
as bright sources, they fall into categories
C, C, and N, respectively, after an overestimated correction.
We also found reasonable agreement between our uncorrected fluxes 
(i.e., before systematic error corrections were made)
and the observed low-energy gamma-ray fluxes from the literature for 
these sources. 
The burster 1E 1724-307 of the Terzan 2 globular cluster is another 
example; 1E 1724-307 was the only 
category A source found in the deep sample.  
1E 1724-307  has rather strong and variable hard X-ray emission 
($\sim$30 mCrab)
\citep[see also Fig. 8]{BarretVedrenne94} in the 20-100 keV band, 
and was not considered a bright source in the current pass through
the data.
We did not alter the systematic corrections to
the flux
for the four sources in the catalog, but it is likely,
pending other tests, that these sources  should be 
included in 
the bright source model for a future pass through the dataset.
%SLX 1735-269, another strong hard X-ray emitter 
%(Barret \etal 2000), was unfortunately omitted from our catalog completely.  

Next we rigorously examined the 
category B subset of sources, a total of 29 objects, which we claimed are 
definite detections.   These sources 
are gathered for inspection in Table 5.  
The (model-dependent) average fluxes (9.1y) in the 20-100 keV
band are between 1.2 and 9.6$\times$10$^{-3}$ 
photons cm$^{-2}$s$^{-1}$,  
%For a source with a Crab-like spectrum, the 
%corresponding intensities are 
%roughly between 5 and 25 mCrab.  
%Note that the average statistical error varies by a 
%factor of 2.  
with detection significances, including systematic error, varying 
between 10 and 52$\sigma$.  
Unlike Table 4, we list both the uncorrected 
average flux and error, and the interpolated
systematic correction based on the source location with respect to the Galactic 
plane.  

Eighteen of the 29 sources were within 10$^{\circ}$ of the Galactic plane and thus had corrections
to the given flux average and error.   Correction factors to the 
average flux for the category B sources ranged as high as 150\% in the GCR.
As indicated in Eq. (2), if
F$_{cor}$ $>$ 0, it was subtracted from the statistical average. 
If F$_{cor}$ $<$ 0, it was added to the 
statistical average.  As can be seen in column 4 of Table 4,
the net effect of the systematic error corrections for the deep sample
was to 
render a number of 
category B and C sources as a lower category, and some 
of the category I sources as category N. 
Of the 29 sources in category B, 27 did not require recategorization
after systematic corrections were applied.  
  
On the other hand, there were a few cases where addition of a large correction 
factor to the average flux of a 
category C or N (non-detection) source resulted in a total corrected flux level
and significance equivalent to a category B source.  Two of these appear in Table 5: 
4U 1820-303, a category B(C) source, and 4U 1916-053, a category B(N) source.  
A previous survey for hard X-ray emission
in the BATSE database \citep{Bloser96} reported 2$\sigma$ upper limits 
of 30 mCrab for both 4U 1820-303 and 4U 1916-053 in the 20-100 keV band 
using 10 days of data. Since that time, \citet{Bloser00a, Bloser00b} followed
up the BATSE results with observations of both 4U 1820-303 
and 4U 1916-053 with RXTE, as did  
\citet{Church98} using {\it BeppoSax}.  Hard tails were observed
in both
sources.  In our deep sample, we report a 20-40 keV flux over 9.1y in Table 5
from 4U 1820-303 of 11$\pm$0.8 mCrab (uncorrected) in the 20-40 keV band, 
and a corrected flux of 23$\pm$1.1 mCrab.  The corrected flux
is a reasonable factor of 4-5 lower than a recent RXTE measurement of the 
hard X-ray flux by \citet{Bloser00b}.
4U 1916-053 showed no significant 
detection (0.9$\pm$0.7 mCrab) (uncorrected) in the 20-40 keV band, but 
a corrected flux of 38$\pm$1.0 mCrab, reasonably close to the RXTE 
and {\it BeppoSax}
results, but about a factor of 2 higher.  Since it is unlikely that
the source spends most of the time at flux levels more
than that observed in the pointed observations, this suggests that the magnitude
of the systematic correction for 4U 1916-053 is over-estimated. 
The sign of $F_{cor}$, i.e., the sense of the correction, though, to
render a significant positive measurement, is correct.
%Thus, in the absence of systematic error, we should be picking up stronger 
%low-energy gamma-ray emission from both sources. 
Of course,   
without
specific knowledge of variability of the hard tails in these objects, it is not 
possible to make a definitive conclusion, but it is possible that 
rather large systematic error corrections could be suspect in our model.  
This is not
too surprising, since 
we are under sampling a rather complex and rapidly changing residual flux as 
a function of sky location as shown in the top frame of Fig 12. 
We are also correcting the 4-band fluxes, which span the energy range
between 20 and 430 keV, with a correction deduced for the 20-100 keV band. 

To investigate the impact of the correction factors more generally, 
we computed the relative change in 
significance for all sources
in our sample before and after the correction was applied.  This is shown 
as a histogram in Fig. 14. The distribution is dominated by
the simple multiplicative correction of the second moment (centered at -0.3), 
but in cases where the uncorrected, statistically-derived
flux was very small relative to the systematic correction to the flux, 
$F_{cor}$/$F$ in Eq. (2) was consequently very large. These are  
represented by the outlying wings of the histogram.
We therefore
made symmetric cuts on the distribution of
fractional changes in
significance about the median value at -1.05 on the low end of the 
distribution 
and 0.45 on the high
end in Fig. 14.  24 of the 179 sources had a fractional change in
significance that fell outside of the vertical dashed lines.
We also found that the absolute value of the significance was increased 
by application of the systematic error correction for 16 of the 24 sources 
beyond the cuts, whereas for most sources, the significance was reduced.
We therefore decided it was better to quote the uncorrected
relative Crab rates in the four bands for these 24 sources in 
Table 4, including the two category B sources discussed above.
The 4-band fluxes with no systematic corrections to the statistically-derived
fluxes are denoted by the four band rates in parenthesis.    
The judgment here is that the uncorrected 4-band fluxes for these objects
are closer to the true 9.1y averages.

In the following sections we illustrate the robustness of the 
category B source detections using periodicity searches and 
Earth occultation imaging. It is important to establish the reliability
of the EOT method at these low flux levels. For example,
detection
of the black hole candidate LS 5039 \citep{McSwain01} in the low-energy gamma-ray
band has not been reported previously
in the literature, and 4U 1812-12, was reported only for the first time
in \citet{Barret03}.

\subsection{Periodicity Search of Light Curves}

X-ray binaries sometimes exhibit periodic or qua\-si-periodic intensity
changes, and in most cases, the 
observed periods correspond to the orbital period of the binary ($P_{orb}$).  
Super-orbital periods ($P_{long}$) also have been 
observed in accreting neutron star systems, and have been attributed to disk 
precession \citep{WijersPringle99}, and/or periodic reversals of the spin direction 
\citep{Bildsten97}.  \citet{Laycock03} have investigated periodic
behavior in HMXBs in the BATSE EOT database. Here we
use the identification of a period or quasi-period in the category B sources 
to add confidence to the detection of the fluxes presented in Table 5.

Searches for periodic behavior in unevenly sampled data from the EOT
are accomplished using the
Lomb-Scargle algorithm for Fourier transforms \citep{Scargle82, 
HorneBaliunas86},
and epoch folding techniques in specific cases.
We first performed Fourier transforms to create power density
spectra (PDS) of the 
9.1-year one-day average light curves to 
search for periods in the flux variations from all sources. 

The PDS were scanned in the frequency range
between 0.002 d$^{-1}$ (500 days, 6.6 cycles) and 0.5d$^{-1}$ 
(the Nyquist
frequency of evenly sampled data).  A simple threshold for reporting
of a peak was
set at a significance, or false alarm probability (FAP) \citep{Press92}, 
of 0.01 (FAP=1 is roughly the level of Poisson noise) for the Lomb periodogram. The FAP
is related to the PDS peak power {\it z} \citep{Press92} as
\begin{equation}
FAP(\>z) = 1 -(1-e^{-z})^M
\end{equation}
or if the probability is $\ll$1, $FAP(\>z)\approx Me^{-z}$.
{\it M} is the number of independent frequencies and is related to 
the total number of flux values {\it N} as  
\begin{equation}
M = N \frac{\kappa \eta}{2}
\end{equation}
The oversampling factor $\eta$ was set
at a value of 4, and the maximum (high end) frequency $\kappa$ for the PDS was
set at 2, so that {\it M}=4{\it N}.  It should be noted that the FAP is
strictly a meaningful quantity for a PDS 
of Gaussian distributed data that is normalized by the variance
of the dataset. As we have shown earlier
for EOT light curves, the second assumption is clearly violated.  Thus the 
FAPs should only be taken as an indication of relative significance.

In Table 6 we list the search results from the PDS analysis of the
category A and B sources.  We limited ourselves to peaks in the PDS
that corresponded to orbital and super-orbital periods  
found in the literature. The far column indicates
the FAP computed for the power at the peak frequency.  
It can be seen that the FAP ranges from relative large values $\sim$0.01 (closer
to the Poisson noise level but very significant) to very small 
values for persistent, eclipsing category A sources
such as Cen X-3, Vela X-1, OAO 1657-415, and
4U 1700-377 (see also Fig. 5), as well as some of the sources
with superperiods such as LMC X-4 and HER X-1. 
The reported
uncertainty in the centroid frequency is an approximate error based on the
oversampling factor and not a formally derived error.  The quoted error can be reduced by proper fitting
of the PDS and subtraction of the sinusoidal signal from the light curve,
and the variance recomputed \citep{HorneBaliunas86}.  Since we are 
interested only in confirming 
detections of the category B sources, this analysis was not performed. 

The only category B sources with detected 
periods using FFTs were SMC X-1, LMC X-4, and 4U 1538-522. All are 
eclipsing HMXBs.  Both 
SMC X-1 \citep{Wojdowski98}
and LMC X-4 \cite{Laycock03} were previously known to show orbital and 
superorbital-related
variations in the BATSE data.  4U 1538-522 was already detected by BATSE via
its pulsed emission \cite{Bildsten97}, but this is the first reported 
EOT detection of its orbital period of 3.728d. 
%Several of the bright, non-eclipsing 
%category A sources in Table 6 show  
%quasi-periods.  For example, GX 339-4 exhibited broad power centered at
%229d, which is roughly twice the reported interval of 450d (15mo)
%between outbursts seen between 1991 and 1994 (Harmon \etal 1994).  
%The power at 450d was reduced considerably due to the marked change
%in behavior of low-energy gamma-ray flux at different times (see 
%Fig. 6).  GRS 0834-430, a transient pulsar, exhibited 
%a sequence of outbursts, some of which were near the derived orbital
%period of the binary (Wilson \etal 1997).  
%There are also evidences of
%long term, quasi-periodic variations in sources such as
%Cen X-3 (140d), Cyg X-1 (150d), and 1E 1740-29 (600d),
%that have been noted by other observers.  
 
The FFT analysis was performed only on the 1-day resolution light
curves with a simple cutoff for the FAP and is complicated by low
frequency noise, CGRO orbital precession, and other effects. 
For example,
\citet{Laycock03} reported a marginal result for the orbital
period of the HMXB pulsar 4U 1907+097 (8.3745d), a category C(B) source, 
that did not show up in the PDS search.  Since only a few category B 
sources were detected in the PDS, we carried out
epoch folding of the single step (twice per CGRO orbit sampling) 
light curves of at known periods for
selected category B sources. 
%For example, we have reported detection of the 14-day period
%in a category C source LSI+61\_303 (Zhang et al. 1996a) via epoch folding.

In Fig 15(a), we show
9.1y single step light curves of the three category B sources folded at 
the {\it $P_{orb}$}
given in column 2 of Table 6 and the light curve of 4U 1907+097 folded
at 8.3745d.   In Fig 15(b), we show single step
folded light curves for sources
with the reported super-orbital periods in column 3 of Table 6.
4U 1538-522, 
LMC X-4, and SMC X-1 all exhibit sharp eclipse profiles, as well
as prominent super-orbital variations for the latter two sources.  
We also epoch-folded single step
4-band rates searching for reported periods from soft X-ray and optical 
measurements among several of the 
remaining category B sources (some perhaps binary-orbit
related) in X Per \citep[250d,][]{DelgadoMarti01}, 1E 1145-614
\citep[14.335d,][]{RayChakrabarty01}, GX 349+2 \citep[21.7h,][]{Wachter97}, 
GX 354+0 \citep[63d, 72d, and 78d,][]{Kong98}, 
4U 1820-303 \citep[171d,][]{ChouGrindlay01}, LS 5039 \citep[4.117d,][]{McSwain01}, 
SS 433 \citep[$\sim$13d, binary period, 65d, jet nodding, $\sim$164d, jet 
precession,][]{Eikenberry01}, 4U 
1916-053 \citep[199d,][]{Bloser00b}, 4U 2127+119  
\citep[17.112h,][]{Ilovaisky93} and Cyg X-2 \citep[9.84, 40.4, 53.7,
61.3, and 68.8d,][]{Paul00}. 
Among these 10 sources, no convincing periodicities were seen in the
epoch folded light curves.
Given that the 
average fluxes of the HMXB eclipsers 4U 1538-522, LMC X-4, and SMC X-1 
are comparable to the other category B sources in Table 5,
it is somewhat surprising that
other binaries, 
which have reported periods in X-rays, did not show recognizable 
periodic variations in the
epoch folded light curves.

We point out that several of the category B \\ LMXBs
are high inclination,
compact, ultra-short period binaries known as ``dippers". These close
binaries have orbital periods of only minutes to hours. 
The actual dips and/or
eclipses are very short, $\sim$few minutes, and are 
observed as absorption-induced changes in the soft X-ray flux
($\sim$keV) \citep{Church98}.  
%It occurs when the hot bulge caused by the accretion stream from 
%the companion 
%contacting the accretion disk intercepts the observed continuum flux from the central 
%X-ray source.  
Such short duration absorption dips
are not accessible using the BATSE EOT; however, the reported 
superorbital periods are easily within the 
timescales accessible for epoch folding, yet none were seen.  
More work should be done
with these sources, but the apparent lack of periodicity in the LMXBs
(6 out of 6) is interesting.  
Assuming the reported X-ray and optical modulations are
predominantly associated with disk precession as in the HMXBs, such
effects do not appear to modulate the intensity 
of the observed gamma-ray flux in our category B LMXBs.

\subsection{Earth Occultation Imaging of Category B Sources}

Due to the limited success of the periodicity searches in the category B
light curves, we employed Earth occultation imaging to localize the
gamma-ray flux from these sources.
The occultation imaging technique \citep{Zhang93} allows 
the construction of a 2-D image of a region of the
sky to an accuracy of 1-2$^{\circ}$ and has been used to identify
new transients where the source location was previously unknown
\citep[e.g., GRO J1655-40,][]{Zhang94}. For more details of the method,
see \citet{Harmon02} and references therein.    

%The process requires a forward transform from the image space 
%(sky pixels) to the data space (count rates vs. time) followed
%by an image reconstruction using maximum entropy methods.  
%The first
%step produces a transform file for a particular day.  In the second 
%stage the days are combined and 
%The total integration time for an image
%has an upper limit set by computation requirements of about 15 days.
Most of the 29 category B sources of the EOT catalog are strong enough 
that we might expect to detect them in a 15-day integration
($\sim$6.5$\sigma$ for
a 25 mCrab source in the 20-100 keV band).
A non-de\-tec\-tion, however, does not imply that the EOT signal
does not come from this source, because the application and interpretation
of the imaging technique is complex and depends on more than the strength of the
source signal and spectral signature. 

%The success of the imaging technique in extracting a signal for 
%a source of a given intensity is affected primarily by two factors: 
%the  presence of other sources in the
%sky around the source and the limb geometry of the source relative to the
%Earth as it enters and leaves its shadow.   Proximity to bright sources
%or to sources that are not included in the image fit can prevent
%the source signal from being resolved in the image.  A limb geometry with the
%rising and setting limbs of the source lying at small angles to each other
%leads to poor signal reconstruction because the quality of the 
%image depends on sampling over a wide range of available limb projections.  

Each of the category B sources was examined over at least one 
time interval to look for an excess in the image at the source location.
A suitable limb geometry was chosen by displaying the rising and setting limbs
over an entire CGRO precession period.  This period will contain all possible
limb geometries for a given source and allow the selection of the longest
consecutive time interval with near-orthogonal rising and setting limbs
for the best possible spatial resolution.
%Some sky regions 
%have an unfavorable geometry (near parallel limbs or lack of occultation at
%high declinations) during much of the CGRO precession period while some
%will allow integration over more than 15 days.
Images were generated for ten separate source locations within
a 10$^{\circ}$ by 10$^{\circ}$ field of view.  The technique is most efficient for imaging
sources in the center of the field, but a true signal should appear when the
source is offset from the center.  
Offsets up to 1$^{\circ}$ in each direction were considered 
and images were produced for LAD energy channels 1 through 3.  
  
Although most of the category B sources exhibited at least 
some variability, the only flux-related
selection of integration time intervals was made for those sources that are
eclipsing binary systems or systems with a superorbital period.  
The RXTE ASM light curves (\url{http://xte.mit.edu/ASM\_lc.html}) 
allowed the selection of time periods when the X-ray emitter was not eclipsed by
its companion and was not occulted by a warped accretion disk in those systems
with a known superorbital period.    

Table 7 summarizes the imaging results. 22 of the 29 category B
sources yielded at least one credible image.    
%A lack of success does not imply the flux detected using the EOT is not 
%from the
%source,  the imaging technique being limited by systematics rather than 
%flux-limited.  
The difficulty in obtaining a good limb geometry
is apparent in some of the sources where the contours appear elongated.
In others, the noisiness of the sky was responsible for the lack of 
image contours associated with the category B source.
Figures 16(a-d) show some of the noteworthy images for the different
types of astrophysical objects seen by BATSE.  The AGN 3C 273 has a spectrum 
hard enough to be imaged in the 100-300 keV energy range as seen in Fig 16(a).
Figures 16(c) and 16(d) are both taken in the 50-100 keV energy range and show 
good source-centered images of the supernova remnant PSR 1508-59 and the X-ray 
binary EXO 0748-676 respectively. Some source images suffer from source 
confusion.  In particular 4U 1702-429 and 4U 1705-440 (the latter a strong 
low-energy gamma-ray source in its hard state \citep{BarretOlive02}) are close
to both each other and to the category A source OAO 1657-415 but
separate contours delineate the sources in at least some of the grid
positions.  Another case is that of 
4U 1812-12, which is close to GX 17+2, a category A source, but has a much
harder spectrum. Even when it is offset from the center of the 
image, the contours may encompass both sources in channel 1,
but GX 17+2 drops out in channel 2 and the contours tighten around
4U 1812-12. Although these cases show poorly resolved contours, the
spatial resolution is sufficient to support the reported detections of these
sources. The ambiguity often disappears when the source of interest is in the 
center of the field-of-view, as can be seen in Fig 16(b) where the flux from 
4U1812-12 (also called Ser X-2) swamps that from GX 17+2 even in the 20-50 keV 
range.

\section{Use of the Deep Sample Results}

The BATSE EOT deep sample has interesting implications for future studies.
The deep sample can serve
as a guide/comparison tool to an unbiased survey with other instruments such  
as INTEGRAL \citep{Winkler01} and {\it Swift} \citep{Barthelmy00}.  There 
also remains the possibility of an unbiased survey with BATSE \citep{Shaw01,  
Shaw03}.  This catalog
serves as a first critical analysis of the entire BASTE dataset, 
and these results are made available on the world wide web for potential
applications, some perhaps unforeseen.

The spectral
information from BATSE were largely unexplored in this work.
For example, it was not practical for us to present an analysis of the 
long-term spectra in the full 16 channel data that are available.  However,
we made limited comparisons of 
our 4-band results to those of 
the HEAO 1 survey \cite{Levine84}, long-term flux averages ($\sim$5 yrs) of the 
RXTE ASM compiled by
\citet{Grimm02}, and to pointed instrument observations for AGNs and LMXB
neutron stars from the {\it BeppoSAX} and RXTE. This can be done using
the conversion factors given in Table A3, and the Crab Nebula, which averages about 75 c/s
in the 2-12 keV ASM band. Allowing for the differing observation 
intervals involved, we find that our 4-band relative rates
compare well in cross-checks with these results.  
This suggests that the BATSE data 
could be used at a next level to obtain duty fractions for these sources 
in various energy bands.

We find from the category B analysis that some types of sources such as 
supernova remnants and AGNs show hard photon emission even into the 
160-430 keV band as shown in Table 5.
Accreting pulsars, in contrast, have relatively hard spectra near
the BATSE threshold and
sharp spectral cutoffs at higher energy consistent with our 
understanding of their 
emission mechanisms.   Across the ASM (2-12 keV) and  
BATSE (20-40 keV) bands, the fluxes reported
for the accreting pulsars are within a few $\sigma$ 
agreement for the two instruments.  
The high duty fraction LMXB neutron stars with hard X-ray emission from 
pointed-instrument observations
\citep{Barret00} have fluxes that are also comparable with the 
averages given in Table 5.    

We caution users not to over interpret some results, such as the sources
in category C considered as possible detections.
Some could indeed be real detections, but represent the 
limits of the
sensitivity of the current version of the BATSE EOT, and the fluxes
are dominated by the systematic corrections.  For example, a 
few X-ray stars \citep[see also]{White94} appear
in this category (none appeared in categories A or B and most are
in category N), but the fluxes are too weak for testing with the methods 
employed in Sec. 4.
We stress that the category C results definitely can suffer from
source confusion and other systematic errors that could easily 
produce the low-level observed fluxes.  
As is, the 4-band relative   
rates in Table 4 for the category C and N sources should be used only
to derive upper limits on the 9.1y averages.

%Incorporation of the Southampton
%background
%model into OCC\_LITE, as planned, may in fact allow us to extend EOT sensitivity,
%particularly away from the Galactic bulge, to considerably lower flux levels.   

\acknowledgements

This research has made use of the SIMBAD data\-base, operated at CDS, Strasbourg, France. This 
work was supported by the Goddard Space Flight Center Compton Gamma Ray Observatory Science Support 
Center, which is funded by the NASA Office of Space Science. We also 
acknowledge the availability of RXTE ASM data courtesy of the RXTE Science
Support Center.  The authors thank
students Patrick Pecot (Georgia Tech), Lindsay Waite (Vanderbilt), 
and Peter Fimognari (University of Alabama in Huntsville), for 
assistance with spectral analyses and period searches.

\onecolumn
\begin{deluxetable}{lcc}
\tabletypesize{\small}
%\rotate
\tablewidth{0pt}
\tablenum{1}
\tablecolumns{6}
\tablecaption{OF\_W Input Parameters}
\tablehead{\colhead{Parameter}  & \colhead{Description} &\colhead{Value}} 
\startdata
$2\tau$ &  Length of CONT data fitting window &  220s \\
$\theta$$_{cut}$ & Aspect angle cutoff for ``source-pointed" LADs & 60$^{\circ}$ \\ 
$\theta$$_{int}$ & Aspect angle cutoff for interfering sources & 70$^{\circ}$ \\
$\delta$ & Exclusion time for interfering source step & 10s \\
$\lambda$ & Fraction of data packets required in fitting window & 0.95 \\
$\phi$$_{cr}$ & Photon flux minimum level for interfering sources
 & 0.01 ph cm$^{-2}$s$^{-1}$\tablenotemark{a} \\
\enddata
\tablenotetext{a}{Flux value is for the energy band 20-100 keV with
the assumed spectral model for the source (see Sec. 4)}
\end{deluxetable}

\begin{deluxetable}{lccllccc}
%\tabletypesize{\tiny}
\rotate
\tablewidth{0pt}
\tablenum{2}
\tablecolumns{8}
\tablecaption{Occultation (OF\_W) Persistent (P) Sources\tablenotemark{a}}
\tablehead{\colhead{Source Name}  & \colhead{Type} &
\colhead{Fit Range} & \colhead{Model} &
\colhead{Param Value\tablenotemark{b}} & \colhead{Time Range of Fit} & \colhead{Flux (20-100 keV)} &
\colhead{$\chi^{2}$/DOF} \\
\colhead{} & \colhead{} & \colhead{(keV)} & \colhead{} & \colhead{} & 
\colhead{(TJD)} & 
\colhead{(erg cm$^{-2}$s$^{-1}\times$10$^{-8}$)} & \colhead{}} 
\startdata
 1E1740-29   & LMXB?,BHC & 20-600 & PL   & 2.57$\pm$0.23       & 10008-10014 & 0.347 & 13.70/9  \\
 3C273       & QSR       & 20-350 & PL   & 1.7 (1.57$\pm$0.25) & 9517-9521   & 0.077 & 10.06/7  \\
 4U0614+091  & LMXB,NS   & 20-100 & PL   & 2.24$\pm$0.70       & 10005-10008 & 0.162 & 1.67/3   \\
 4U1700-377  & HMXB,NS   & 20-150 & OTTB & 25 (30.7$\pm$4.6)   & 8701-8705   & 0.325 & 3.43/4   \\
 4U1705-440  & LMXB,NS   & 20-150 & PL   & 3.17$\pm$0.45       & 9136-9140   & 0.274 & 1.15/4   \\
 CENX-3      & HMXB,PSR  & 20-150 & OTTB & 10 (10.4$\pm$1.8)   & 9048-9053   & 0.220 & 6.13/4   \\
 CEN\_A      & SY2       & 20-500 & PL   & 2.0 (2.15$\pm$0.10) & 8899-8903   & 0.240 & 4.93/8   \\
 CRAB        & SNR       & 20-200 & PL   & fl* (2.09$\pm$0.02) & 10266-10272 & 1.732 & 5.80/4   \\
\enddata
\tablenotetext{a}{Plots and XSPEC input files for these spectra are available on
\url{http://gammaray.msfc.nasa.gov/batse/occultation/}.}
\tablenotetext{b}{In column 5, parameter values with errors result from spectral fits
described in this table. Parameter values without errors result from historical
spectral fits and were used to generate flux histories. If only a single value
is listed, the spectral model for flux histories was allowed to default to a
power law with photon index = 3.0. For the Crab and Cyg X-1, fl* denotes that
the power law index was allowed to float when flux histories were generated.}
\end{deluxetable}

\begin{deluxetable}{lcclccccc}
\tabletypesize{\scriptsize}
\setlength{\tabcolsep}{0.03in}
\rotate
\tablewidth{0pt}
\tablenum{3}
\tablecolumns{9}
\tablecaption{Occultation (OF\_W) Transient (T) Sources\tablenotemark{a}}
\tablehead{
\colhead{Name}  & \colhead{Type} &
\colhead{Outburst Times}  &  \colhead{Fit Range} &
\colhead{Model} & \colhead{Param Value\tablenotemark{b}} & \colhead{Time Range of Fit} &
\colhead{Flux (20-100 keV)} & \colhead{$\chi^{2}$/DOF} \\
\colhead{}  & \colhead{} &
\colhead{(TJD)}  & \colhead{(keV)} & 
\colhead{} & \colhead{} & \colhead{(TJD)} & \colhead{(erg cm$^{-2}$s$^{-1}\times$10$^{-8}$)} &
\colhead{}} 
%\tablehead{\multicolumn{6}{c}{Pulsed Transients}}
\startdata
%{\bf\small Pulsed Transients}        & \nodata & \nodata & \nodata & \nodata  & \nodata & \nodata & \nodata \\
\multicolumn{9}{l}{\bf\scriptsize Pulsed Transients} \\
\hline \\
2S1417-624    & PSR & 9600-9708, 11542-11578                 & 20-130 & OTTB & 27 (19.1$\pm$2.3)   & 9645-9650   & 0.400 & 7.64/4  \\
4U0115+634    & PSR & 8363-8375, 9483-9523, 10040-10072,     & 20-130 & OTTB & 16.1$\pm$3.3        & 10042-10046 & 0.226 & 6.97/4  \\
\nodata	  & \nodata & 10084-10093, 11228-11280              &\nodata &\nodata&   \nodata           &    \nodata  &\nodata&\nodata  \\ 
4U1145-619    & PSR & 8866-8878, 9054-9078, 9230-9270,       & 20-130 & PL   & 3.77$\pm$0.28       & 10356-10361 & 0.344 & 3.24/4  \\
\nodata	    &\nodata& 9424-9445, 9610-9622, 9805-9811,       &\nodata &\nodata&   \nodata          & \nodata     &\nodata&\nodata  \\
\nodata	    &\nodata& 10186-10198, 10354-10363, 10546-10553  & \nodata& \nodata&  \nodata          & \nodata     &\nodata&\nodata  \\
\enddata
\tablenotetext{a}{Plots and XSPEC input files for these spectra are available on
\protect\url{http://gammaray.msfc.nasa.gov/batse/occultation/}.}
\tablenotetext{b}{
In column 6, parameter values with errors result from spectral
fits described in this table. Parameter values without errors result from 
historical spectral fits and were used to generate flux histories. If only a 
single value is listed, the spectral model for flux histories was allowed to 
default to a power law with photon index = 3.0.}
\end{deluxetable}

\begin{deluxetable}{lllcccccccccc}
\tabletypesize{\tiny}
\rotate
\tablewidth{0pt}
\tablenum{4}
\tablecolumns{13}
\tablecaption{BATSE Earth Occultation Deep Sample Results\tablenotemark{a}}
\tablehead{
\colhead{Name}  & \colhead{Alt Name} &
\colhead{Type} & \colhead{Category\tablenotemark{b}}  & 
\multicolumn{2}{c}{($\alpha$,$\delta$)\tablenotemark{c}}   &
 \multicolumn{2}{c}{({\it l,b})\tablenotemark{c}} &
\colhead{Significance} & 
\multicolumn{4}{c}{Flux (mCrab)\tablenotemark{d,e}} \\
\colhead{}  & \colhead{} &
\colhead{} & \colhead{}  & 
\multicolumn{2}{c}{(deg)}   &
 \multicolumn{2}{c}{(deg)} &
\colhead{} & 
\colhead{20-40} & \colhead{40-70} &
\colhead{70-160} & \colhead{160-430 keV}}
\startdata
SMCX-3 & SMC X-3 & HMXB,PSR & C 
 &  13.02 & -72.43 & 302.86 & -44.69 &    4.0(   5.6)
 &    4.3 $\pm$    1.0 &    0.9 $\pm$    0.8 &   -0.5 $\pm$    0.8 &   -8.4 $\pm$    2.4 \\
SMCX-2 & SMC X-2 & HMXB,PSR & C 
 &  13.64 & -73.68 & 302.63 & -43.44 &    4.1(   5.7)
 &    3.6 $\pm$    1.0 &    1.0 $\pm$    0.8 &    0.6 $\pm$    0.8 &   -1.5 $\pm$    2.4 \\
\bf XTEJ0111-7317 & \bf XTE J0111-732 & \bf HMXB,PSR & \bf \bf A\bf $^{\dagger}$
 & \bf  17.80 & \bf -73.29 & \bf 300.97 & \bf -43.76 & \bf    1.2(   1.7)
 & \bf    2.1 $\pm$    2.0 & \bf    3.0 $\pm$    1.7 & \bf    4.0 $\pm$    1.8 & \bf    0.8 $\pm$    5.2 \\
\bf SMCX-1 & \bf SMC X-1 & \bf HMXB,PSR & \bf \bf B\bf  
 & \bf  19.27 & \bf -73.44 & \bf 300.42 & \bf -43.56 & \bf   28.7(  40.1)
 & \bf   24.0 $\pm$    0.8 & \bf    9.1 $\pm$    0.7 & \bf   -0.6 $\pm$    0.8 & \bf   -6.8 $\pm$    2.1 \\
\bf 4U0115+634 & \bf \nodata & \bf HMXB,PSR & \bf \bf A\bf $^{\dagger}$
 & \bf  19.63 & \bf  63.74 & \bf 125.93 & \bf   1.03 & \bf   13.4(  18.4)
 & \bf    9.9 $\pm$    0.9 & \bf    4.8 $\pm$    0.7 & \bf    1.1 $\pm$    0.7 & \bf    3.4 $\pm$    2.1 \\
TTARI & TT ARI & Nova-like star & N 
 &  31.72 &  15.30 & 148.53 & -43.80 &   -1.2(  -1.7)
 &   -0.4 $\pm$    0.7 &   -0.6 $\pm$    0.7 &   -2.2 $\pm$    0.7 &   -4.2 $\pm$    1.8 \\
LSI+61\_303 & \nodata & HMXB,Radio Star & C 
 &  40.13 &  61.23 & 135.68 &   1.09 &    5.2(   9.5)
 &    3.0 $\pm$    0.9 &    2.3 $\pm$    0.7 &    3.3 $\pm$    0.7 &    5.1 $\pm$    2.1 \\
\bf NGC1275 & \bf \nodata & \bf SY & \bf \bf B\bf  
 & \bf  49.95 & \bf  41.51 & \bf 150.58 & \bf -13.26 & \bf   10.7(  15.0)
 & \bf    7.4 $\pm$    0.8 & \bf    3.0 $\pm$    0.7 & \bf    3.4 $\pm$    0.7 & \bf    4.0 $\pm$    2.0 \\
\enddata
\tablenotetext{a}{Daily 4-band light curves in Crab units for the 83 
category A \& B sources and FITS files containing daily 20-100 keV fluxes 
and 16-channel spectra for all 179 sources are available on 
\url{http://gammaray.msfc.nasa.gov/batse/occultation/}.}
\tablenotetext{b}{A$^{\dagger}$ denotes a category A transient 
source, otherwise the source is considered a persistent source.}
\tablenotetext{c}{Listed positions should be used for reference only. These 
positions were used in the analysis, and many are based on historical 
measurements and are often only accurate to within a few arcminutes. 
Location errors of a few arcminutes do not significantly affect 
flux measurements with the BATSE EOT.}
\tablenotetext{d}{Conversion factors from mCrab units to photon and energy 
fluxes are given in Appendix B, Table B2.}
\tablenotetext{e}{Parenthesis denote cases where the uncorrected flux was 
very small relative to the systematic flux correction. In these cases the 
uncorrected 4-band fluxes are reported because we believe they are closer 
to the true 9.1-year averages. See section 4.3 for details.}
\end{deluxetable}

\begin{deluxetable}{llcccccllll}
\tabletypesize{\small}
\rotate
\tablewidth{0pt}
\tablenum{5}
\tablecolumns{11}
\tablecaption{BATSE Deep Sample Category ``B" Sources}
\tablehead{
\colhead{Name}  &
\colhead{Type} & \colhead{Behavior} &
\colhead{{\it F}}  & \colhead{$F_{cor}$}  &
\colhead{$\delta${\it F}} & \colhead{$\epsilon$$_{cor}$} &
\colhead{(20-40)}  & \colhead{(40-70)} & \colhead{(70-160)} &
\colhead{(160-430)} \\
\colhead{} & \colhead{} & \colhead{} &
\multicolumn{2}{c}{(ph cm$^{-2}$s$^{-1}$)} &
\colhead{(ph cm$^{-2}$s$^{-1}$)} & \colhead{} &
\multicolumn{4}{c}{(mCrab)}}
\startdata
SMCX-1 & HMXB,PSR & P
 &  4.27E-03 &  0.00E+00 &  1.06E-04 & 1.40  
& 24.2$\pm$ 0.6
& 9.2$\pm$ 0.5
&-0.6$\pm$ 0.6
&-.70$\pm$ 1.5
\\
NGC1275 & SY & \nodata
 &  1.84E-03 &  0.00E+00 &  1.23E-04 & 1.40 
& 7.1$\pm$ 0.6
& 3.1$\pm$ 0.5
& 3.3$\pm$ 0.5
& 3.9$\pm$ 1.4
\\
XPER & HMXB,PSR & \nodata
 &  3.90E-03 &  0.00E+00 &  1.17E-04 & 1.40 
& 9.6$\pm$ 0.5
& 12.7$\pm$ 0.5
& 6.5$\pm$ 0.5
&-0.2$\pm$ 1.3
\\
LMCX-4 & HMXB,PSR & P
 &  5.60E-03 &  0.00E+00 &  1.55E-04 & 1.40 
& 20.2$\pm$ 0.6
& 9.4$\pm$ 0.6
& 0.7$\pm$ 0.6
&-8.0$\pm$ 1.5
\\
4U0614+091 & LMXB,NS & P
 &  5.98E-03 & -5.87E-04 &  1.11E-04 & 1.47 
& 20.6$\pm$ 0.6
& 16.6$\pm$ 0.5
& 12.4$\pm$ 0.5
& 8.0$\pm$ 1.4
\\
EXO0748-676 & LMXB,NS & \nodata
 &  3.16E-03 &  0.00E+00 &  1.46E-04 & 1.40 
& 9.2$\pm$ 0.7
& 8.8$\pm$ 0.6
& 5.1$\pm$ 0.6
&-6.1$\pm$ 1.6
\\
VELA & SNR,PSR & \nodata
 &  3.47E-03 & -2.57E-04 &  1.42E-04 & 1.77 
& 9.5$\pm$ 0.6
& 9.3$\pm$ 0.6
& 10.6$\pm$ 0.6
& 11.6$\pm$ 1.6
\\
1E1145-614 & HMXB,PSR & \nodata
 &  6.06E-03 &  8.64E-05 &  1.62E-04 & 1.62 
& 20.5$\pm$ 0.7
& 14.8$\pm$ 0.6
& 4.2$\pm$ 0.7
&-6.5$\pm$ 1.8
\\
\enddata
\tablecomments{Definitions for $F$, $\delta$$F$, $F_{cor}$, and 
$\epsilon$$_{cor}$ are given in Eq. (1).}
\end{deluxetable}

\begin{deluxetable}{lllllll}
\tablecolumns{7}
\tablewidth{0pc}
\tabletypesize{\small}
\tablenum{6}
\tablecaption{Orbital, Superorbital and Quasi- Periods Observed in BATSE EO Light Curves}
\tablehead{\colhead{} &
\multicolumn{3}{c}{Other Instruments/Meth} &  
\multicolumn{3}{c}{Earth Occultation} \\
\cline{2-4} \cline{5-7} \\
\colhead{Name} & \colhead{\it P$_{orb}$ (d)}   & \colhead{\it P$_{long}$ (d)} & 
\colhead{refs} & \multicolumn{2}{c}{\it FFT Peaks}  &
\colhead{FAP}  } 
\startdata
{\bf Category A}        &     &  &      &     &  \\
\hline
%4U 0115+634 & 24.317037 & \nodata\phn & [2] &
%\nodata\phn & \nodata\phn & \nodata\phn \\
%A 0535+262 & 110.3 & \nodata\phn & [2] &
%\nodata\phn & \nodata\phn & \nodata\phn \\
%2S 1417-624 & 42.12 & \nodata\phn & [2] &
%\nodata\phn & \nodata\phn & \nodata\phn \\
%4U 1145-619 & 187 & \nodata\phn & [2] &
%\nodata\phn & \nodata\phn & \nodata\phn \\
%2S 1417-624 & 42.12 & \nodata\phn & [2] &
%\nodata\phn & \nodata\phn & \nodata\phn \\
4U 1700-377 & 3.411581 & \nodata\phn & [4] &
3.411$\pm$0.001 & \nodata\phn & 3.43E-65 \\
%4U 1907+097 & 8.3745  & \nodata\phn & [2] &
%\nodata\phn & \nodata\phn & \nodata\phn \\
CEN X-3 & 2.08706533 & $\sim$140 & [1] &
2.0873$\pm$0.0003 & \nodata\phn & 5.28E-25 \\
CYG X-1 & 5.599829 & 294 & [3,8] &
5.6$^{a}$ & \nodata\phn & \nodata\phn \\
CYG X-3 & 4.792426h & \nodata\phn & [6] &
4.79249h\tablenotemark{a} & \nodata\phn & \nodata\phn \\
EXO2030+375 & 46.0 & \nodata\phn & [7] &
45.9$\pm$0.1 & \nodata\phn & 5.78E-20 \\
GRO J2058+42 & $\sim$110 & \nodata\phn & [2] &
97.5$\pm$0.4 & \nodata\phn & 2.88E-3 \\
GRS 0834-430 & 105.8  & \nodata\phn & [2] &
broad power 97-118 & \nodata\phn & $<$7.48E-6 \\
\enddata
\tablerefs{
[1] \citet{WijersPringle99}
[2] \citet{Bildsten97} 
[3] \citet{PriedhorskyHolt87} 
[4] \citet{Rubin96} 
[5] \citet{Finger99}
[6] \citet{Singh02}
[7] \citet{Wilson02}
[8] \citet{Brocksopp99a}
}
\tablenotetext{a}{Detected only in FFTs of single step data}
\end{deluxetable}

\begin{deluxetable}{lll}
\tablecolumns{3}
\tablewidth{0pc}
\tabletypesize{\small}
\tablenum{7}
\tablecaption{Earth Occultation Imaging Results for Category B Sources}
\tablehead{
\colhead{Source Name}  &   \colhead{$\#$ of Days}   & 
\colhead{Channel/Position Results} } 
\startdata
SMC X-1 & 14 &  visible in all 3 channels, some positions \\
NGC 1275&       14      & visible in all 3 channels, some positions \\
XPER    &       15 &    slight indication of excess in some positions \\
LMC X-4 &       15 &    strong in ch 1-2, visible in 3, most positions \\
4U 0614+091     & 13    & visible in ch 1 when in center of grid \\
EXO 0748-676 & 13       & strong in ch 1-2, most positions, sometimes in ch 3 \\
Vela pulsar     & 11 &  no convincing image -- noisy sky around source \\
1E 1145-614     & 14    & slight indication of excess in some positions, ch 1 \\
\enddata
\end{deluxetable}

\clearpage
\begin{appendix}
\twocolumn
\section{Special Notes on Transient Activity in the BATSE Database}

The available literature on the results of BATSE detection and monitoring for
transient sources is not always complete.  The difficulty of identifying
new sources, in particular, directly stems from the 
limited spatial resolution and 
sensitivity of the BATSE EOT.  As discussed in
Harmon \etal (2002), our ability to spatially resolve point sources on the sky is
limited by the geometry of the Earth's limb at the time of measurement,
coupled with the slow change of this geometry ($\sim$days) as the spacecraft 
orbit precesses.  This is compounded by the sometimes brief duration
of transient outbursts. Some special notes are provided here regarding 
activity that was not as well understood at the time of its occurrence.

{\it EXO 1846-031, GS 1843+00 and GS 1845-024}

These sources, occupying a relatively small (few-degree) region on the sky,
were all detectable at various times in the BATSE database.  First noted in a 
deep search of the BATSE database using epoch folding, a source with a 241 day 
period was given
the name GRO J1849-03 (Zhang \etal 1996a, 1996b). 
At that time, the regularity of the outbursts suggested that 
GRO J1849-03 was a HMXB, but pulsations
were not detected. Later sensitivity enhancements in BATSE pulsar
analysis, however, revealed 94-s pulsations detected at times of outbursts.
This observation, combined with a pulse arrival timing analysis that
revealed an 
orbital period of $242.18 \pm 0.01$ days, identified GRO J1849--03 with 
GS 1845--024 (Finger \etal 1999). A more precise positional
confirmation was also made with the Wide Field Camera on
{\it BeppoSax} (Soffitta \etal 1998).

BATSE observed short-lived transients with a hard spectrum in 1993
(Harmon \etal 1993) and slightly less than a year later (Zhang \etal 1994) 
in this same region of the sky.  The spectral shape, both times, was 
reminiscent of a BHC.
Unfortunately, the limb geometry was not sufficiently to locate the
source to more than a few degrees in one direction within the few days that
the outbursts were observable.  A follow-up observation two weeks after
the 1994 outburst did not indicate any unusual optical activity in the vicinity of EXO 1846-031 (Grindlay, Garcia, \& Zhao 1994).
The BATSE error boxes were consistent with the reported discovery
location of the black hole transient EXO 1846-031 (Parmar \etal 1993), and
so we tentatively identified the outbursts with this source (Zhang \etal 1994).    

In Table 3, we include the regular outbursts of GS 1845-024, a single 
outburst of the pulsar GS 1843+00 (Wilson \etal 1997b) and the two outbursts
with hard spectra attributed to EXO 1846-031.  

\noindent
{\it Galactic Center Transient GRS 1739-278}

A transient source was first detected on 16 Mar 1996 with {\it Granat}/Sigma 
(Paul \etal 1996) in the Galactic center region with a hard spectrum.
A study of its properties with {\it Granat} (Vargas \etal 1997) indicated that
the source, named GRS 1739-278, probably contained a black hole.
Because of its proximity ($\sim$1$^{\circ}$) to a bright source discovered with
BATSE only a few months earlier (GRO J1744-28 or ``bursting pulsar") 
(Fishman \etal 1995; Kouveliotou \etal 1996), it was difficult to separate the
fluxes from the two sources.  The first observations showing a clear detection
of GRS 1739-278 by RXTE/ASM was on TJD 10143 (1 Mar 1996), although a single
measurement exists on TJD 10101 (19 January 1996) that indicates about a
3$\sigma$ detection.  Greiner, Dennerl, \& Predehl
(1996) made a follow-up observation to the {\it Granat} discovery on TJD 10158
(18 Mar 1996) to get a more accurate location for the X-ray counterpart, and
discovered a substantial X-ray scattering halo.   Due to the delay in arrival
time of the scattered photons, the ROSAT results implied that the source
could have first appeared as early as 1995 November-December.  
%In Fig. , we
%show the light curves obtained in the first iteration of the catalog
%with the EOT for GRS 1739-278 and that of GRO J1744-28. We also show the ASM
%light curve for GRS 1739-278, and the BATSE pulsed flux light curve for 
%GRO J1744-28.  
A very sparsely sampled BATSE light curve for GRS 1739-278 indicated the source was visible as early as $\sim$10035 (14 Nov 1995) at a level of $\sim$200 mCrab
in the 20-100 keV band, and decaying below the source confusion limit
($\sim$50 mCrab) in the Galactic center region by $\sim$10497 (18 Feb 1997).  
These beginning and ending dates for the outburst were entered into the
flare database and Table 3.

{\it RXTE Transients XTE J1550-564, XTE J1859+226, XTE J1118+480,
XTE J1543-568, XTE J0111-7317 and V4641 SGR (Sagittarius)}  

These transients were discovered in outburst by RXTE during the final years 
of the CGRO mission.
Several of these transients were quite bright and  \\ reached 
a Crab or more in the BATSE energy bandpass.  The RXTE/ASM, which was 
designed to quickly locate and monitor new transients, generally was able to 
report these outbursts rapidly, sometimes even before BATSE data containing 
the same transient were received for processing.  It is interesting scientifically, as well as important to our accounting of bright sources,
to examine archival data for these transients, especially prior to the 
launch of RXTE in 1995.

As part of the deep sample, light curves were generated for XTE J1550-564,
XTE J1859+226, XTE J1118+480, XTE J1543-568, XTE J0111-7317 and V4641 SGR.
All showed detectable outbursts as re\-cor\-ded in Table 3.  Examination of
light curves from the beginning of the CGRO mission in April 1991
prior to the known outbursts occurring at various times after December 1995
did not show any emission more than five successive days 
greater than $\sim$100
mCrab intensity in the 20-100 keV band.  Outbursts of lesser duration 
and intensity
could possibly be pre\-sent in the dataset, but require more 
in-depth analysis.

\section{Energy Channel Mapping Scheme, Flux Conversion Factors and
Data Problems}

Pulse height data from the BATSE large area 
detectors were normally acquired into 128 high energy 
resolution (HER) channels.   Due to telemetry limits, the HER data
were mapped into 16 broader 
energy channels and read out at 2.048s to create the continuous (CONT) background data.  
EOT measurements were made with the CONT dataset.
Normally the CONT channel bin edges were kept to a constant set of values, 
that,
coupled with the LAD gain control, kept the effective energy loss 
pulse height response per channel extremely stable.
However, on rare occasions, the 128 to 16 channel mapping was
changed to accommodate a 
greater energy dispersion when
observations of sources with softer
spectra and possible line features 
were desired (e.g., accreting pulsars or soft gamma-ray repeaters).  
Table B1 shows several mapping
schemes (numbered 0, 1, 2, 3, 4a and 4b) that were used during the 
mission and the times in which 
these were applied.  From column 2, it can be seen that 
scheme 2 was used about 90\% of the time.

Schemes 2, 3, 4a and 4b had in common several of the same HER channels for 
which we could obtain
4-energy band data spanning changes in the HER to CONT channel mapping.  
Using the Crab fluxes as a 
standard candle, a tool was created for which the
flux of any source relative to the Crab could be generated in the energy 
bands 20-40, 40-70, 70-160
and 160-430 keV.  The Crab relative rates 
quoted in Tables 4 and 5 can be converted to 
photon number per keV (ph cm$^{-2}$s$^{-1}$ keV$^{-1}$) or 
energy flux per band using the multipliers given in Table B2.  Note that this method of four-band conversion to Crab relative rates cannot be 
used for schemes 0 or 1, and so data before TJD 8406/81139 and for 
the brief intervals TJD 8807/8279 -
8809/64510 and 8812/65817 - 8812/66102 cannot be treated in this manner. 

Table B3 is a list of days when unusual or anomalous
events occurred during the CGRO mission and no Earth occultation 
measurements could be extracted.  
These events produced gaps in most or all source flux histories at the given 
times. Routine events causing loss of measurements
such as source interference, exceeding the critical angle for 
Earth occultation, or telemetry gaps are not included.  
Table B3 gives the times, data condition reported, and the cause of 
the anomaly.

\section{Comparison to JPL Enhanced BATSE Occultation Package (EBOP)}

Skelton \etal (1994), Ling \etal (1996, 2000) and co-workers at the Jet Propulsion 
Laboratory developed an Earth occultation software package over a period
of several years with the intent to generate higher sensitivity
results than can be obtained with the treatment 
described in Harmon \etal (2002).  The JPL package
is known as the Enhanced BATSE Occultation
Package (EBOP) and has been applied to BATSE data independent of
the efforts at MSFC.   Ling \etal (2000) published a compendium of 
measurements (also called a catalog) for the  
1991 May to 1994 October epoch.  We feel it is very important 
to compare the results from EBOP with those of this work, 
since there are
significant discrepancies between what we report and what can be found
in Ling \etal (2000).  

Primarily, there are two major differences between
the JPL method and the method used here. First, EBOP uses a 
semi-physical model for the detector background counting
rates. The model is based on expected contributions of low-energy gamma-ray fluxes
local to the low Earth orbit (LEO) environment.  These include cosmic ray
secondary radiation and activation products from orbital passes through
radiation fields in LEO.   The JPL 
global background model consists of a mix of
the local radiation components and a combination (determined by the fit)
of the BATSE spectroscopic detector (Fishman \etal 1989a,b) counting rates as a predictor of the low-energy background in the
LADs.   Secondly, the extraction of source signals is usually performed 
in one-day segments,
with a single fit including terms for {\it all} sources in the 
EBOP catalog
with no {\it a priori} assumptions of their intensity.  The main goal of
this work is increase sensitivity by fitting much longer intervals of 
background data
than the four minutes typically used in our work,
which does not require a sophisticated background model. 

In Table C1, we compare selected results for a few sources
from the deep sample in Table 4 and the EBOP catalog (Ling \etal 2000).  
To best illustrate the comparison, we selected sources over a range of 
flux intensities, as well as to pick
sources that did not show a large degree of 
variability on timescales of months to years. 
Table C1 shows fluxes in three selected energy bands with centroid 
energies that overlap in EBOP and the Table 4 deep sample. Results from
the deep sample are quoted
with and without systematic error corrections. The EBOP results are
reported as minimum and maximum average fluxes for 
for several $\sim$400 day intervals in Table 3 of
Ling \etal (2000).  The deep sample results are the 9.1y averages from 
this work (Table 4) converted from mCrab to photon flux with 
the multiplicative factors from Table B2.
For the Crab, the two methods 
give very similar results, and confirm the cross-check of the
of the two methods from
Ling \etal (2000).  Sources of moderate
intensity, such as 4U 1700-37 and NGC 4151, show significant discrepancies, 
particularly for the highest of the three energy bands.
For relatively weak sources at the few mCrab level, such as the LMXB Circinus X-1, the supernova
remnant PSR 1509-58, and the Seyfert Galaxy NGC 1275,
our average fluxes are roughly an order of magnitude less than the EBOP
results. For Sct X-1, we only show upper limits ($<$2$\sigma$) (corrected
for systematic error), whereas
the JPL method yielded significant broad-band emission at least 
an order of magnitude higher. 

To make sure that the time interval for which the averages were 
constructed was
not a major factor in the comparison, we generated long term spectra 
for TJD 8393-8800 as
was done in Ling \etal (2000).  Figures 17a and 17b show the broad-band spectra
of Cir X-1 and and PSR 1509-58 using the MSFC method with no systematic
error correction applied, the EBOP
results from Ling \etal (2000), and the Table
4-band average fluxes with systematic error corrections.  The broad-band  
spectra from the JPL and MSFC methods are considerably different
over most of the sensitive energy range of the LADs.  Differences
between the broad-band MSFC data and the 4-band averages for Cir X-1
are mostly caused by the large systematic error correction for
that source.  The findings reflect the same conclusions made using
Table C1.

Ling \etal (2000) discuss sources of error for the Earth occultation
analysis specific to EBOP in some detail and point out that the
high energy
fluxes they see for some sources such as Sct X-1 and NGC 1275 should
be viewed with caution.  Since we
observe considerably lower fluxes for these and other sources in
the 160-430 keV band, it suggests that their concern could be  
justified.  

Here we point out possible reasons for the discrepancy
between our work and the EBOP results.
One possibility concerns unaccounted for components in the
gamma-ray background for which 
the LADs have significant sensitivity, but are not included in the JPL
model.  We have found that, when fitting the background near
occultation steps for more than a few minutes,
there is an increased likelihood for 
slow, very significant variations 
to cause the measured background to be considerably 
different from the assumed model.  These variations are not associated in any
way with occultation steps.  Such effects could be
high energy diffuse emission from the Galactic plane or unusual local
background variations on specific days (e.g., solar activity or 
electron precipitation events) that can affect the computed level of 
the background in the high energy (a few hundred keV) data. 
These variations from the model can be transferred to source terms
which are fit simultaneously with the background. Another
possibility is the assumed set of sources used by the JPL method.  
JPL used a set of 64 sources, all of which are fit simultaneously, regardless
of any intensity or time information.  
Our catalog consisted of 179 sources, however, no more than 10 sources
were fit simultaneously in the MSFC method, and usually only 1-3 per
4-minute fitting window were typical.  This reduces the tendency toward
large scale coupling between source terms, and presupposes that we 
have partial information about what sources a detector is likely to be
seeing.
This may not explain anomalously large fluxes at high energy in
the JPL data, but it can help to explain the tendency to yield fluxes
higher than the MSFC method for lower intensity sources.
Although a finite source set is also a problem for the MSFC method, the 
strategy was to iteratively improve characterization of the sources in the 
catalog based on
average brightness, times of transient activity, and to use a source
set taken from published catalogs.  
Our method therefore 
yields improved results by having a better knowledge of source 
activity each time a pass through the dataset is performed.

\onecolumn
\begin{deluxetable}{ll}
\tablecaption{LIST OF ACRONYMS AND ABBREVIATIONS}
\tablenum{A1}
\tablecolumns{2}
\tablewidth{0pt}
\tablehead{\colhead{Acronym or Abbrev.} & \colhead{Meaning}}
\startdata
AGN & active galaxy or galactic nucleus \\ 
ASM & All-Sky Monitor \\
BATSE & Burst and Transient Source Experiment \\
BH or BHC & black hole candidate \\
BL Lac & BL Lacertae object \\
CGRO & {\it Compton Gamma Ray Observatory} \\
CH & channel \\
CONT &  large area detector continuous data \\
CV & cataclysmic variable \\
%DISCLA & large area detector discriminator data \\
DN & dwarf nova \\
EBOP & JPL Enhanced BATSE Occultation Package \\
EGRET & Energetic gamma-ray Experiment Telescope \\
EXIST & Energetic X-Ray Imaging Survey Telescope \\
%EGS & Stanford electromagnetic cascade and transport code \\
%EKU & Eastern Kentucky University \\
EOT & Earth occultation technique \\
FFT & fast Fourier transform \\
%FOV & field of view \\
%GEANT & Southampton electromagnetic cascade and transport code \\
GCR & Galactic center region \\
%GRB & gamma-ray burst \\
%GSFC & Goddard Space Flight Center \\
%GRIS & Gamma-Ray Imaging Spectrometer \\
HEAO & {\it High Energy Astronomy Observatory} \\
HER & high energy resolution \\
%HEXEL & honeycomb-like aluminum-epoxy composite \\
%HEXTE & High Energy X-Ray Timing Experiment (RXTE) \\
HISGEN & BATSE photon flux history generator \\
HMXB & high mass x-ray binary \\
%ICAO & International Civil Aviation Organization \\
INTEGRAL & {\it International Gamma-Ray Astrophysics Laboratory} \\
JPL & Jet Propulsion Laboratory \\
LAD & BATSE large area detector \\
%LEO & low Earth orbit \\
LMXB & low mass x-ray binary \\
%MEM & maximum entropy method \\
%MJD & Modified Julian Date (Julian Date - 2,400,000.5) \\
MSFC & Marshall Space Flight Center \\
NS & neutron star \\
OCCSYS & automated data flagging software \\
OF\_W & MSFC EOT flux history generator \\
%OCC\_LITE & platform-independent EOT flux history generator \\
%OSSE & Oriented Scintillation Spectrometer Experiment \\
OTTB & optically thin thermal bremsstrahlung spectral fitting model \\ 
%PCA & Proportional Chamber Array \\
PL & power law spectral fitting model \\
PSR & pulsar \\
QSR & quasar \\
ROSAT & German X-Ray Roentgen Satellite  \\
RS CVn & eruptive variable of RS Canum Venaticorum type. \\
RXTE & {\it Rossi X-Ray Timing Explorer} \\
%SAA & South Atlantic Anomaly \\
S/C & spacecraft \\
%SD & BATSE spectroscopy detector \\
SGR & soft gamma-ray repeater \\
SNR & supernova remnant \\
%S/N & signal-to-noise \\
SOD & seconds of day \\
SOI & source of interest \\
SY(1 or 2) & Seyfert galaxy, type 1 or 2 \\
%TDRSS & Tracking \& Data Relay Satellite System \\
TJD & Truncated Julian Date (Julian Date - 2,440,000.5) \\
UVS & ultraviolet source \\
%WFC & Wide Field Camera \\
XRS & X-ray source \\
\enddata
\end{deluxetable}

\begin{deluxetable}{llllcccccccccccccccc}
\tabletypesize{\scriptsize}
\rotate
\tablewidth{0pt}
\tablenum{B1}
\tablecolumns{20}
\tablecaption{BATSE LAD 128-Channel to 16-Channel Mapping (Lookup Table)}
\tablehead{\colhead{Scheme \#} & 
\colhead{\%Time} & \colhead{TJD}  & \colhead{SOD} &
\multicolumn{16}{c}{CONT energy channel} \\
\colhead{}  & \colhead{} &
\colhead{(Days)} & \colhead{(secs)} & \colhead{0}  & \colhead{1} &
\colhead{2}  & \colhead{3} &
\colhead{4}  & \colhead{5} &
\colhead{6}  & \colhead{7} &
\colhead{8}  & \colhead{9} &
\colhead{10} & \colhead{11} &
\colhead{12} & \colhead{13} &
\colhead{14} & \colhead{15}}

\startdata
0 & 0.2 & 8361 & 0     & 0 & 13 & 15 & 17 & 19 & 22 & 25 & 30 & 37 & 49 & 64 & 74 & 89 & 100 & 111 & 125 \\
1 & 1.2 & 8367 & 0     & 0 & 9 &  11 & 14 & 18 & 22 & 26 & 30 & 38 & 49 & 65 & 75 & 90 & 98 &  106 & 122 \\
2 & 12.0 & 8406 & 81139 & 0 & 7 &  9  & 11 & 14 & 18 & 23 & 28 & 36 & 49 & 65 & 75 & 90 & 98 &  106 & 122 \\
0 & 0.1  & 8807 &  8279 & 0 & 13 & 15 & 17 & 19 & 22 & 25 & 30 & 37 & 49 & 64 & 74 & 89 & 100 & 111 & 125 \\
2 & 0.1 & 8809 & 64511 & 0 & 7 &  9  & 11 & 14 & 18 & 23 & 28 & 36 & 49 & 65 & 75 & 90 & 98 &  106 & 122 \\
0 & 0.0 & 8812 & 65817 & 0 & 13 & 15 & 17 & 19 & 22 & 25 & 30 & 37 & 49 & 64 & 74 & 89 & 100 & 111 & 125 \\
2 & 17.7 & 8812 & 66103 & 0 & 7 &  9  & 11 & 14 & 18 & 23 & 28 & 36 & 49 & 65 & 75 & 90 & 98 &  106 & 122 \\
3 & 0.1 & 9400 & 59319 & 7 & 9 &  11 & 14 & 18 & 20 & 23 & 25 & 28 & 32 & 36 & 49 & 65 & 75 &  90  & 122 \\
2 & 19.3 & 9419 & 60041 & 0 & 7 &  9  & 11 & 14 & 18 & 23 & 28 & 36 & 49 & 65 & 75 & 90 & 98 &  106 & 122 \\
4a& 3.5 & 10062 & 71275&122 &7 &  8  &  9 & 10 & 11 & 12 & 14 & 16 & 18 & 23 & 28 & 36 & 49 &  75  & 90  \\
2 & 8.6 & 10178 & 85302& 0 & 7 &  9 &  11 & 14 & 18 & 23 & 28 & 36 & 49 & 65 & 75 & 90 & 98 &  106 & 122 \\
4b& 1.2 & 10465 & 83335& 0 & 7 &  8  &  9 & 10 & 11 & 12 & 14 & 16 & 18 & 23 & 28 & 36 & 49 &  75  & 122 \\
2 & 14.4& 10504 & 80991& 0 & 7 &  9 &  11 & 14 & 18 & 23 & 28 & 36 & 49 & 65 & 75 & 90 & 98 &  106 & 122 \\
4b& 2.1 & 10983 & 71907& 0 & 7 &  8  &  9 & 10 & 11 & 12 & 14 & 16 & 18 & 23 & 28 & 36 & 49 &  75  & 122 \\
2 & 0.1 & 11053 & 51166& 0 & 7 &  9 &  11 & 14 & 18 & 23 & 28 & 36 & 49 & 65 & 75 & 90 & 98 &  106 & 122 \\
4b& 0.1 & 11055 & 54394& 0 & 7 &  8  &  9 & 10 & 11 & 12 & 14 & 16 & 18 & 23 & 28 & 36 & 49 &  75  & 122 \\
2 & 18.2 & 11085 & 74663& 0 & 7 &  9 &  11 & 14 & 18 & 23 & 28 & 36 & 49 & 65 & 75 & 90 & 98 &  106 & 122 \\
\enddata
\end{deluxetable}

\begin{deluxetable}{lllll}
\tablecaption{Conversion Factors for 4-Energy Band Data in Table 4}
\tablenum{B2}
\tablecolumns{4}
\tablewidth{0pt}
\tablehead{\colhead{HER Channels} & \colhead{Energy Band} & 
\colhead{Log Avg} & \colhead{Multiplier} & \colhead{Multiplier} \\
\colhead{} & \colhead{(keV)} & \colhead{(keV)} & \colhead{(ph cm $^{-2}$s$^{-1}$keV$^{-1}$)} 
& \colhead{(ergs cm$^{-2}$s$^{-1}$)}}

\startdata
7 - 11 & 20-40 & 28.3 & 8.508 $\times$ 10$^{-6}$ & 7.567 $\times$ 10$^{-12}$ \\
11 - 18 & 40-70 & 52.9 & 2.323 $\times$ 10$^{-6}$ & 5.830 $\times$ 10$^{-12}$ \\
18 - 36 & 70-160 & 105.8 & 5.514 $\times$ 10$^{-7}$ & 8.079 $\times$ 10$^{-12}$ \\
36 - 75 & 160-430 & 262.3 & 6.261 $\times$ 10$^{-8}$ & 6.887 $\times$ 10$^{-12}$ \\    
\enddata
\tablecomments{Lower HER channel is the lower energy edge of the corresponding 
CONT channel and the higher HER channel is the corresponding upper edge of the 
corresponding CONT channel as given in Table B1.  Example: HER channels 7 - 11,
corresponds to CONT channels 1 - 2 in scheme 2 and CONT channels 1 - 4 in 
scheme 4a.}
\end{deluxetable}

%\begin{deluxetable}{lll}
%\tabletypesize{\scriptsize}
%\tablecaption{List of Mission Anomalies/Data Loss Events Affecting BATSE 
%Earth Occultation Results}
%\tablenum{B3}
%\tablecolumns{3}
%\tablewidth{0pt}
%\tablehead{\colhead{TJD/SOD} & \colhead{Condition} & \colhead{Cause of Anomaly}}
%\startdata
%8361 & HKG read, status=9 & no data for OCCULT5 mask \\ 
%8588 & observing plan read error, status=36 & ? \\
%8807/03065 - 66377, 68578 - 71280  & HV off & S/C power supply problem \\
%8807/66300 - 8809/53160 & LAD gains out of balance &  S/C power supply (cont.) \\
%9111 & HV off & reboost activities \\
%9112 & observing plan read error & reboost-related \\
%9114 & HV off & reboost activities \\
%9115 & observing plan read error & reboost-related \\
%9153 & HV off & reboost activities \\
%9336 & observing plan read error & reboost-related \\ 
%10415/16347 - 10418/69320 & LAD gains out of balance & flight software upset \\
%11001/11000-50000 & LAD gains out of balance & SAA passage w/HV on \\
%11001/50000-86400 & LADs 0,6 gains out of balance & SAA passage w/HV on \\
%11046/74756 - 11047/28002 & HV off & loaded bad solar ephemeris \\
%11047/30897 -11050/77609 & HV off & bad solar ephemeris \\  
%11354-11355 &     & HGA problem \\
%11518-11521 &      & CGRO gyro 3 failure \\
%\enddata
%\end{deluxetable}

\begin{deluxetable}{lll}
\tabletypesize{\scriptsize}
\tablecaption{List of Mission Anomalies/Data Loss Events Affecting BATSE
Earth Occultation Results}
\tablenum{B3}
\tablecolumns{3}
\tablewidth{0pt}
\tablehead{\colhead{TJD/SoD} & \colhead{Condition} & \colhead{Cause of
Anomaly}}
\startdata
8449/05500-12000 & some LAD gains out of balance & SAA passage w/HV on \\
8490/03000-35000 & some LAD gains out of balance & SAA passage w/HV on \\
8588 & observing plan read error & ? \\
8741/80355 - 8742/11519 & data crash & flight software problem \\
8742/63376 - 8742/83566 & data crash & flight software problem \\
8807/03065-66377, 68578-71280 & HV off & S/C power supply problem \\
8807/66300 - 8809/53160 & LAD gains out of balance & S/C power supply
(cont.) \\
8812/65733-65964, 68510-71490 & HV off & change S/C power supply \\
8812/65960-68512 & LAD gains out of balance & S/C power supply (cont.) \\
9075/57503-62797 & data crash & flight software problem \\
9075/79431 - 9076/09796 & data crash & flight software problem \\
9110/84820 - 9112/54665 & HV off & reboost activities \\
9113/81403 - 9115/53135 & HV off & reboost activities \\
9152/84296 - 9154/16964 & HV off & reboost activities \\
9336/49077-65666 & HV off & reboost activities \\
9336/65666-75000 & some LAD gains out of balance & reboost activities \\
10014/03349-15300 & LAD gains out of balance & SAA passage w/HV on \\
10014/15300-65700  & LAD 3 gain out of balance & SAA passage w/HV on; LAD 3
reset \\
10014/15300 - 10016/10008  & LAD 6 gain out of balance & SAA passage w/HV
on; HV reset \\
10416/16897-19200 & bad rates & recovery from flight software upset \\
10417/70332-71000 & bad rates & recovery from flight software upset \\
10416/16897 - 10418/69320 & LAD gains slightly out of balance & flight
software upset \\
10900 - 11690 & LAD 7 gain decreased & AGC caused HV to reach maximum value \\
11001/11000-50000 & LAD gains out of balance & SAA passage w/HV on \\
11001/50000 - 11002/30400 & LADs 0,6 gains out of balance & SAA passage
w/HV on \\
11002/30400 - 11003/46570 & LAD 0 gain out of balance & SAA passage w/HV on \\
11003/46570-50500 & LADs 0,2,6 gains out of balance & HV reset \\
11046/74756 - 11047/28002 & data lost & bad solar ephemeris \\
11047/30897-71816 & data lost & bad solar ephemeris \\
11047/71816 - 11050/77609 & HV off & bad solar ephemeris \\
11307/04563-31857 & data lost & high-gain antenna problems \\
11354/03716 - 11359/72432 & data lost & high-gain antenna problems \\
11518/55813 - 11522/00089 & data lost & gyro failure \\
\enddata
\end{deluxetable}
\begin{deluxetable}{llllcc}
\tablecaption{Comparison of JPL EBOP and MSFC Catalog Results}
\tablenum{C1}
\tabletypesize{\scriptsize}
\rotate
\tablewidth{0pt}
\tablecolumns{6}
\tablehead{\colhead{Source} & \colhead{Energy Band}  & \colhead{EBOP} & 
\colhead{Energy Band} & \colhead{MSFC} &
\colhead{MSFC/w cor.} \\
\colhead{} & \colhead{(keV)} & \colhead{ph cm$^{-2}$s$^{-1}$keV$^{-1}$} &
\colhead{(keV)} & 
\colhead{ph cm$^{-2}$s$^{-1}$keV$^{-1}$} & 
\colhead{ph cm$^{-2}$s$^{-1}$keV$^{-1}$}}
\startdata
CRAB & 45-55 & 2.58$-$3.10$\times$10$^{-3}$ & 40-70 & 2.36$\times$10$^{-3}$ & 2.38$\times$10$^{-4} $ \\
     & 98-123 & 5.10$-$5.60$\times$10$^{-4}$ & 70-160 & 5.61$\times$10$^{-4}$ & 5.64$\times$10$^{-4}$ \\
     & 230-313 & 6.14$-$7.01$\times$10$^{-5}$ & 160-430 & 6.47$\times$10$^{-5}$ & 6.50$\times$10$^{-5}$ \\
4U1700-37 & 45-55 & 4.89$-$5.39$\times$10$^{-4}$ & 40-70 & 3.49$\times$10$^{-4}$ & 3.48$\times$10$^{-4}$ \\
     & 98-123 & 3.99$-$4.34$\times$10$^{-5}$ & 70-160 & 3.52$\times$10$^{-5}$ & 3.51$\times$10$^{-5}$ \\
     & 230-313 & 1.70$-$4.44$\times$10$^{-6}$ & 160-430 & 8.39$\times$10$^{-7}$ & 8.39$\times$10$^{-7}$ \\
NGC4151 & 45-55 & 9.93$-$16.1$\times$10$^{-5}$ & 40-70 & 9.80$\times$10$^{-5}$ & 9.80$\times$10$^{-5}$ \\
     & 98-123 & 2.18$-$2.88$\times$10$^{-5}$ & 70-160 & 2.04$\times$10$^{-5}$ & 2.04$\times$10$^{-5}$ \\
     & 230-313 & 2.69$-$4.08$\times$10$^{-6}$ & 160-430 & 1.31$\times$10$^{-6}$ & 1.31$\times$10$^{-6}$ \\
CIRX-1 & 45-55 & 6.66$-$7.87$\times$10$^{-5}$ & 40-70 & 8.13$\times$10$^{-6}$ & $<$6.97$\times$10$^{-6}$ \\
     & 98-123 & 1.91$-$2.48$\times$10$^{-5}$ & 70-160 & 1.87$\times$10$^{-6}$ & $<$1.65$\times$10$^{-6}$ \\
     & 230-313 & 4.99$-$7.18$\times$10$^{-6}$ & 160-430 & $<$2.69$\times$10$^{-7}$ & $<$3.44$\times$10$^{-7}$ \\
NGC1275 & 45-55 & 4.59$-$5.82$\times$10$^{-5}$ & 40-70 & 6.97$\times$10$^{-6}$ & 6.97$\times$10$^{-6}$ \\
     & 98-123 & 1.26$-$1.70$\times$10$^{-5}$ & 70-160 & 1.87$\times$10$^{-6}$ & 1.87$\times$10$^{-6}$ \\
     & 230-313 & 3.60$-$4.20$\times$10$^{-6}$ & 160-430 & 2.50$\times$10$^{-7}$ & 2.50$\times$10$^{-7}$ \\
PSR1509-58 & 45-55 & 5.23$-$8.77$\times$10$^{-5}$ & 40-70 & 2.37$\times$10$^{-5}$ & 1.60$\times$10$^{-5}$ \\
     & 98-123 & 1.48$-$2.61$\times$10$^{-5}$ & 70-160 & 6.67$\times$10$^{-6}$ & 4.52$\times$10$^{-6}$ \\
     & 230-313 & 4.48$-$5.97$\times$10$^{-6}$ & 160-430 & 8.95$\times$10$^{-7}$ & 6.07$\times$10$^{-7}$ \\
SCTX-1 & 45-55 & 1.53$-$1.67$\times$10$^{-4}$ & 40-70 & $<$4.65$\times$10$^{-7}$ & $<$2.32$\times$10$^{-6}$ \\
     & 98-123 & 2.31$-$2.91$\times$10$^{-5}$ & 70-160 & $<$1.49$\times$10$^{-6}$ & $<$1.43$\times$10$^{-6}$ \\
     & 230-313 & 4.26$-$5.50$\times$10$^{-5}$ & 160-430 & 2.50$\times$10$^{-7}$ & $<$4.51$\times$10$^{-7}$ \\
\enddata
\end{deluxetable}
\clearpage

\section*{FIGURE CAPTIONS}

\figcaption{Comparison of sensitivity (5$\sigma$) in photons 
cm$^{-2}$s$^{-1}$keV$^{-1}$ 
for BATSE, the 1977-1979 HEAO 1 A-4 sky survey (Levine \etal 1984) and the 
survey concept mission EXIST (Grindlay \etal 2001, 2003). Nominal energy   
bins for the two historical instruments are shown.  An approximate 
representation of the Crab Nebula spectrum is shown for reference.}

\figcaption{Aitoff-Hammer projection of the sky in Galactic coordinates 
showing the locations of the 179 sources in the BATSE Earth occultation 
catalog.  The sources are labeled by object type.}

\figcaption{(a) The {\it Compton Gamma Ray Observatory} 
showing placement of the BATSE detector modules. The detector modules on the top
four corners of the spacecraft are numbered 0,2,4, and 6, proceeding clockwise
from the detector facing the reader above the high gain antenna. Similarly the 
bottom four modules are numbered 1,3,5, and 7.(b) Major components
of the BATSE detector modules (one of eight).}
%\label{fig:cgro.ps}}
%\label{fig:module.ps}

\figcaption{Flow diagram for processing of BATSE Earth Occultation
data. Part (a) shows major processing steps beginning with components of 
the BATSE data archive on the far left (see text) leading to
raw (counts per detector) rate histories. The raw histories were then
converted to photon fluxes (20-100 keV) using the BATSE detector response, 
and 
also to intensities relative to the Crab Nebula flux in four flux bands
(20-40, 40-70, 70-160 and 160-430 keV). 
Part (b) shows the final processing 
steps (epoch folding, FFT analysis, deriving average flux and uncertainty)
in generating the tables in the catalog.  The ``Crab model" allows
conversion of raw count rates directly into Crab relative fluxes (see text).
Iteration \#{\it n} refers to a single pass through the entire 9.1y
dataset, which can be repeated to increase the size of the sky sample
and to reduce systematic error.}

\figcaption{Eclipse profiles in the 20-40 keV band for the five 
bright persistent HMXBs 4U 1700-377, Cen X-3, Her X-1, OAO 1657-415 and
Vela X-1 in the BATSE dataset.  
Published ephemerides 
are used to eliminate fitting terms for these binaries while
eclipses are in progress. 
Table 6 indicates the 
orbital periods used in epoch folding of the 
occultation flux histories shown in the figure.}

\figcaption{Flux histories (20-100 keV, 1991 April - 2000 June) 
for four persistent black hole binary systems observed with BATSE.}

\figcaption{Flux histories (20-100 keV, various times) for
five transient black hole binary systems.}

\figcaption{Flux histories (20-100 keV, 1991 April - 2000 June) for
five low mass X-ray binary neutron star systems.}

%\figcaption{(a) Flux histories (20-100 keV, 1991 April - 2000 June) 
%for six persistent accreting X-ray pulsars. 
%(b) Flux histories (20-100 keV, various times) for four transient 
%X-ray pulsars.} 
\figcaption{(a) Flux histories (20-100 keV, 1991 April - 2000 June) for six persistent
accreting X-ray pulsars. For purposes of this catalog, a persistent source is 
defined as consistently exceeding 0.01 photons cm$^{-2}$ s$^{-1}$ (20-100 keV, 
$\sim 35$ mCrab). Two of these systems GX 1+4 ($P_{\rm spin} \simeq 120$ s) and
Her X-1 ($P_{\rm spin} \simeq 1.24$ s) have a low-mass companion, while the rest,
Cen X-3 ($P_{\rm spin} \simeq 4.8$ s), GX 301--2 ($P_{\rm spin} \simeq 681$ s),
OAO 1657--415 ($P_{\rm spin} \simeq 10.4$ s), and Vela X-1 ($P_{\rm spin} 
\simeq 283$ s) have a high-mass evolved supergiant companion.
(b) Flux histories (20-100 keV, various times) for four transient X-ray pulsars.
For purposes of this catalog, a transient source is defined as exceeding 0.01 
photons cm$^{-2}$ s$^{-1}$ (20-100 keV, $\sim 35$ mCrab) for an identifiable,
but limited period of time. All four sources shown, EXO 2030+375 ($P_{\rm spin}
\simeq 42$ s), GRO J1944+26 ($P_{\rm spin} \simeq 15.8$ s), GRS 0834--430 
($P_{\rm spin} \simeq 12.3$ s), and GS 1845--02 ($P_{\rm spin} \simeq 94.8$ s),
have an identified Be star companion.}
%EXO 2030+375 (top) and GS 1845--02 (bottom)
%are Be/X-ray binaries that have outbursted for nearly every periastron passage
%during the BATSE mission. GRO J1944+26 and GRS 0834-430 (center panels) are more
%typical Be/X-ray binaries, undergoing short series of outbursts. Each point plotted is
%a 10 day average EOT measurement. \label{fig:pulsed_sources}}

\figcaption{Flux histories (20-100 keV, 1991 April - 2000 June) for four galaxies with 
gamma ray active nuclei.}

\figcaption{Sky grid, consisting of 162 points, used for determination of
location-dependent systematic error.  Grid spacing is described in the text.}
%\label{fig:gal162_grid.ps}}

\figcaption{(Top) Measured averages in photons cm$^{-2}$s$^{-1}$ (20-100 keV) and 
(Bottom) width (in $\sigma$s or standard deviations)
of the 9.1y flux histories for the grid locations shown in Fig. 11 
as a function of Galactic longitude.}
%\label{fig:gal_plots2.ps}}

%\figcaption{Intensity of systematic deviations from statistical zero as a
%function of Galactic latitude and longitude.  Solid points indicate positive
%deviations and open circles denote negative deviations.}

\figcaption{Distribution of 9.1y flux averages (20-100 keV) in order of 
brightness before (filled circles)
and after correction with systematic error model. Symbols denote source
categories used for the deep sample results.}

\figcaption{Histogram of fractional changes in flux significance of the
deep sample when systematic errors were applied.  The peak in the histogram 
at $\sim$-0.3 corresponds to cases where the 
correction to the flux is zero, and the correction is dominated by
the correction to the flux uncertainty.  Systematic corrections to
the average fluxes in Table 4 were only applied when the computed 
fractional change was between the two vertical dashed lines.}

\figcaption{(a) Flux histories in the indicated energy band for the three category B sources 
(LMC X-1, SMC X-1 and 4U 1538-522) detected in the FFT analysis
and epoch folded at the appropriate orbital periods as given in Table 6. 
We also show 
the epoch folded history result of the category C(B) source 4U 1907+097  
detected in the BATSE EOT light curves by Laycock \etal (2003).
(b) Flux histories in the indicated energy band for the two category B sources 
LMC X-4 and SMC X-1 epoch folded at the appropriate
superorbital periods as given in Table 6.  Note the effect of the
superorbital period shift in SMC X-1, which smears the profile over 
long times.}

\figcaption{Images of selected category B sources: (a) 3C 273 (20-50 keV),
an AGN, (b) 4U 1812-12 (Ser X-2) (20-50 keV), a LMXB, (c) PSR 1509-58 
(50-100 keV), a SNR, and (d) EXO 0748-676 (20-50 keV), a LMXB.}

\figcaption{Broadband spectra for (a) the LMXB Cir X-1 and (b) 
the SNR PSR 1509-58 from the JPL EBOP (open squares)
and MSFC EOT measurements (filled circles) for Truncated Julian Dates
8393-8800.  For comparison, we also show the 9.1y average 4-band fluxes from
Table 4, with systematic errors applied, and converted to photon fluxes
using the multipliers in Table B2.}

\newpage
\plotone{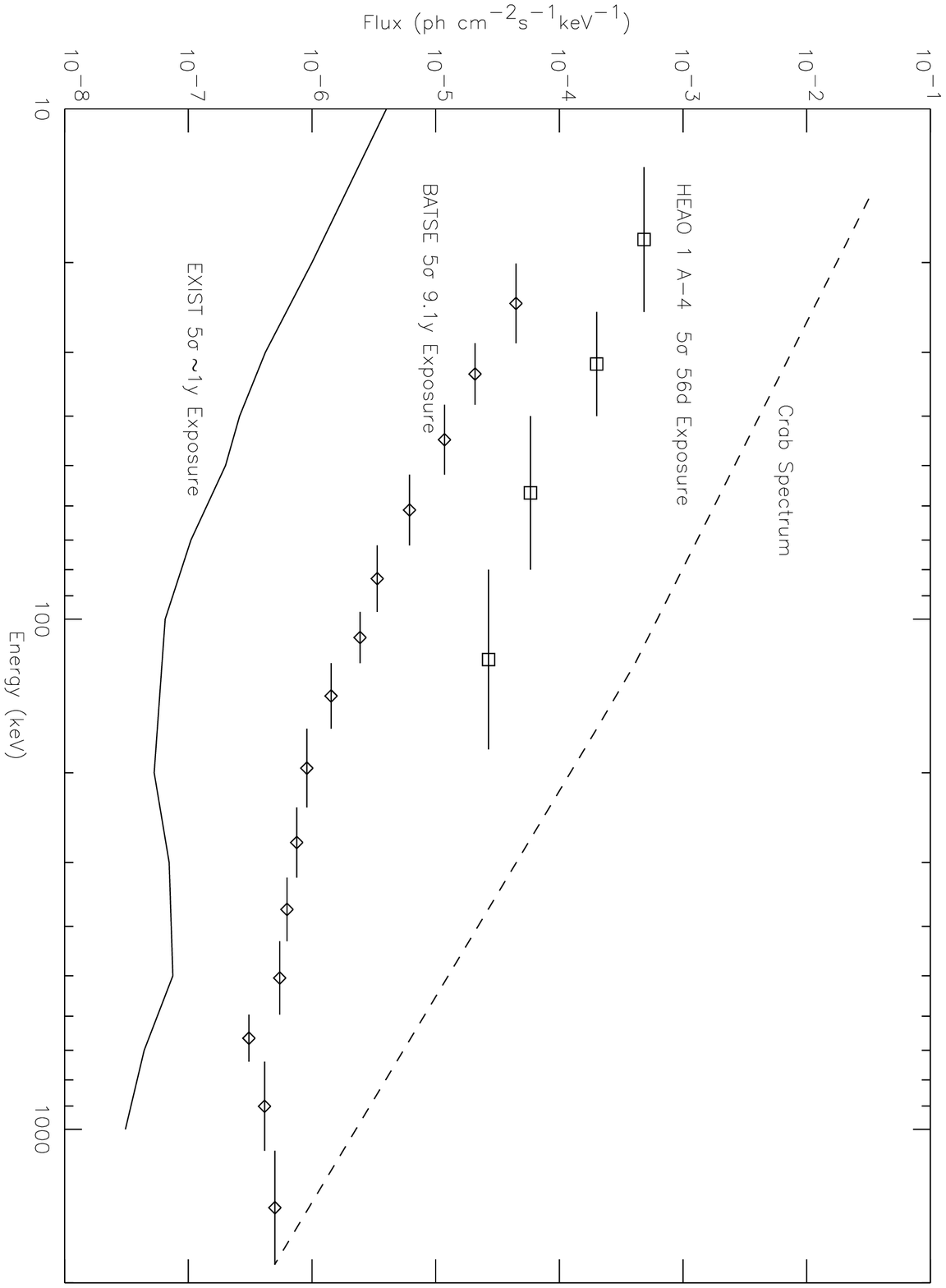}

\newpage
\epsscale{0.8}
\plotone{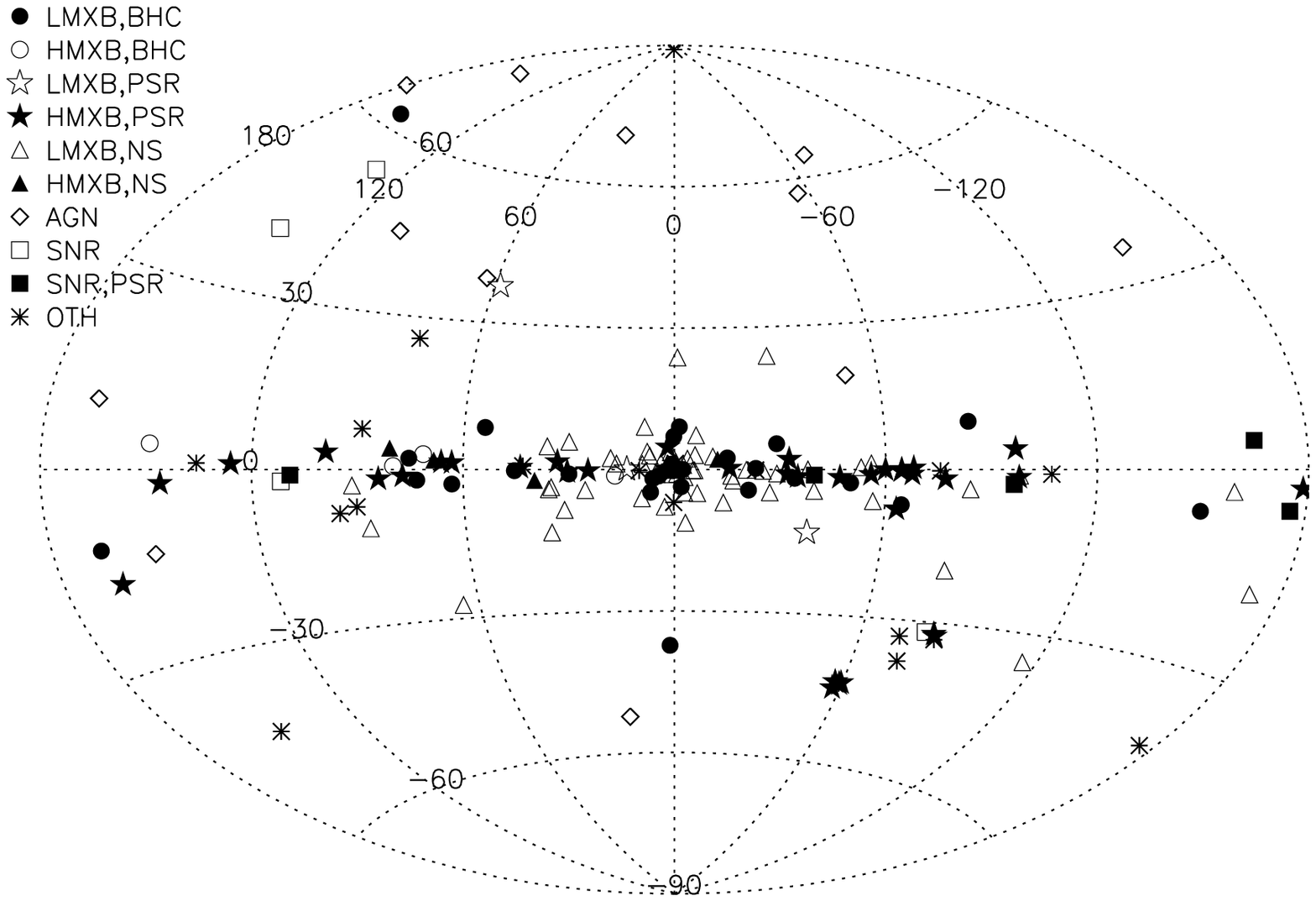}
\plotone{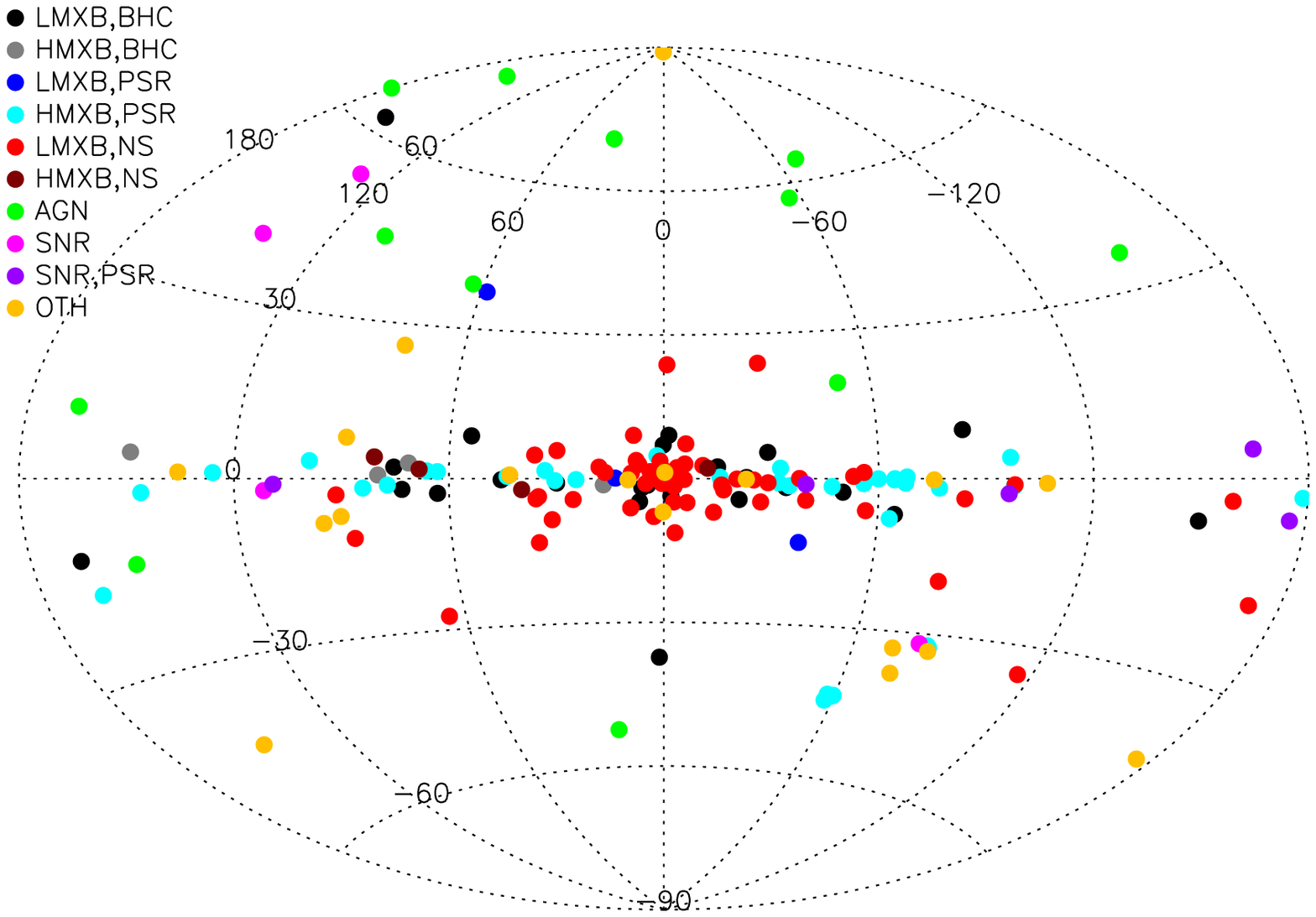}

\newpage
\plotone{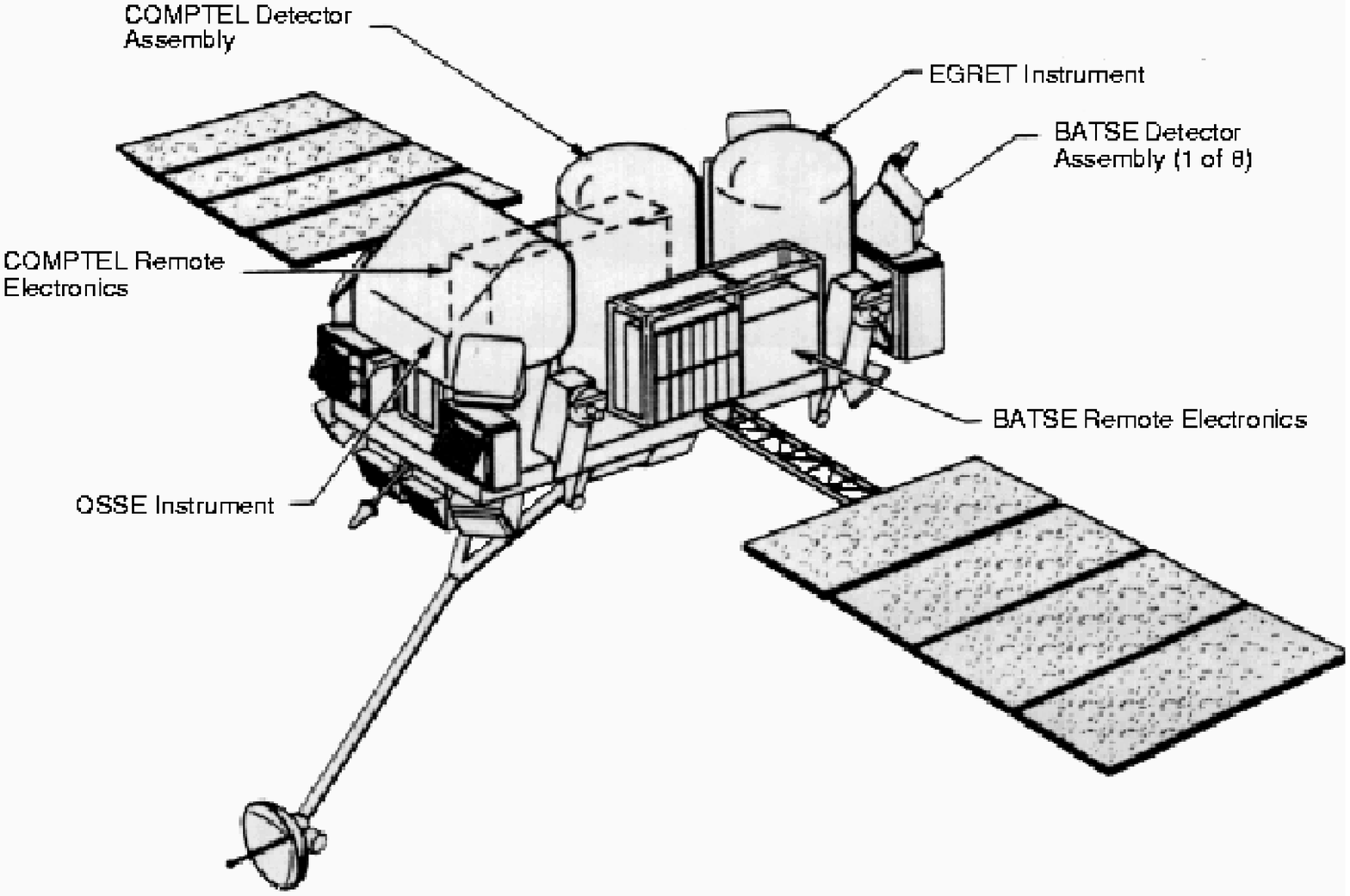}
\plotone{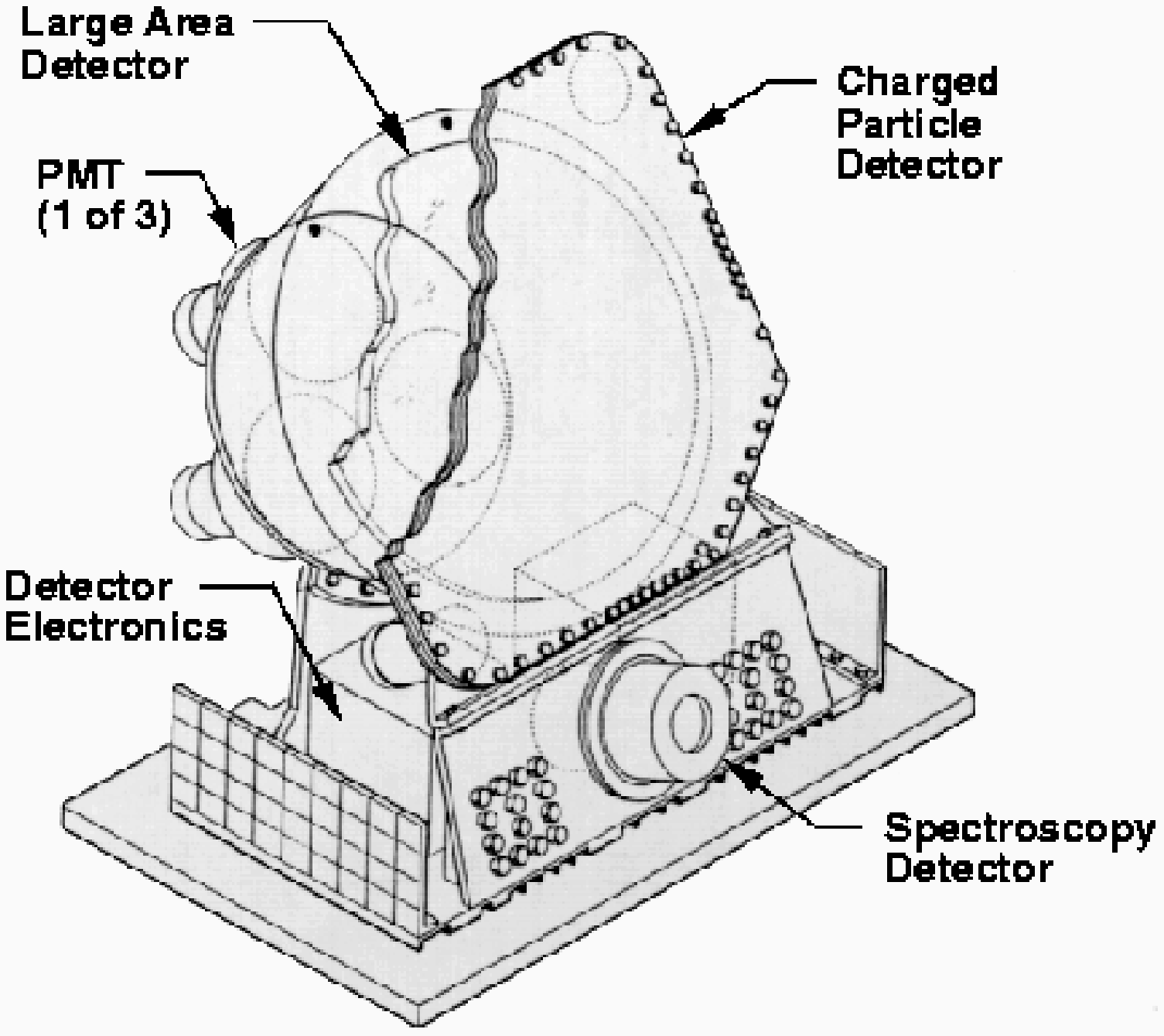}

\newpage
\epsscale{1.0}
\plotone{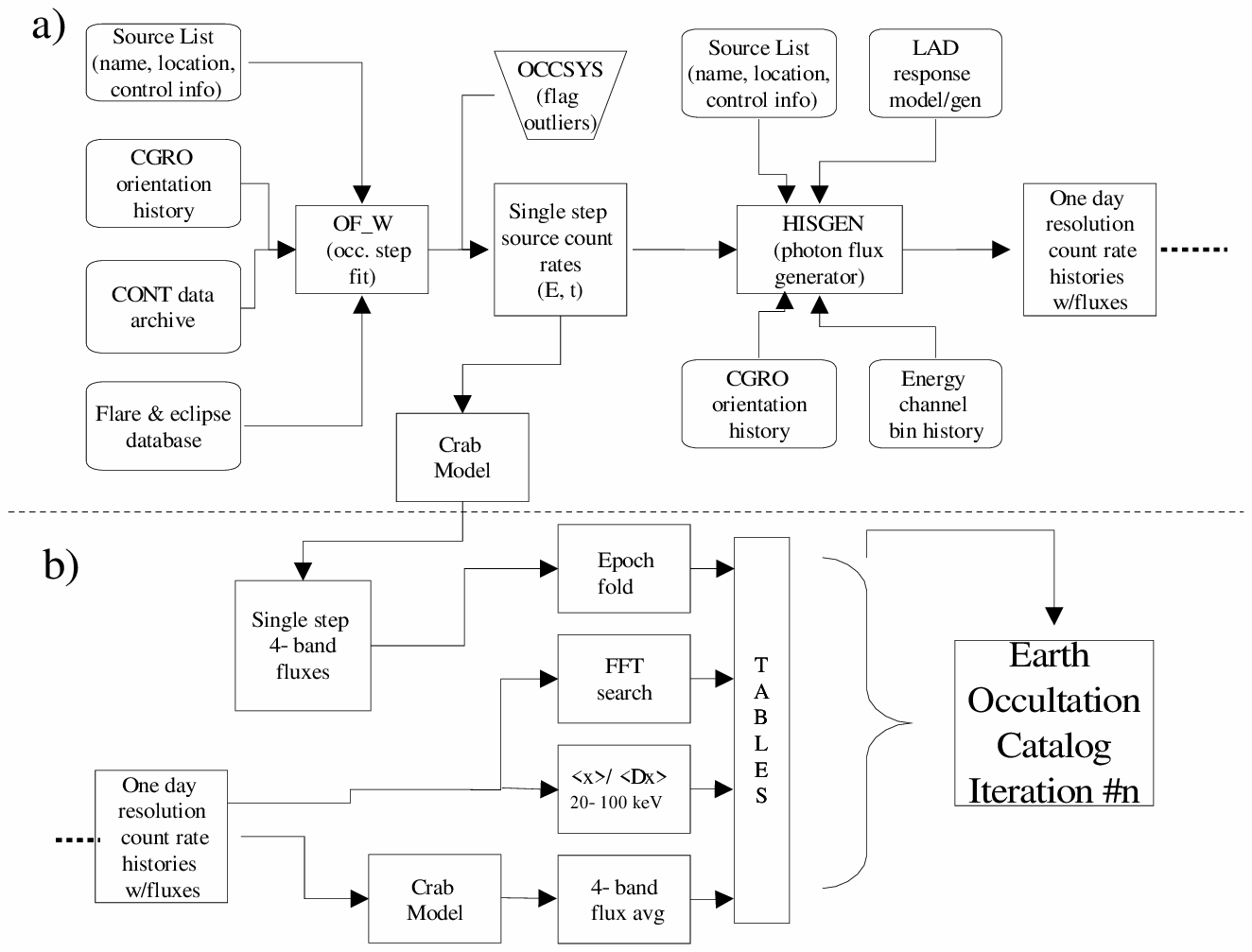}

\newpage
\plotone{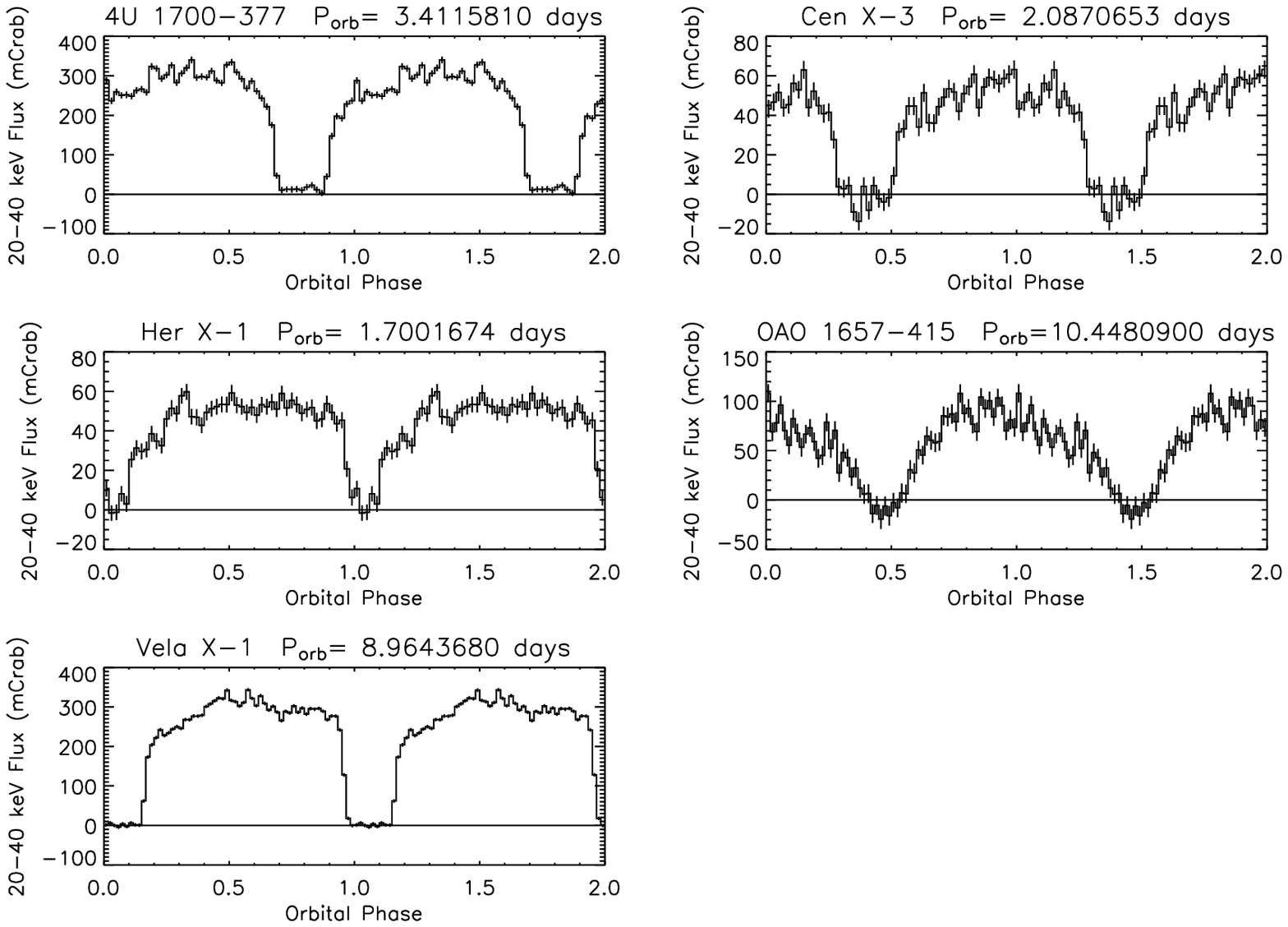}

\newpage
\plotone{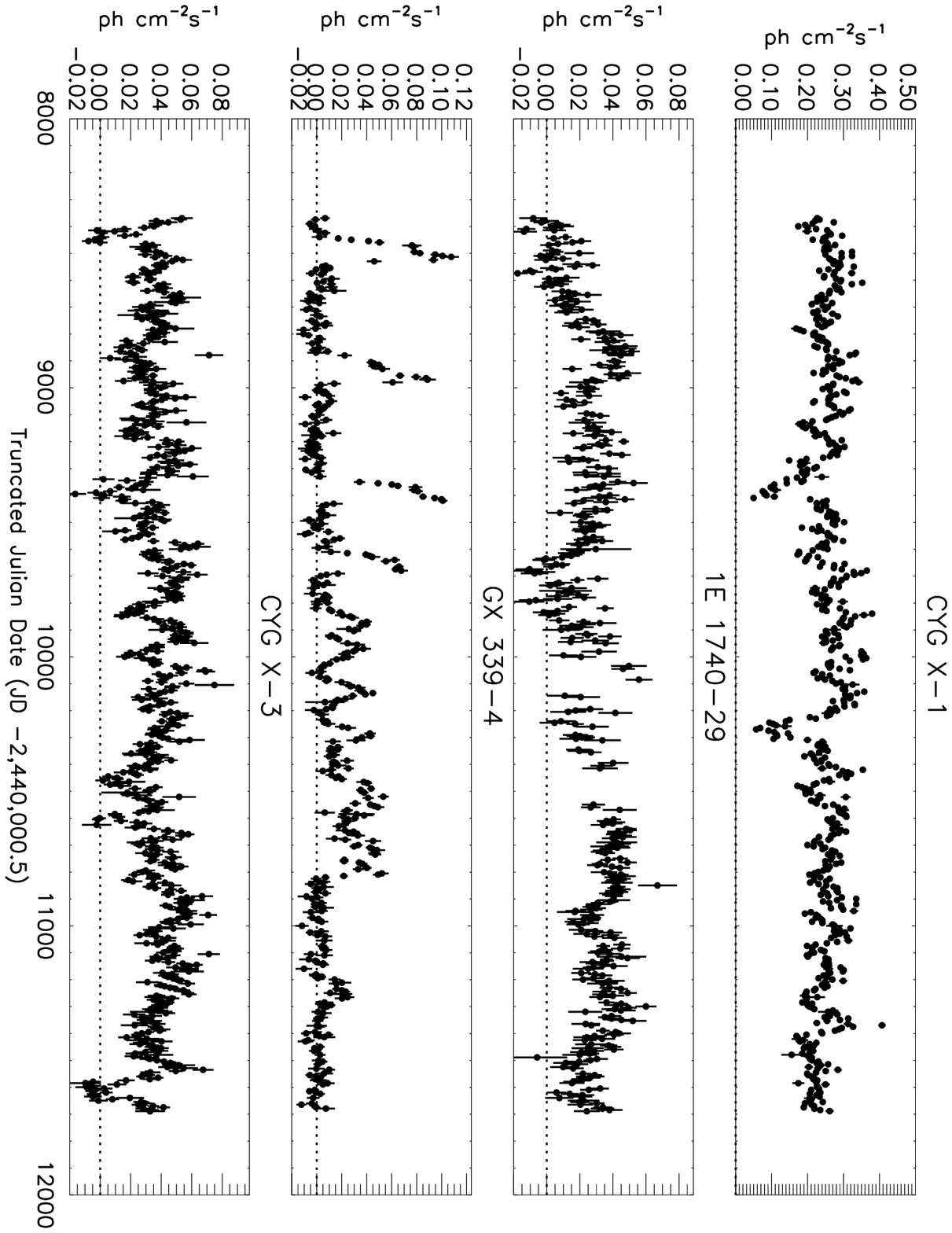}

\newpage
\plotone{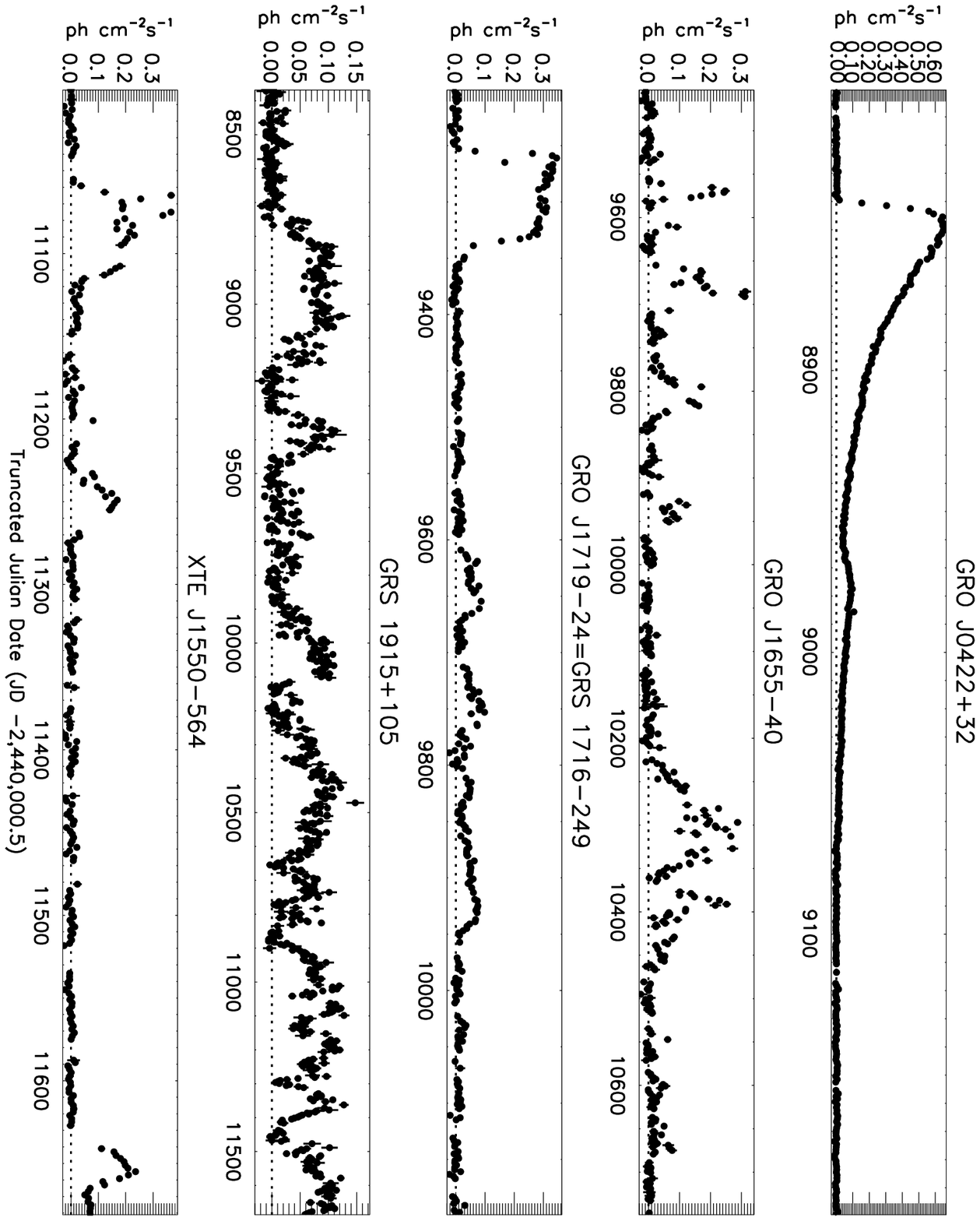}

\newpage
\plotone{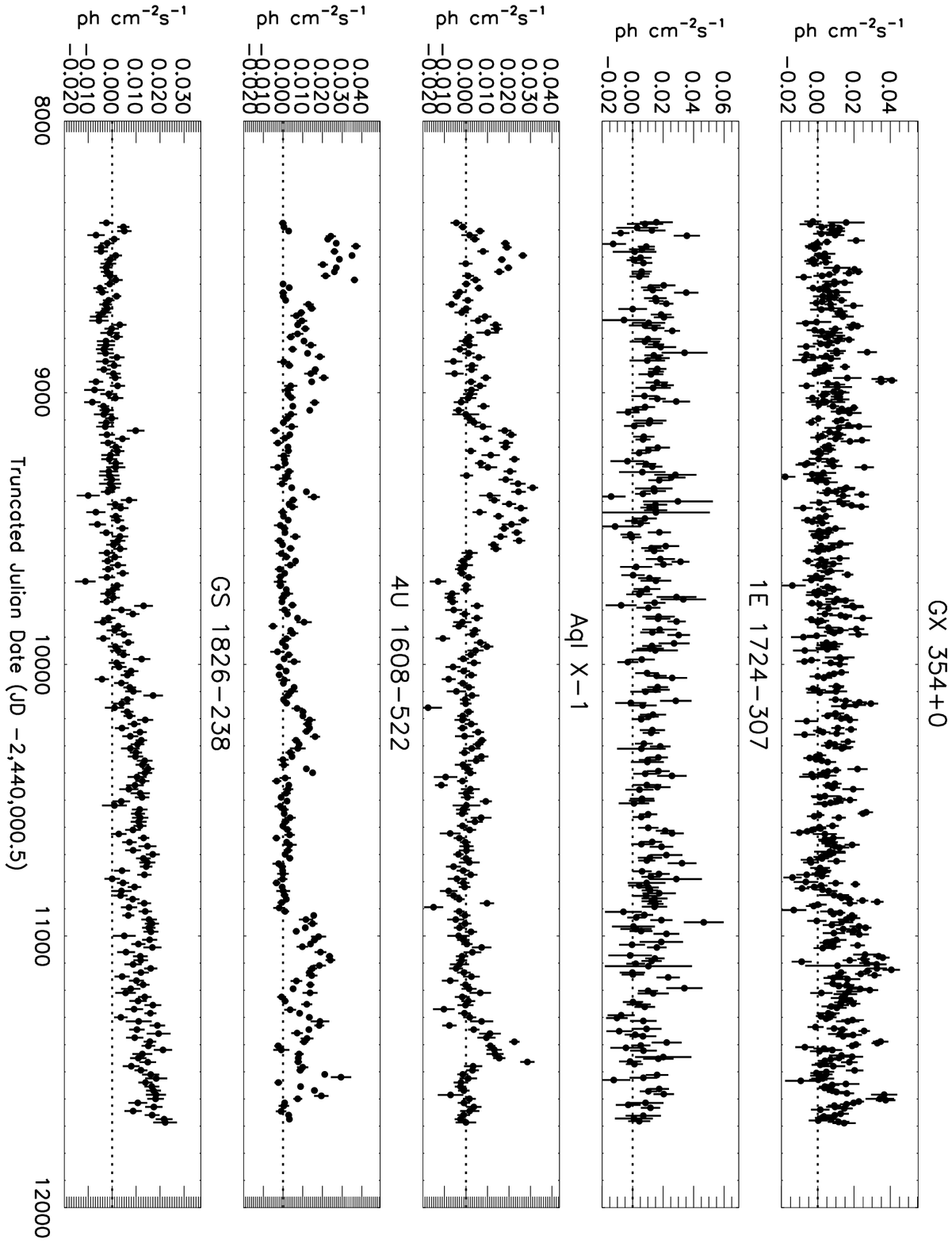}

\newpage
\epsscale{0.8}
\plotone{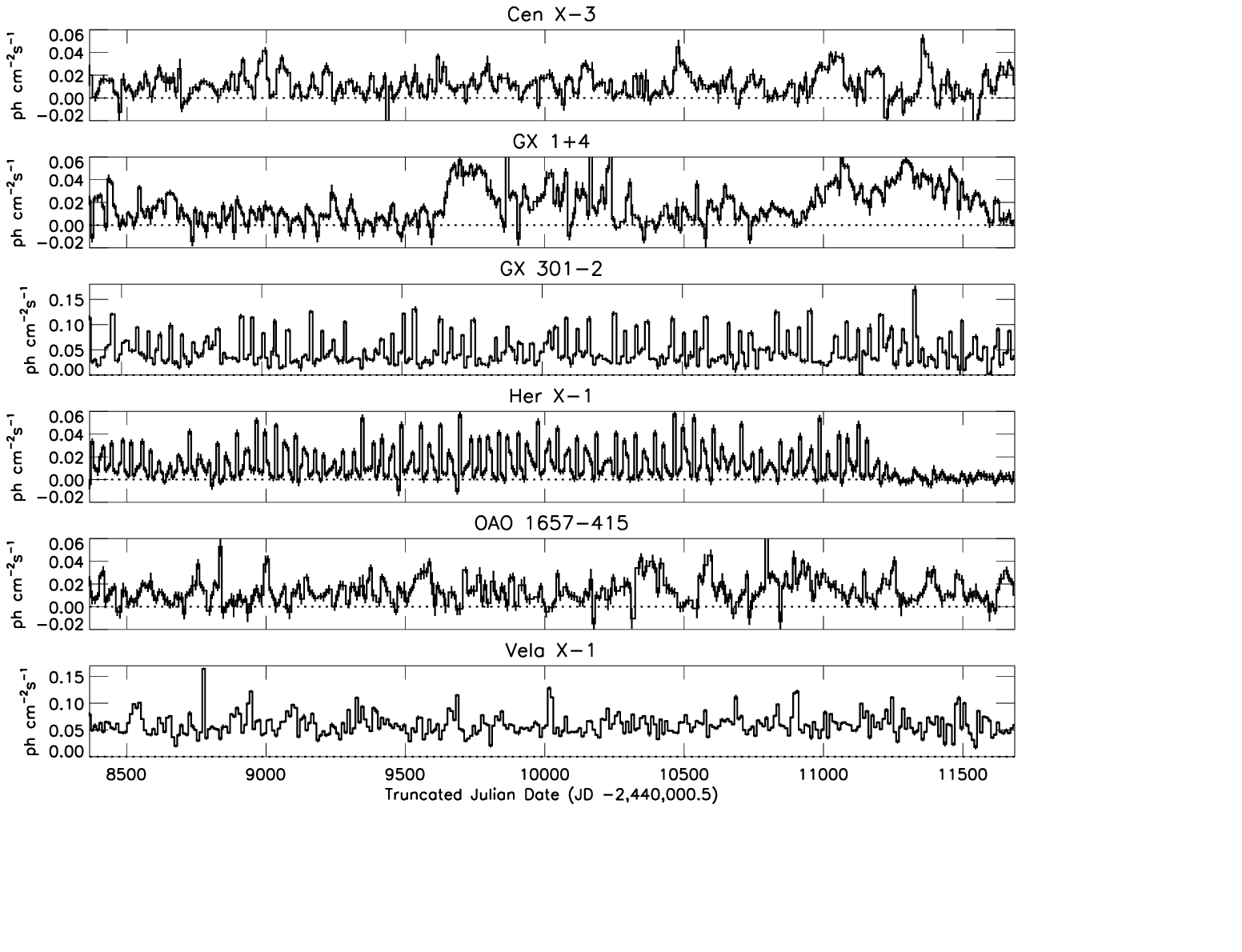}
\plotone{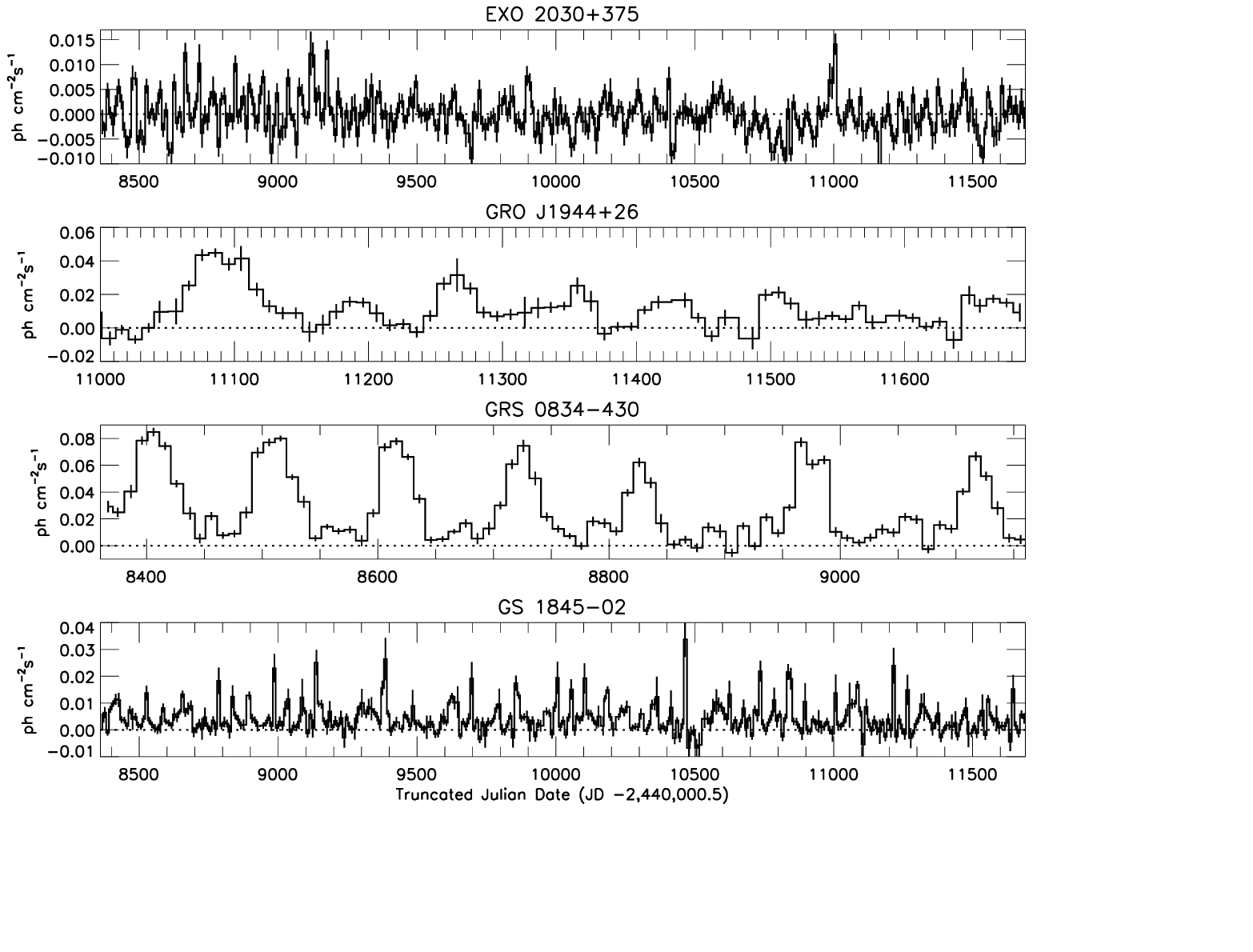}

%\newpage
%\plotone{f9b.ps}

\newpage
\epsscale{1.0}
\plotone{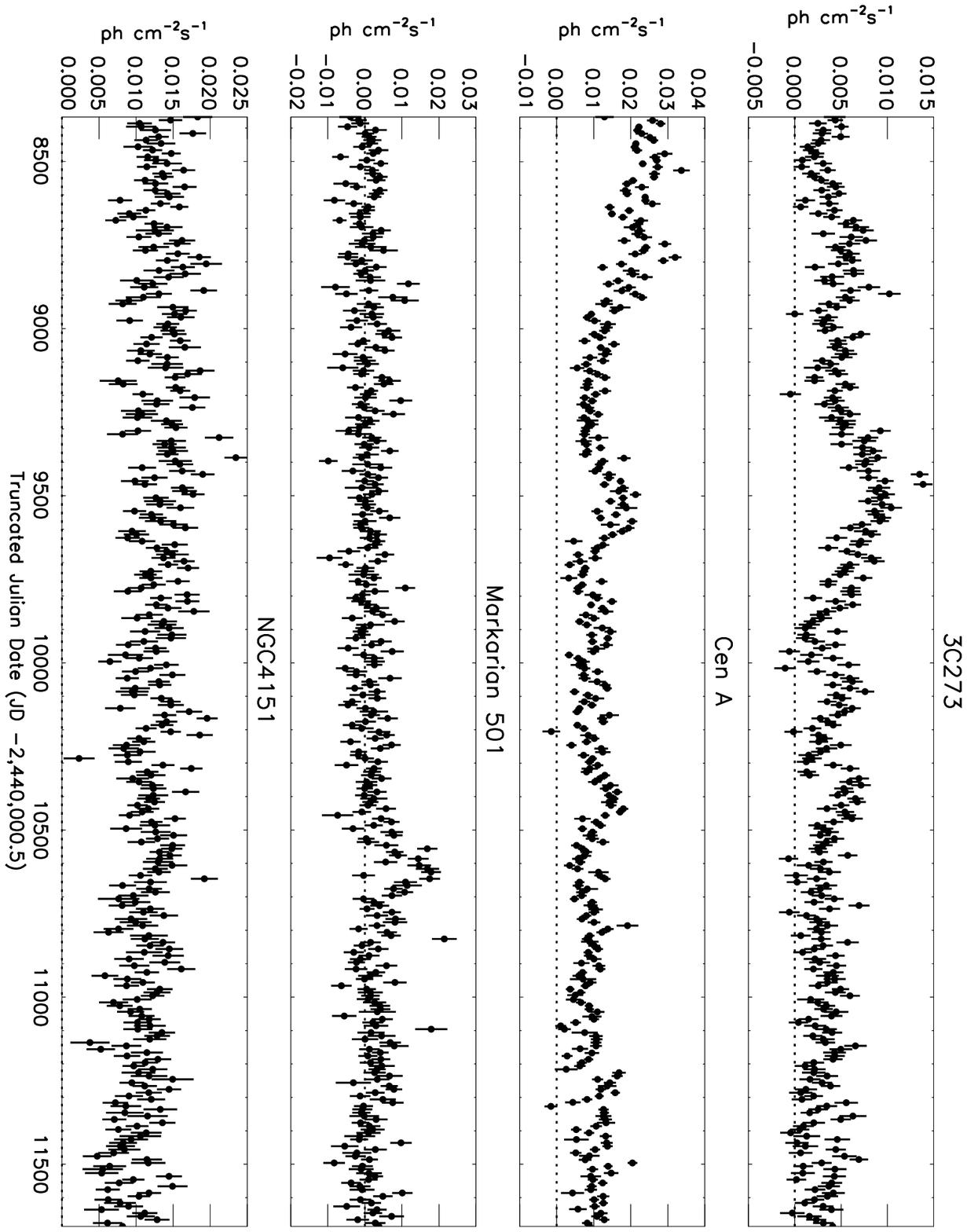}

\newpage
\plotone{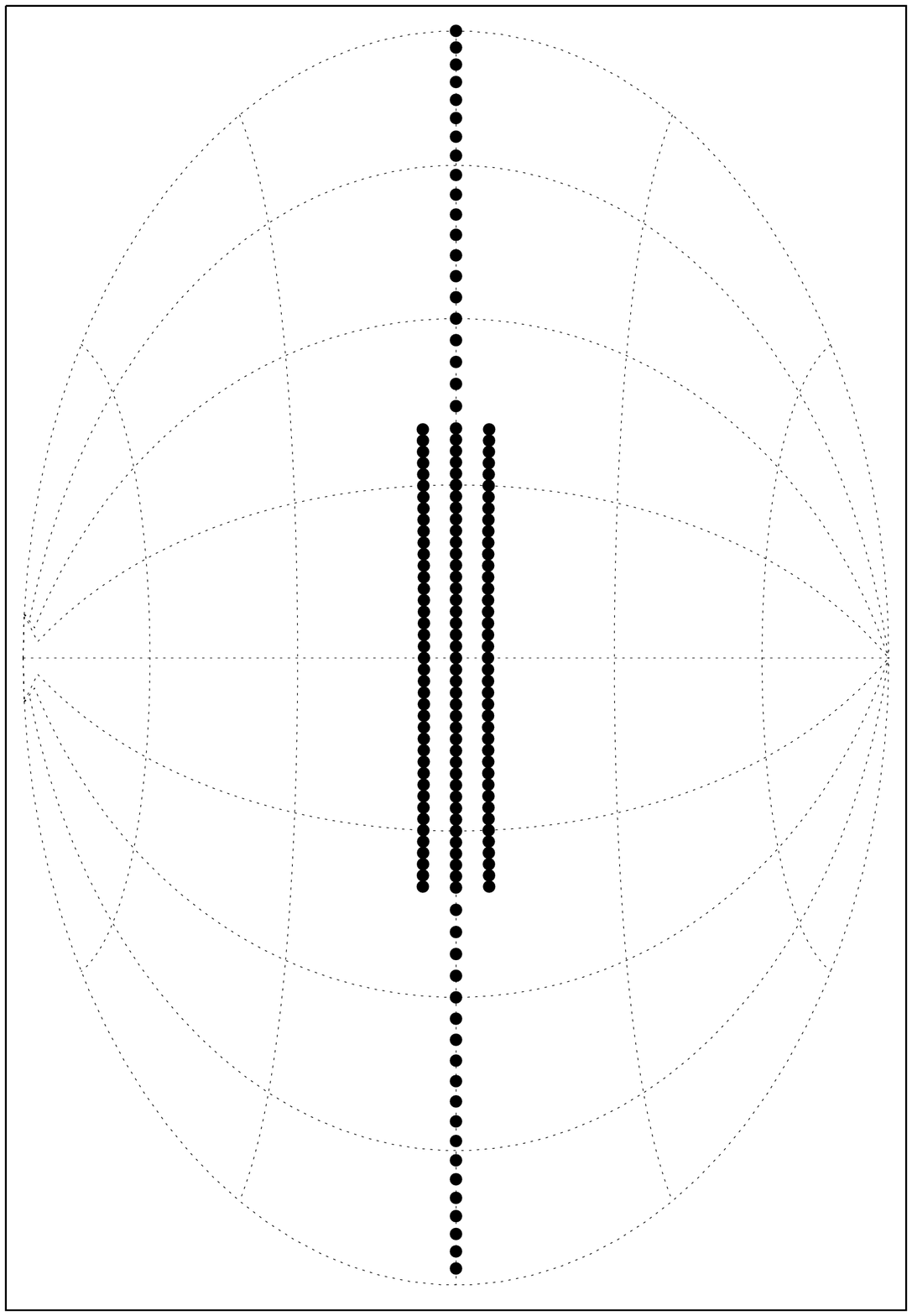}

\newpage
\plotone{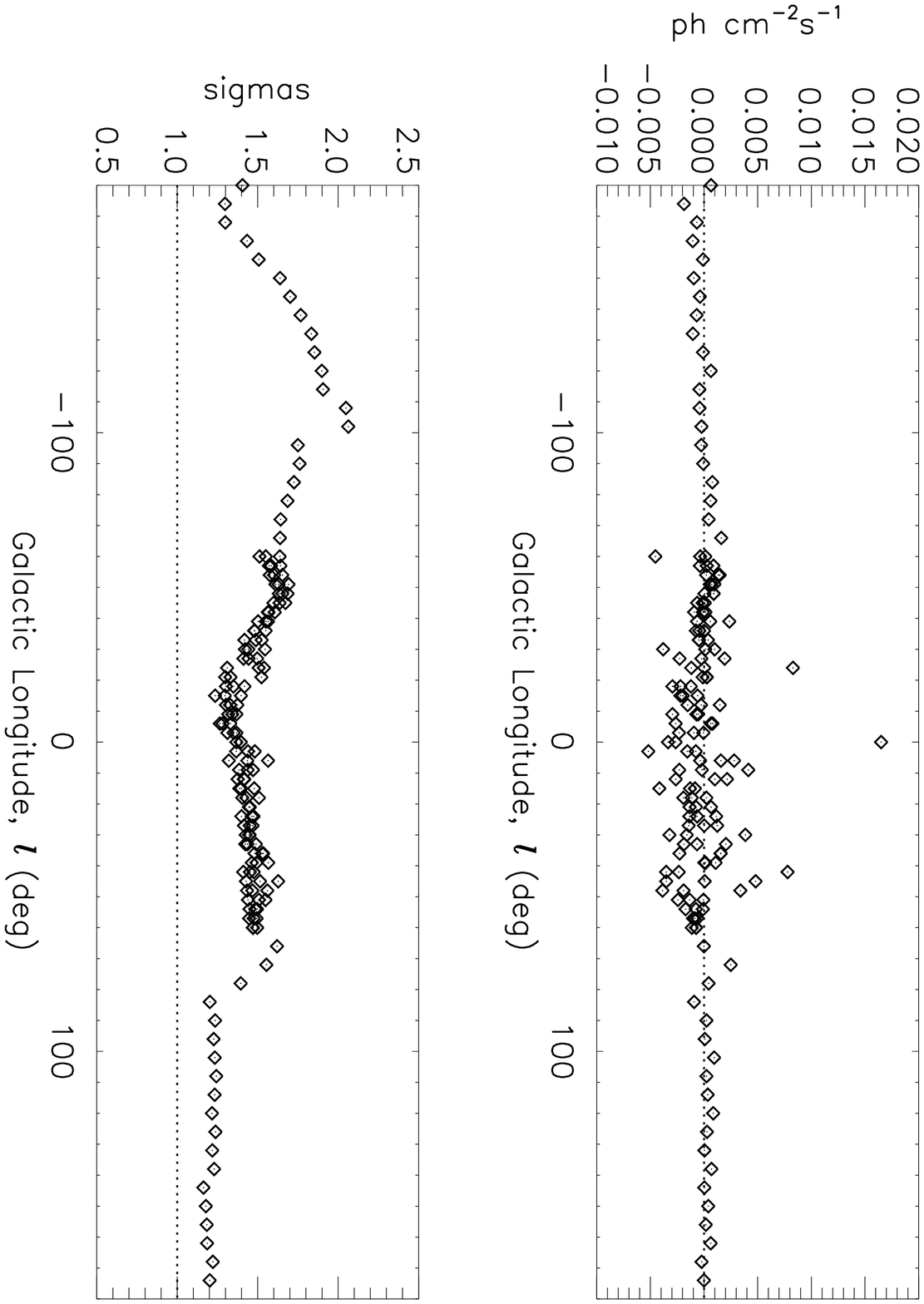}

\newpage
\plotone{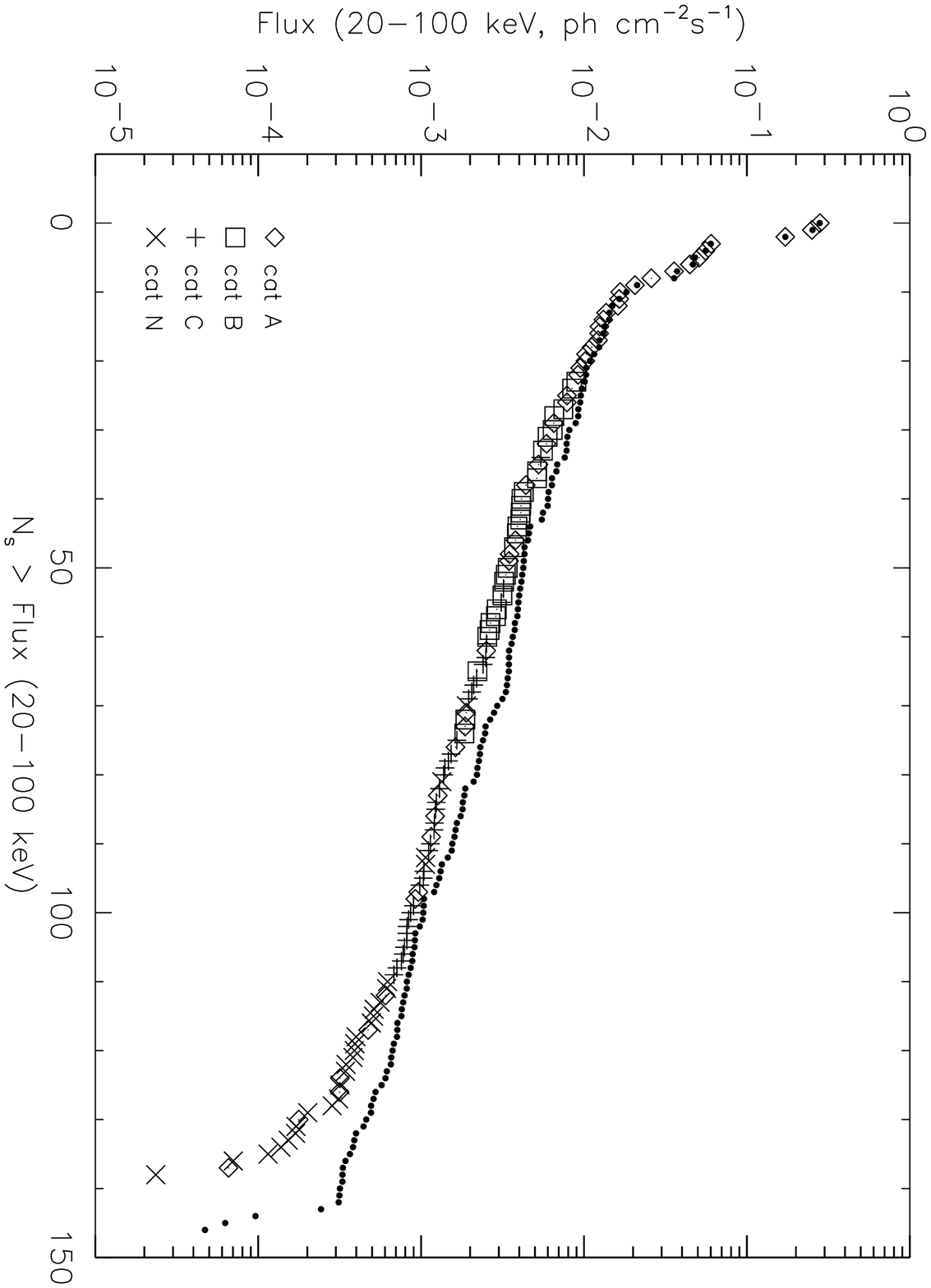}

\newpage
\plotone{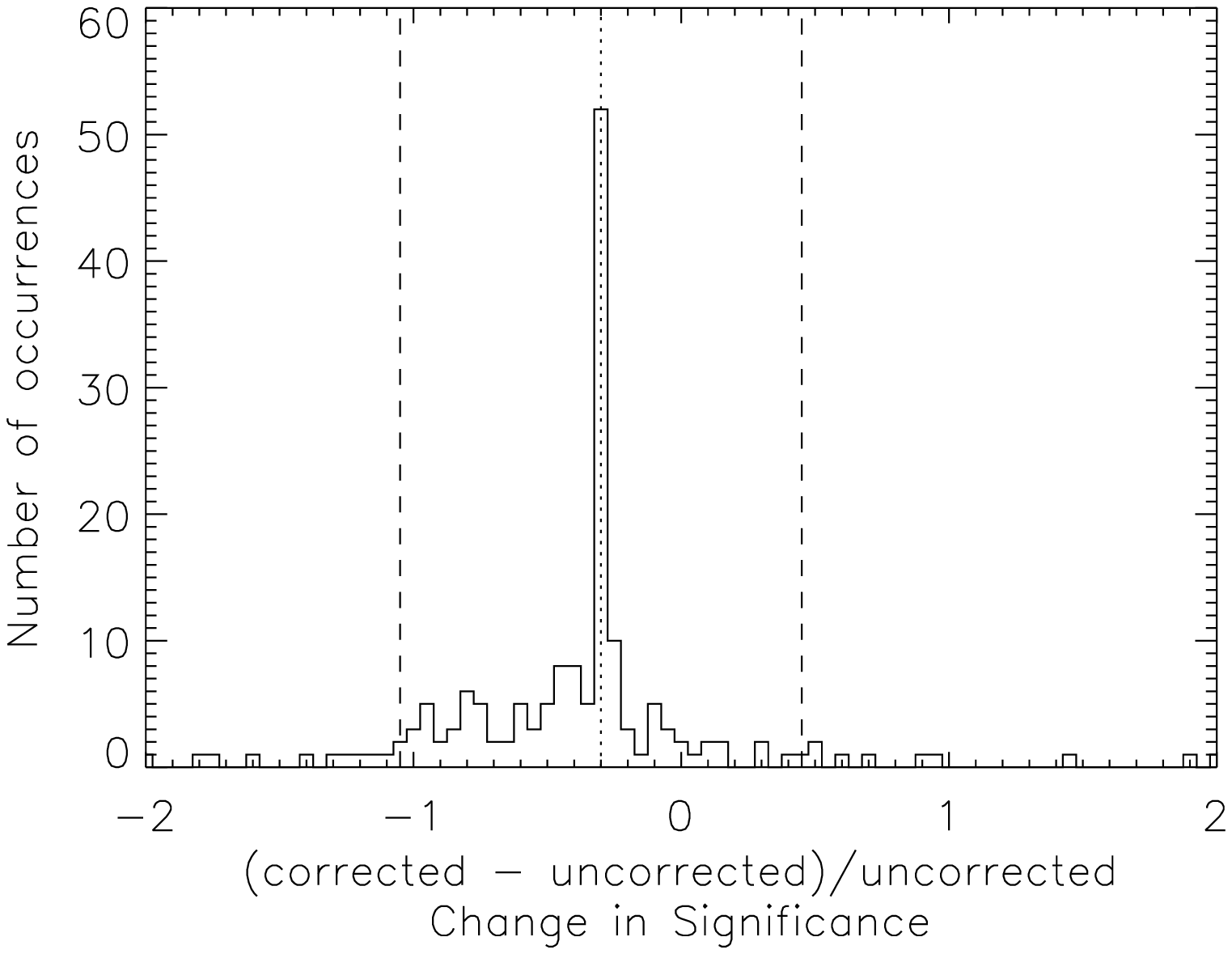}

\newpage
\plotone{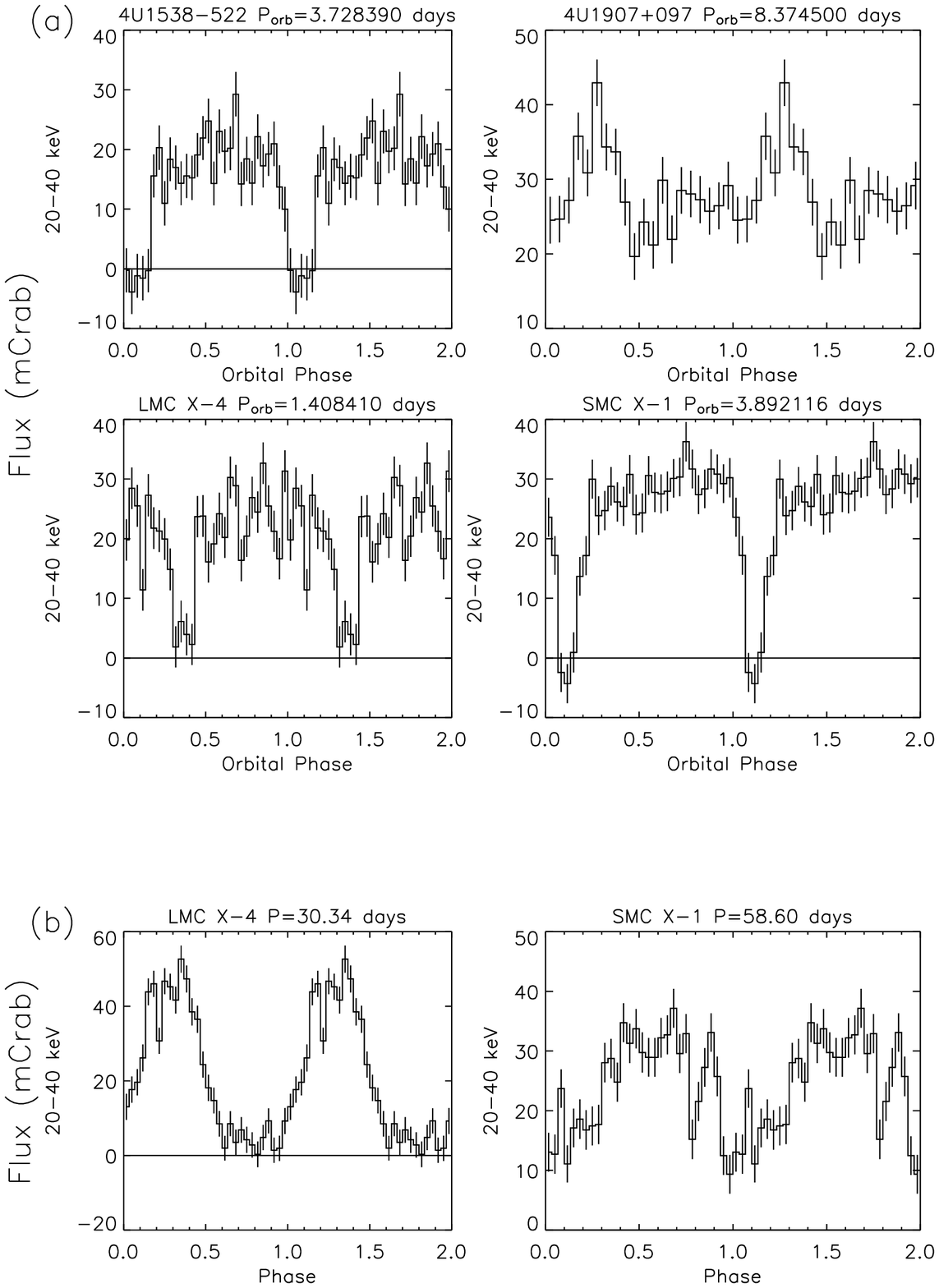}

\newpage
\plottwo{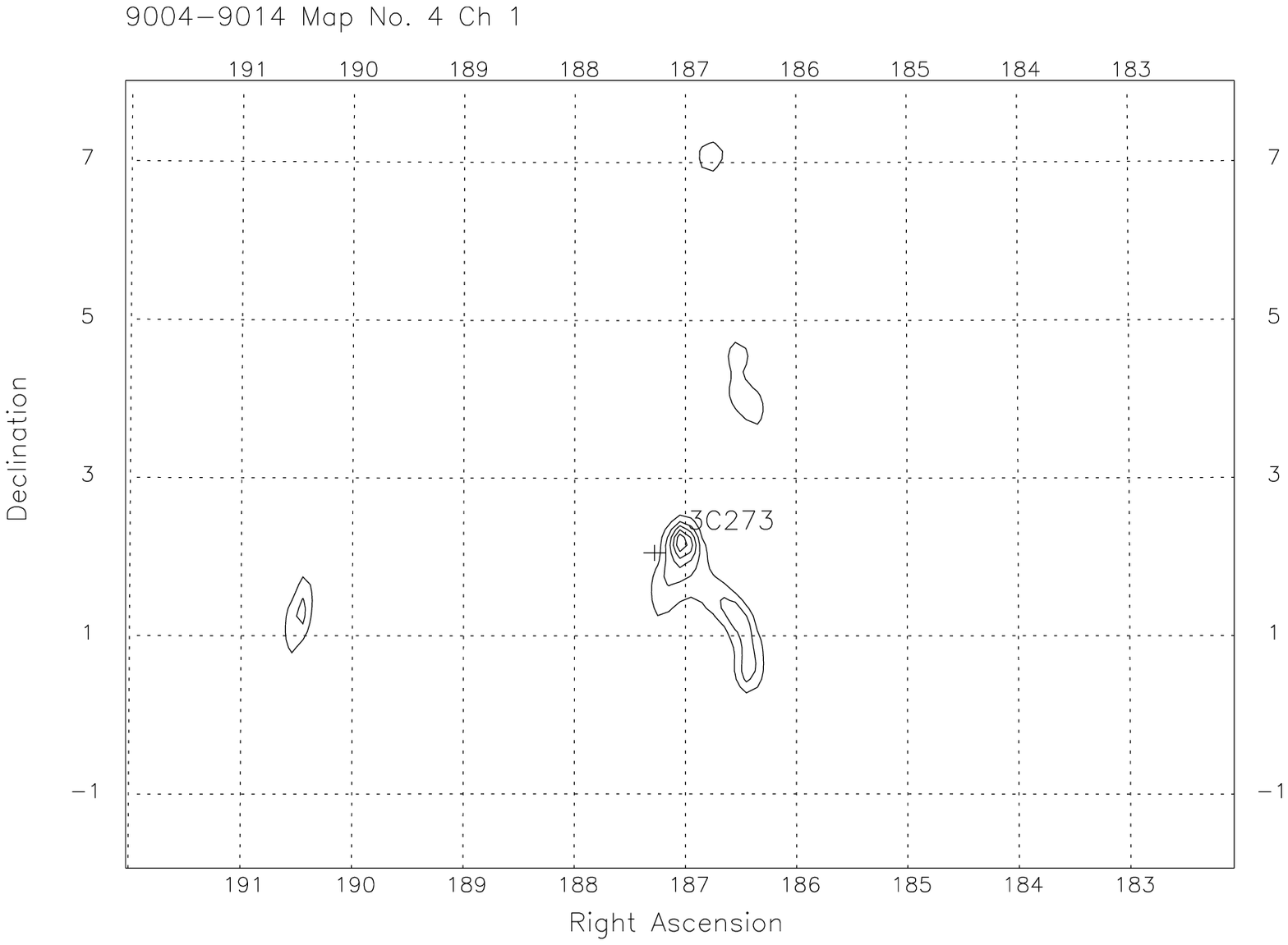}{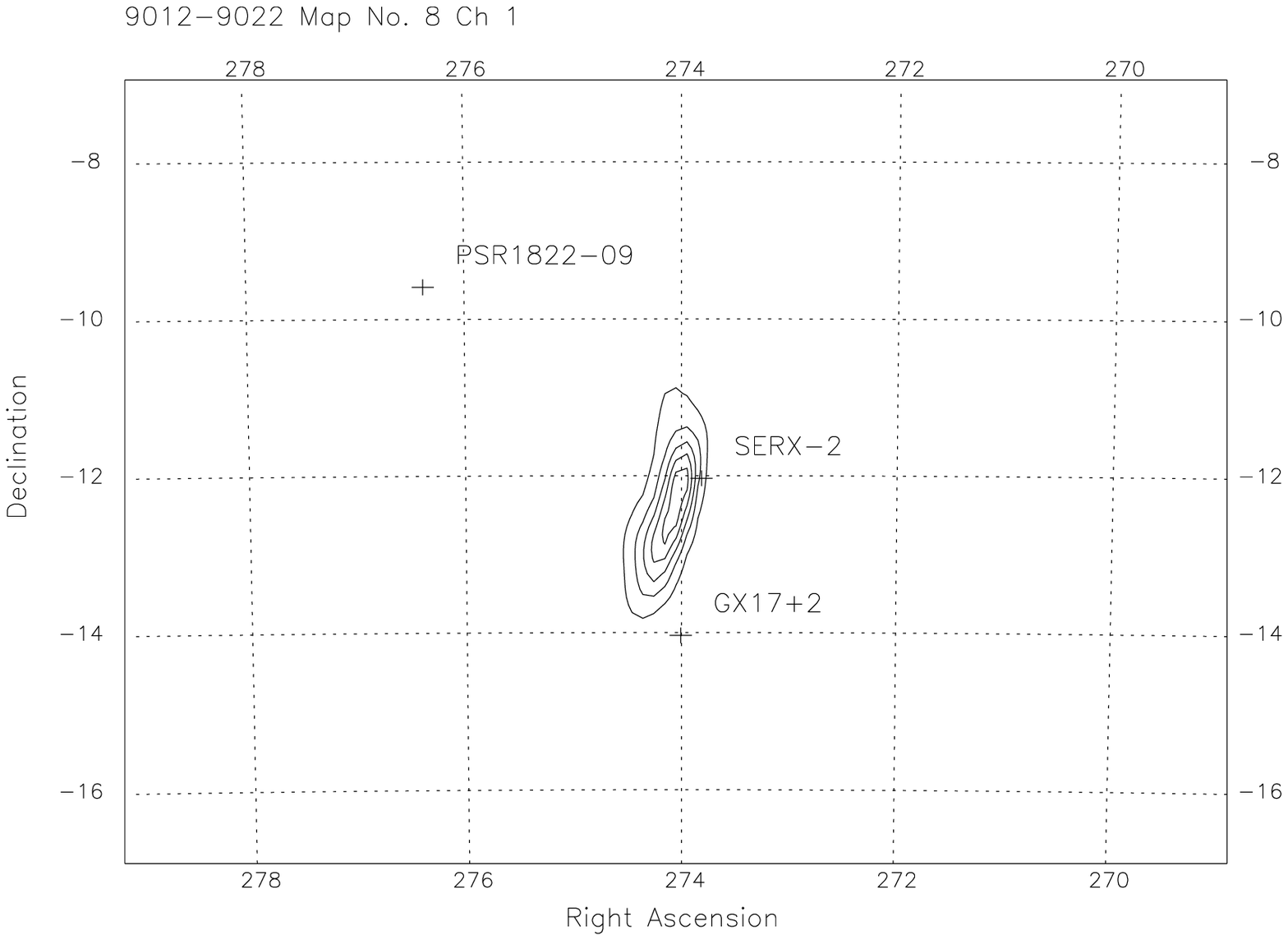}

%\newpage
%\plotone{f16b.ps}

\epsscale{2.2}
\plottwo{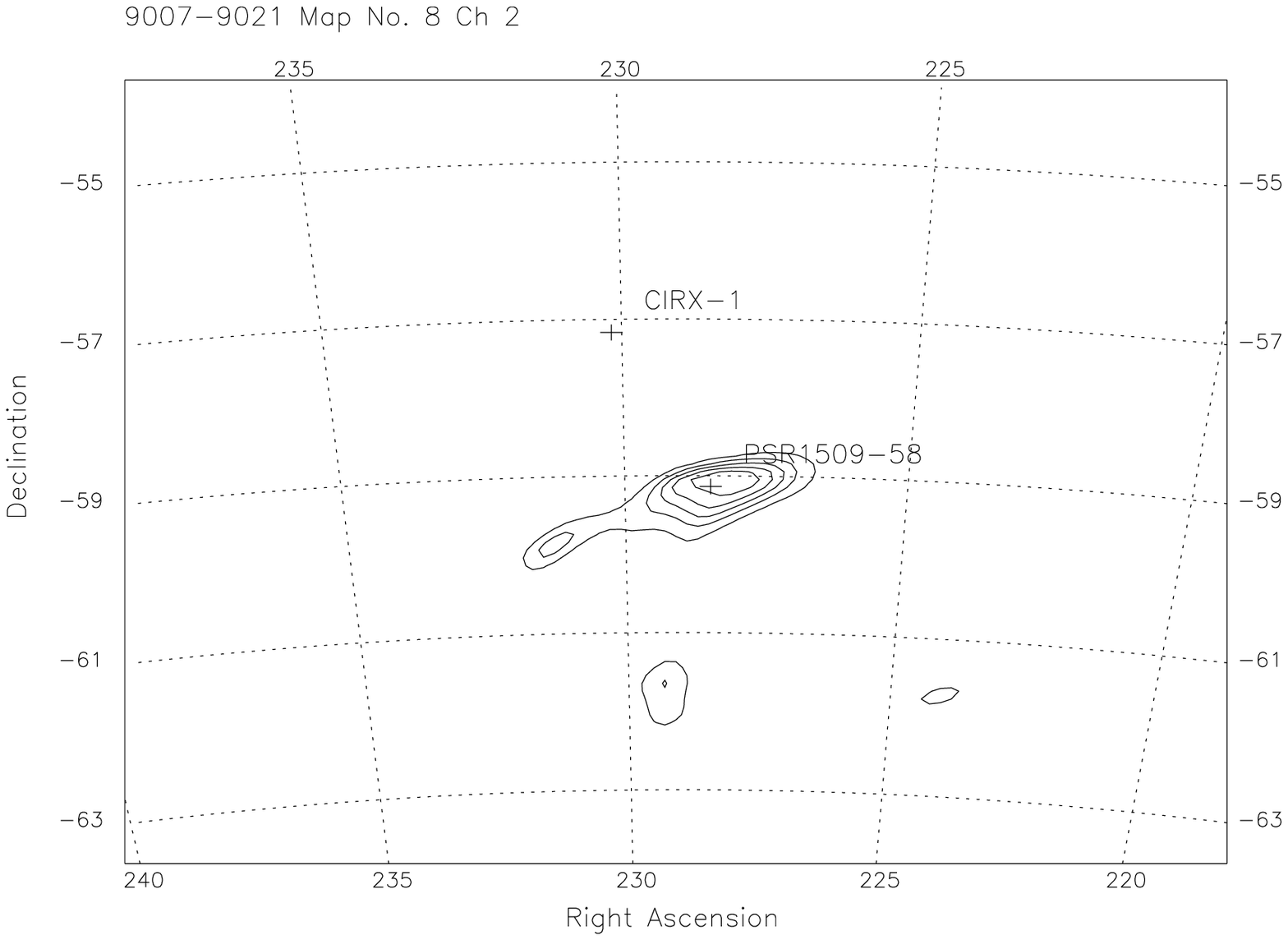}{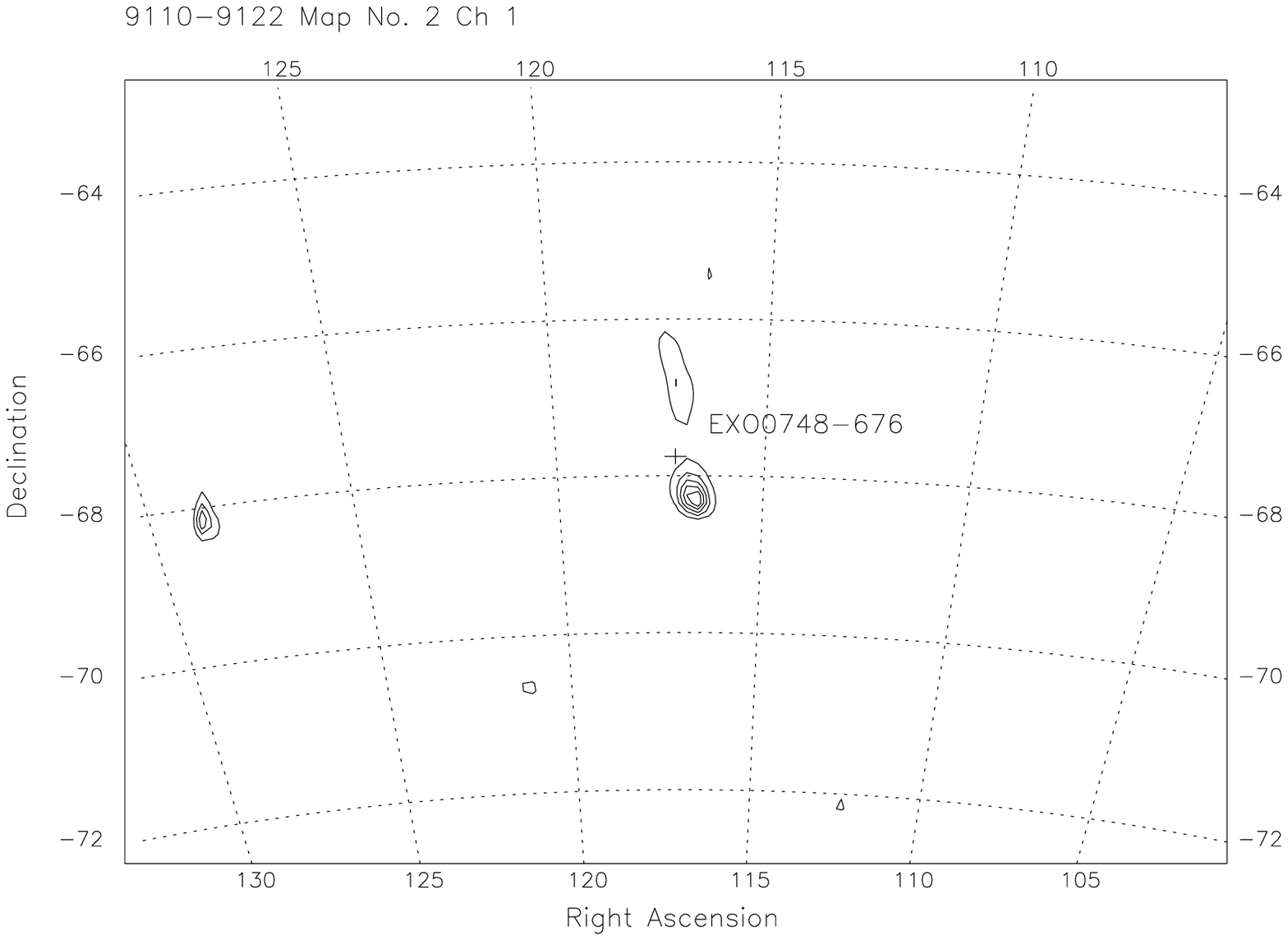}

%\newpage
%\plotone{f16d.ps}

\newpage
\epsscale{6.0}
\plottwo{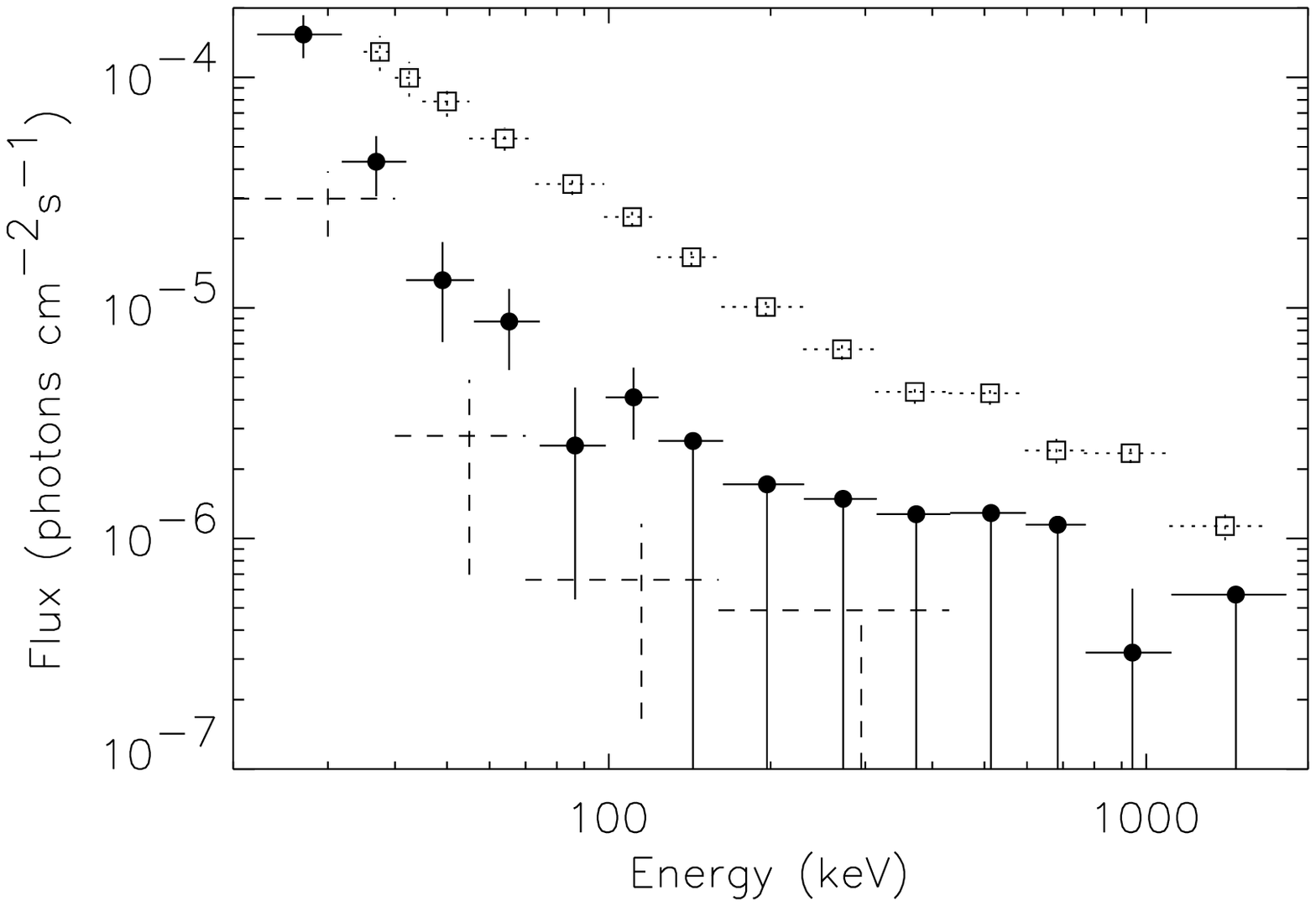}{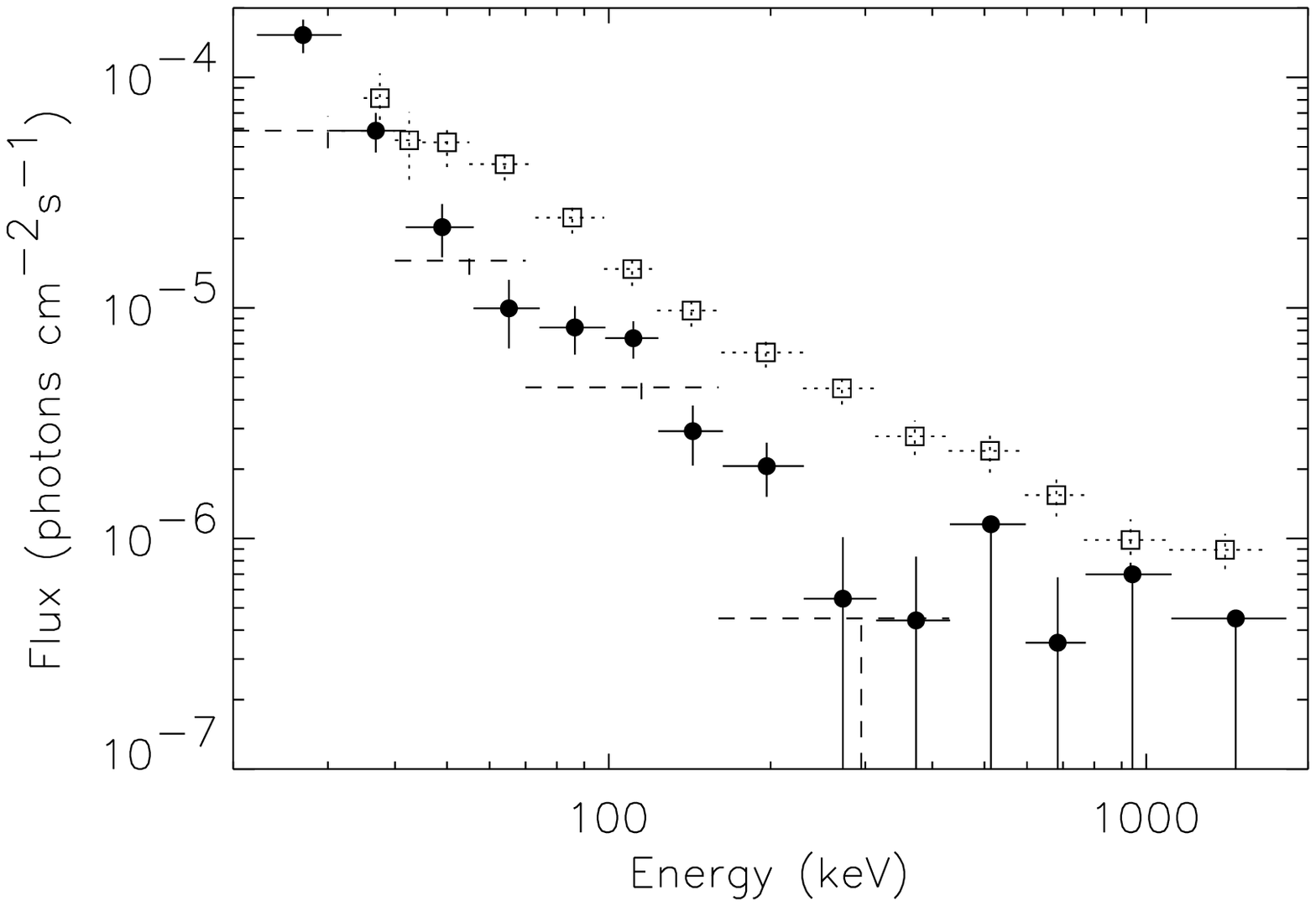}

%\newpage
%\plotone{f17b.ps}
\end{appendix}
\end{document}